\documentclass[11pt,a4paper]{article}
\pdfoutput=1

\usepackage{jheppub}
\usepackage{amsmath}
\usepackage{amssymb}
\usepackage{graphics}
\usepackage[active]{srcltx}
\usepackage{pdfsync}
\usepackage{shuffle}
\usepackage{slashed}
\usepackage{hyperref}
\usepackage{subfigure}

\newcommand{\bea}{\begin{eqnarray}}
\newcommand{\eea}{\end{eqnarray}}
\newcommand{\nn}{\nonumber \\}

\def\W #1{\widetilde{#1}}

\def\eref#1{(\ref{#1})}

\def\a{{\alpha}}

\def\b{{\beta}}

\allowdisplaybreaks

\title{Universal Interpretation of Hidden Zero and $2$-Split of Tree-Level Amplitudes Using Feynman Diagrams, Part $\mathbf{I}$: ${\rm Tr}(\phi^3)$, NLSM and YM}
\author[a]{Kang Zhou}

\affiliation[a]{Center for Gravitation and Cosmology, College of Physical Science and Technology, Yangzhou University,\\
No.180, Siwangting Road, Yangzhou, 225009, P.R. China}

\emailAdd{zhoukang@yzu.edu.cn}

\date{\today}
\abstract{In this paper, we propose a universal diagrammatic interpretation of hidden zeros and $2$-splits of tree-level amplitudes. Originally developed for ${\rm Tr}(\phi^3)$ amplitudes in our previous work, this interpretation is now extended to tree-level amplitudes in Nonlinear sigma model (NLSM) and Yang-Mills (YM) theories. The interpretation is based on a certain factorization behavior of Feynman diagrams under specific kinematic constraints, which we term shuffle factorization along a specific line (SFASL). This mechanism allows us to separate Feynman diagrams along specific lines after summing over shuffle permutations. When applied to NLSM and YM amplitudes, we perform proper extensions of the SFASL used in the ${\rm Tr}(\phi^3)$ case. Through the SFASL, the interpretation for the hidden zeros and $2$-splits of tree amplitudes of ${\rm Tr}(\phi^3)$, NLSM, and YM can be unified as: the hidden zeros are ascribed to the on-shell condition $k_j^2=0$ of a massless particle, while the $2$-splits are caused by separating each Feynman diagram along two lines, akin to unzipping two zippers.
}


\begin{document}

\maketitle \flushbottom

\section{Introduction}
\label{sec-intro}

In recent research on scattering amplitudes, the development of the surfaceology framework \cite{Arkani-Hamed:2023lbd,Arkani-Hamed:2023mvg,Arkani-Hamed:2023jry,Arkani-Hamed:2024nhp,Arkani-Hamed:2024vna,
Arkani-Hamed:2024yvu,Arkani-Hamed:2024tzl,Arkani-Hamed:2024nzc,Arkani-Hamed:2024pzc,Backus:2025njt}, along with the discovery of  novel properties of tree-level amplitudes---namely, hidden zeros and new factorizations near these zeros---from surfaceology, represents significant progress. In \cite{Arkani-Hamed:2023swr}, it was first shown via the kinematic mesh and surfaceology that tree-level amplitudes for ${\rm Tr}(\phi^3)$, NLSM, and YM models vanish on special loci in kinematic space---a property termed hidden zeros. Upon appropriately relaxing the kinematic conditions, a new factorization behavior emerges near these zeros, where each tree amplitude decomposes into three amputated currents, without taking residue at a pole. In subsequent work, hidden zeros have been extended to tree amplitudes in a broader range of models, including gravity (GR), special Galileon (SG), Dirac-Born-Infeld (DBI), as well as YM and GR with specific higher-derivative corrections \cite{Rodina:2024yfc,Bartsch:2024amu,Li:2024qfp,Zhang:2024efe,Huang:2025blb,Zhou:2025tvq}. Furthermore, another more fundamental factorization behavior near hidden zeros, where the amplitude decomposes into two currents, has been discovered and is referred to as $2$-split \cite{Cao:2024gln,Cao:2024qpp,Arkani-Hamed:2024fyd}. The previously observed three-current factorization in \cite{Arkani-Hamed:2023swr} can be derived from this $2$-split. The $2$-split behavior also applies to the aforementioned series of models.
For further research on hidden zeros and $2$-split, see \cite{Guevara:2024nxd,Cao:2025hio,Zhang:2025zjx,Zhang:2026dcm,Saha:2026ftv,Azevedo:2025vxo,
CarrilloGonzalez:2026lnu,Zhou:2024ddy,Feng:2025ofq,Feng:2025dci,Li:2025suo,Zhou:2025xly,Zhou:2026isc}.

Despite being mathematically flawless, our understanding of hidden zeros and $2$-split is far less thorough and conceptual than our understanding of traditional factorization on poles, which is dictated by locality and unitarity. Therefore, exploring hidden zeros and $2$-split from various perspectives and delving deeper into their physical picture constitutes an important direction. This paper represents an attempt in this direction.

Current existing interpretations of hidden zeros and $2$-splits mainly include the following: those based on the kinematic mesh and surfacology \cite{Arkani-Hamed:2023swr}; those based on the Cachazo-He-Yuan (CHY) formalism \cite{Cao:2024gln,Cao:2024qpp,Zhang:2024efe}; those based on the BCFW on-shell recursion relations \cite{Feng:2025ofq}; and those based on the universal expansions of amplitudes \cite{Huang:2025blb,Feng:2025dci,Zhou:2025tvq,Zhang:2026dcm}. This work, however, focuses on another interpretation developed in our previous work \cite{Zhou:2024ddy,Feng:2025dci} based on Feynman diagrams. This interpretation relies on a special factorization pattern exhibited by Feynman diagrams during the summation process, which we refer to as shuffle factorization along a specific line (SFASL)\footnote{A similar shuffle factorization also appears in recent studies of cosmological wavefunctions \cite{Li:2026gns}.}. This mechanism causes each Feynman diagram to be separated along this certain line, much like unzipping a zipper. In our previous work \cite{Zhou:2024ddy}, we used the SFASL to interpret the hidden zeros and $2$-split of tree ${\rm Tr}(\phi^3)$ amplitudes. In this paper, we extend this interpretation to the NLSM and YM cases\footnote{In \cite{Zhou:2024ddy}, we also attempted to use the SFASL to interpret the hidden zeros and $2$-split of YM amplitudes. However, that attempt was not successful. In particular, the treatment of the quartic vertices of YM theory was incorrect in \cite{Zhou:2024ddy}. Therefore, this paper also includes corrections to part of that work.}.

The motivation for extending this SFASL-based interpretation of hidden zeros and $2$-split to other models consists of the following three points:
\begin{itemize}
\item (1) Among all currently known hidden zeros and $2$-splits, the kinematic conditions for realizing $2$-splits can be obtained by slightly relaxing those for zeros. The high similarity of the kinematic conditions suggests that hidden zeros and $2$-splits may share the same underlying mechanism. The work of \cite{Zhou:2024ddy} precisely meets this expectation, as both are interpreted through an unique mechanism---SFASL. Therefore, it is natural to expect that the SFASL (with proper generalization) also applies to amplitudes of other physical models possessing hidden zeros and $2$-splits.
\item (2) When encountering a new physical model, if we wish to quickly determine whether its tree amplitudes exhibit hidden zeros and $2$-split, we need to understand the general requirements that hidden zeros and 2-split impose on the physical model. In practice, a physical model is more often presented in terms of its Lagrangian or Feynman rules rather than in forms such as the CHY formalism. The SFASL, being a mechanism based on Feynman diagrams and Feynman rules, is better suited than other interpretations for deriving constraints on Lagrangians and Feynman rules. If most of the known hidden zeros and $2$-splits can be interpreted by SFASL---not only in ${\rm Tr}(\phi^3)$ model---then, although the SFASL is a sufficient rather than necessary condition for the hidden zeros and $2$-splits, it nevertheless governs the hidden zeros and 2-splits of a large class of amplitudes. In this way, we can translate the problem into constraints imposed by SFASL on Lagrangians and Feynman rules, thereby deriving a general criterion for determining whether the amplitudes of a given physical model exhibit the hidden zeros and 2-split.
\item (3) In a recent work \cite{Zhou:2026ukg}, we extended the SFASL of ${\rm Tr}(\phi^3)$ diagrams at tree-level to the loop-level, and discovered the new hidden zeros and 2-split of ${\rm Tr}(\phi^3)$ Feynman integrands, which differ from those found in the literature at loop-level \cite{Backus:2025hpn,Arkani-Hamed:2024fyd}. A question then arises: given that the hidden zeros and $2$-split at tree-level apply to a wide class of models, can the aforementioned loop-level hidden zeros and $2$-split also be extended from ${\rm Tr}(\phi^3)$ to other models? Since we uncovered the loop-level hidden zeros and $2$-split through the SFASL, extending them to other models via the SFASL is the most direct and natural approach. However, applying the SFASL to other models at loop-level presupposes that the SFASL is satisfied by those models at tree-level.
\end{itemize}

Based on the above motivations, in this paper we generalize the SFASL to Feynman diagrams of NLSM and YM. In ${\rm Tr}(\phi^3)$, every vertex is cubic and provides a trivial constant. When extending to the general case, we allow vertices to become of higher-point and to carry non-trivial interaction forms. As will be explained later, when vertices are no longer restricted to cubic, the pattern of shuffle permutations needs to be extended. Furthermore, in the SFASL for the NLSM, Lorentz invariants arising from vertices will be included; whereas the SFASL for the YM will also involve Lorentz indices carried by vertices.

When higher-point vertices participate in the shuffle permutation, a type of vertices that may be called the mixed vertices will hinder the realization of the SFASL. We will show that the contributions from mixed vertices are canceled by certain terms in unmixed vertices through a very simple mechanism. For YM, there is another factor that hinders the SFASL, caused by the polarization vector carried by a special external gluon. We will introduce a viewpoint that decomposes spacetime into two mutually orthogonal subspaces to resolve this issue. As will be explained, the idea of orthogonal subspaces is completely consistent with the kinematic conditions for realizing the SFASL, hidden zeros, and $2$-split.

Note that although the formulation of the SFASL varies slightly among different models, its manifestation through Feynman diagrams is completely identical. Through the generalized SFASL, the hidden zeros and $2$-splits of NLSM and YM amplitudes can be interpreted in a universal manner, in exactly the same way as in the ${\rm Tr}(\phi^3)$ case: each hidden zero is ultimately reduced to the on-shell condition $k_j^2=0$ of a massless particle, while each $2$-split is understood as separating each Feynman diagram along two specific lines.

The remainder of this paper is organized as follows. In section \ref{sec-phi3-SFASL}, we review the SFASL for ${\rm Tr}(\phi^3)$ diagrams, as well as the interpretation of the hidden zeros and $2$-split of ${\rm Tr}(\phi^3)$ amplitudes based on this SFASL. Furthermore, we also discuss the extension of shuffle permutations when considering NLSM and YM amplitudes. In section \ref{sec-NLSM}, we study the SFASL for NLSM diagrams, and use it to interpret the hidden zeros and $2$-split of NLSM amplitudes. Subsequently, in section \ref{sec-YM}, we generalize the SFASL to YM diagrams, and interpret the hidden zeros and $2$-split of YM amplitudes via it. We end in section \ref{sec-conclu} with a brief conclusion and discussion.

\section{Shuffle factorization along a specific line}
\label{sec-phi3-SFASL}

For readers' convenience, we give a brief review of some ingredients in our previous work \cite{Zhou:2024ddy}, including the SFASL for Feynman diagrams of ${\rm Tr}(\phi^3)$ model, and the corresponding interpretation of hidden zeros and $2$-split of tree ${\rm Tr}(\phi^3)$ amplitudes. Then, we will discuss the appropriate extension of shuffle permutations to vertex general configurations.

Before starting, it is worth clarifying the ${\rm Tr}(\phi^3)$ model under consideration.
The ${\rm Tr}(\phi^3)$ model describes cubic interactions among colored massless scalars. Its Lagrangian takes the form
\bea
{\cal L}_{{\rm Tr}(\phi^3)}={\rm Tr}(\partial\phi)^2+g\,{\rm Tr}(\phi^3)\,,
\eea
where $\phi$ is an $N\times N$ matrix, with one index in the fundamental representation of $SU(N)$ and the other in the anti-fundamental. After stripping off coupling constants, the color-ordered tree amplitudes in this model involve only massless scalar propagators.

\subsection{Shuffle factorization along a specific line (SFASL) of tree ${\rm Tr}(\phi^3)$ amplitudes}
\label{subsec-phi3-SFASL}

In this subsection, we introduce the mechanism we refer to as shuffle factorization along a specific line (SFASL) in the summation over Feynman diagrams, for the tree-level ${\rm Tr}(\phi^3)$ amplitudes.

\begin{figure}
  \centering
   \includegraphics[width=10.5cm]{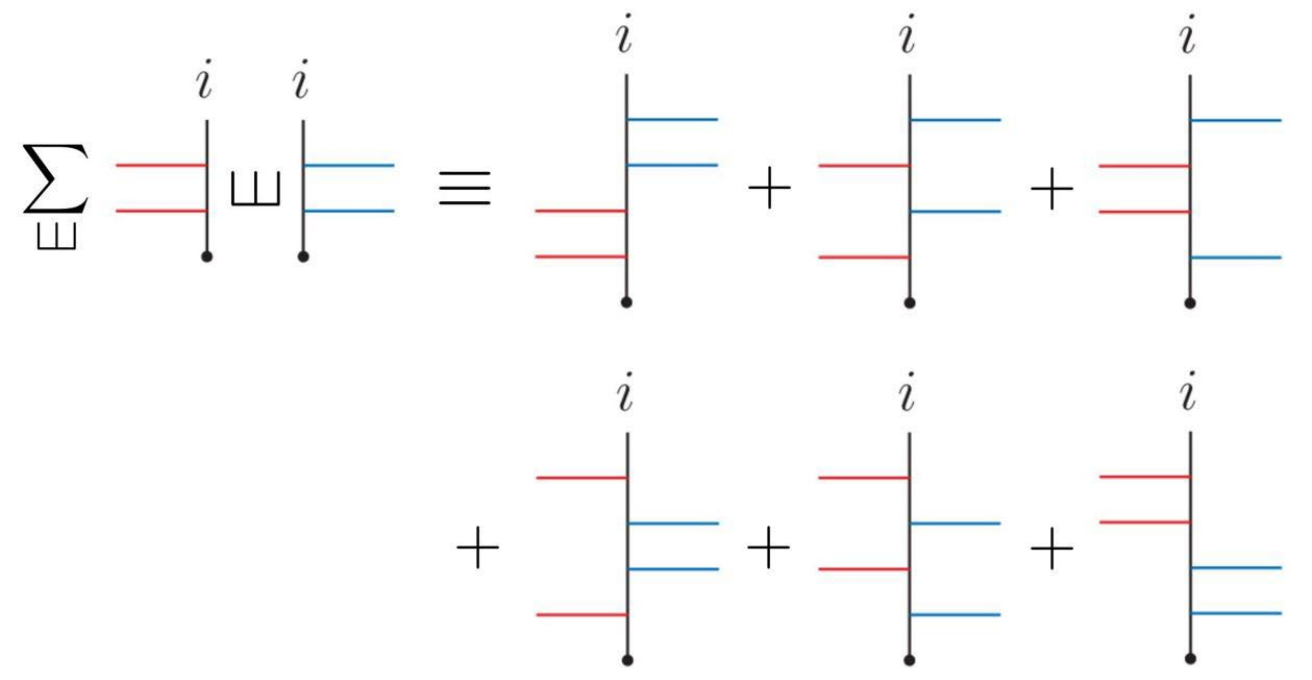} \\
  \caption{The meaning of summing over shuffle permutations. The red lines are $A$-lines, while the blue lines are $B$-lines.}\label{Fig2}
\end{figure}

For a specific line $L_{(i,\bullet)}$ in a Feynman diagram---where $i$ is a massless external line satisfying $k_i^2=0$, and $\bullet$ is an interaction vertex---we divide the lines attached to $L_{(i,\bullet)}$ into two sets, called $A$-lines and $B$-lines, and draw them separately on the two sides of $L_{(i,\bullet)}$. Each $A$-line or $B$-line can be either an internal line or an external line. We will consider summing over all shuffle permutations of the $A$-lines and $B$-lines along $L_{(i,\bullet)}$. A shuffle permutation is a permutation that preserves the relative order of all $A$-lines as well as the relative order of all $B$-lines, as illustrated by the example in Fig.\ref{Fig2}. Notice that all vertices discussed above are cubic vertices of the ${\rm Tr}(\phi^3)$ model.

\begin{figure}
  \centering
   \includegraphics[width=8cm]{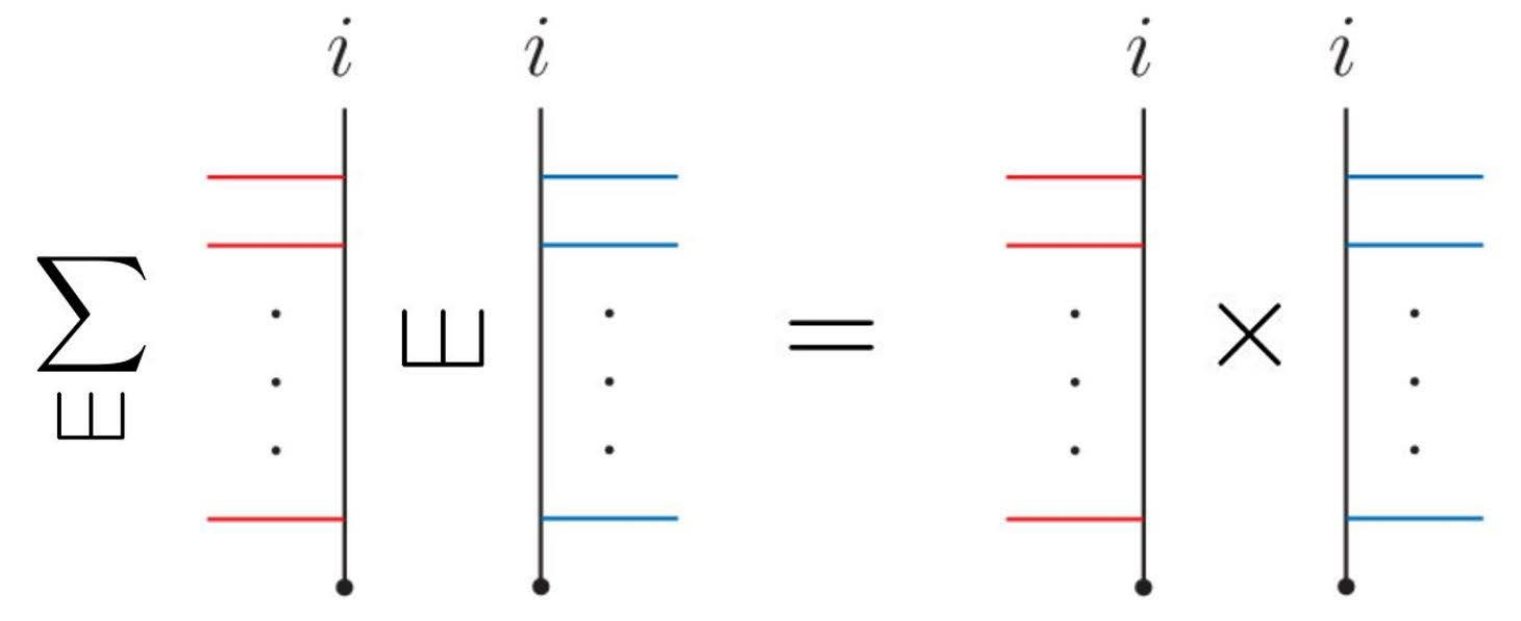} \\
  \caption{Shuffle factorization along the line $L_{i,\bullet}$. The red lines are $A$-lines, while the blue lines are $B$-lines.}\label{Fig1}
\end{figure}

\begin{figure}
  \centering
   \includegraphics[width=10cm]{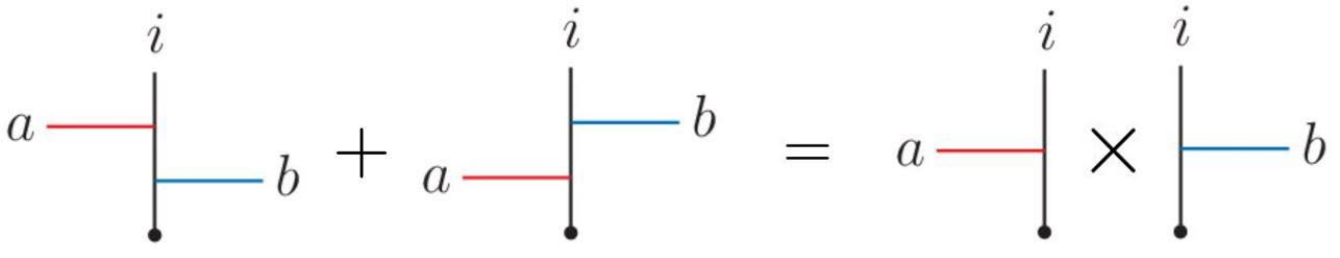} \\
  \caption{The first example of Fig.\ref{Fig1}. The red lines are $A$-lines, while the blue lines are $B$-lines.}\label{Fig3}
\end{figure}

\begin{figure}
  \centering
   \includegraphics[width=14cm]{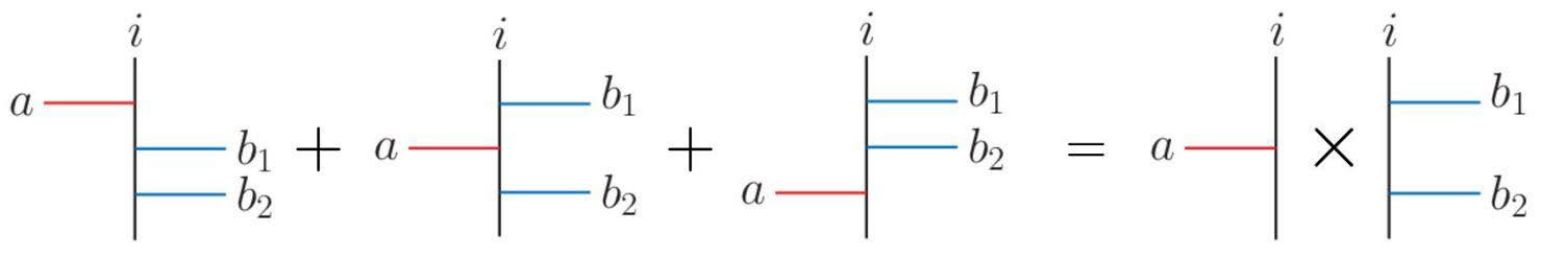} \\
  \caption{The second example of Fig.\ref{Fig1}. The red lines are $A$-lines, while the blue lines are $B$-lines.}\label{Fig4}
\end{figure}

The shuffle factorization along a specific line (SFASL) states that, when momenta carried by $A$-lines and $B$-lines satisfy
\bea
k_{\hat{a}}\cdot k_{\hat{b}}=0\,,~~\label{kine-condi-shuffle}
\eea
for any $A$-line $\hat{a}$ and $B$-line $\hat{b}$, then the result of summing over shuffle permutations factorizes as in Fig.\ref{Fig1}.
Two simple examples of such SFASL are given in Fig.\ref{Fig3} and Fig.\ref{Fig4}, respectively.
In the first example, by plugging the observation
\bea
s_{iab}\,\xrightarrow[]{\eref{kine-condi-shuffle}}\,\big(2\,k_i\cdot k_a+k_a^2\big)+\big(2\,k_i\cdot k_b+k_b^2\big)=s_{ia}+s_{ib}\,,~~~\label{key-observation}
\eea
one can verify
\bea
{1\over s_{ia}}\,{1\over s_{iab}}+{1\over s_{ib}}\,{1\over s_{iab}}\,\xrightarrow[]{\eref{kine-condi-shuffle}}\,{1\over s_{ia}}\,\times\,{1\over s_{ib}}\,.~~\label{meaning-fig3}
\eea
This is precisely the factorization in Fig.\ref{Fig3}. Throughout this paper, we adopt the usual notation
\bea
s_{\pmb S}=k_{\pmb S}^2\,,~~~~~~~~k_{\pmb S}=\sum_{\a=1}^s\,k_\a\,,
\eea
for any set $\pmb S=\{s_1,\cdots,s_s\}$.
In the second example, using the previous result \eref{meaning-fig3}, as well as the observation
\bea
s_{iab_1b_2}\,\xrightarrow[]{\eref{kine-condi-shuffle}}\,\big(2\,k_i\cdot k_a+k_a^2\big)+\Big(2\,k_i\cdot (k_{b_1}+k_{b_2})+(k_{b_1}+k_{b_2})^2\Big)=s_{ia}+s_{ib_1b_2}\,,~~\label{key-observation-2}
\eea
one can verify
\bea
{1\over s_{ia}}\,{1\over s_{iab_1}}\,{1\over s_{iab_1b_2}}+{1\over s_{ib_1}}\,{1\over s_{iab_1}}\,{1\over s_{iab_1b_2}}
+{1\over s_{ib_1}}\,{1\over s_{ib_1b_2}}\,{1\over s_{iab_1b_2}}\,\xrightarrow[]{\eref{kine-condi-shuffle}}\,{1\over s_{ia}}\,\times\,\Big({1\over s_{ib_1}}\,{1\over s_{ib_1b_2}}\Big)\,,
~~~\label{meaning-fig4}
\eea
which is exactly the meaning of Fig.\ref{Fig4}.


For latter convenience, we express the summation over shuffle permutations along the line $L_{(i,\bullet)}$ as
\bea
\sum_{\shuffle(p,q)}\,\prod_{t=1}^{p+q}\,{1\over D_t^{(i,\bullet)}}\,,
\eea
where $1/D_t^{(i,\bullet)}$ are propagators along $L_{(i,\bullet)}$, $p$ and $q$ are numbers of $A$-lines and $B$-lines, respectively. For instance,
summations in \eref{meaning-fig3} and \eref{meaning-fig4} can be organized as
\bea
{1\over s_{ia}}\,{1\over s_{iab}}+{1\over s_{ib}}\,{1\over s_{iab}}&=&\sum_{\shuffle(1,1)}\,\prod_{t=1}^{1+1}\,{1\over D_t^{(i,\bullet)}}\,,\nn
{1\over s_{ia}}\,{1\over s_{iab_1}}\,{1\over s_{iab_1b_2}}+{1\over s_{ib_1}}\,{1\over s_{iab_1}}\,{1\over s_{iab_1b_2}}
+{1\over s_{ib_1}}\,{1\over s_{ib_1b_2}}\,{1\over s_{iab_1b_2}}&=&\sum_{\shuffle(1,2)}\,\prod_{t=1}^{1+2}\,{1\over D_t^{(i,\bullet)}}\,.
\eea
Based on the above notation, the SFASL in Fig.\ref{Fig1} is expressed as
\bea
\sum_{\shuffle(p,q)}\,\prod_{t=1}^{p+q}\,{1\over D_t^{(i,\bullet)}}\,\xrightarrow[]{\eref{kine-condi-shuffle}}\,\Big(\prod_{\a=1}^p\,{1\over s_{ia_1\cdots a_{\a}}}\Big)\,\times\,
\Big(\prod_{\b=1}^q\,{1\over s_{ib_1\cdots b_{\b}}}\Big)\,,~~\label{fac-propa-gen}
\eea
where the $A$-lines and $B$-lines are encoded as $\{a_1,\cdots, a_p\}$ and $\{b_1,\cdots,b_q\}$, respectively.

The observation in \eref{key-observation} and \eref{key-observation-2} can be extended to the general case as
\bea
s_{i\{a\}\{b\}}\,\xrightarrow[]{\eref{kine-condi-shuffle}}\,\Big(2\,k_i\cdot k_{\{a\}}+k_{\{a\}}^2\Big)+\Big(2\,k_i\cdot k_{\{b\}}+k_{\{b\}}^2\Big)=s_{i\{a\}}+s_{i\{b\}}\,,~~\label{key-observation-general}
\eea
where $\{a\}$ is an arbitrary set of $A$-lines, and $\{b\}$ is an arbitrary set of $B$-lines. The above relation will be used frequently in the rest of this paper.

\subsection{From SFASL to hidden zero and $2$-split}
\label{subsec-0and2split-phi3}

The general proof of the SFASL in \eref{fac-propa-gen} can be found in our previous work \cite{Zhou:2024ddy,Feng:2025dci,Zhou:2026ukg}. In this paper, we do not intend to repeat this proof; instead, we skip the proof and directly interpret the hidden zeros and $2$-split of tree-level amplitudes from the perspective of SFASL.

\subsubsection{Hidden zero}

We begin with the hidden zeros. For any amplitude, one can always choose a pair of external legs $(i,j)$, and divide the remaining external legs into two sets $\pmb A$ and $\pmb B$. For a color ordered amplitude, such as a tree ${\rm Tr}(\phi^3)$ amplitude, a natural choice is $\pmb A=\{i+1,\cdots,j-1\}$ and $\pmb B=\{j+1,\cdots,i-1\}$, according to the color ordering. Then we require the external momenta to satisfy
\bea
k_a\cdot k_b=0\,,~~~~{\rm for}~\forall\,a\in\pmb A\,,\,b\in\pmb B\,.~~\label{kine-condi-0-scalar}
\eea
In each connected Feynman diagram, one can always find a line $L_{(i,j)}$ connecting external legs $i$ and $j$ together. Such a diagram can then be thought of as planting trees onto the line $L_{(i,j)}$. The kinematic condition \eref{kine-condi-0-scalar} forces the condition for SFASL in \eref{kine-condi-shuffle} to be satisfied, if we regard lines attached to $L_{(i,j)}$ from the $\pmb A$-side and $\pmb B$-side as $A$-lines and $B$-lines, respectively.

\begin{figure}
  \centering
   \includegraphics[width=12cm]{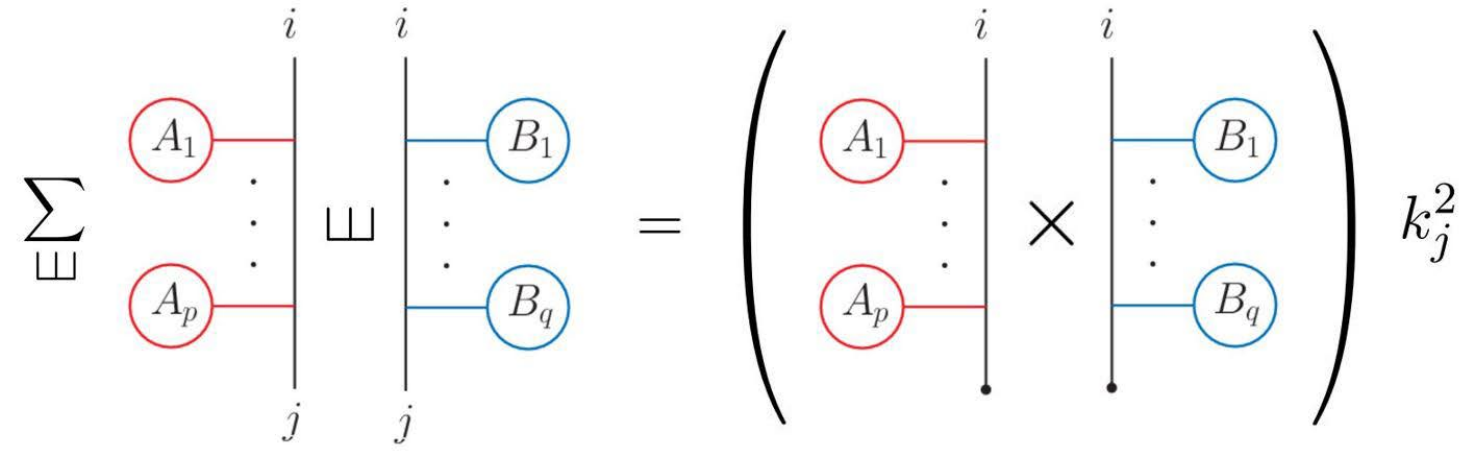} \\
  \caption{Diagrammatical interpretation for hidden zero. Each red circle or blue circle represents the BG current generated by the corresponding subset.}\label{Fig5}
\end{figure}

Then the SFASL in \eref{fac-propa-gen} and Fig.\ref{Fig1} leads to the behavior shown in Fig.\ref{Fig5}. This figure is understood as follows. We divide $\pmb A$ and $\pmb B$ into their subsets as $\pmb A=\{A_1,\cdots, A_p\}$ and $\pmb B=\{B_1,\cdots,B_q\}$, respectively. For a given division, the summation over Feynman diagrams is naturally separated into two steps. The first step generates Berends-Giele (BG) currents corresponding to each subset $A_\a$ or $B_\b$---with $\a\in\{1,\cdots,p\}$ and $\b\in\{1,\cdots,q\}$, which are represented by red or blue circles in Fig.\ref{Fig5}. Based on this diagrammatic representation, in the following discussion we will also frequently refer to the contribution from a given subset $A_\a$ or $B_\b$ as a block. Each block is connected to $L_{(i,j)}$ via an $A$-line or a $B$-line. The second step of the summation is precisely the summation over shuffle permutations of these $A$-lines and $B$-lines. Note that summing over shuffle permutations along the line $L_{(i,j)}$ differs from the situation in SFASL in Fig.\ref{Fig1} and \eref{fac-propa-gen}: the endpoint of $L_{(i,j)}$ is an external leg $j$, while the endpoint of $L_{(i,\bullet)}$ is a vertex $\bullet$. The external line $j$ causes the absence of the propagator $1/D^{(i,\bullet)}_{p+q}$. This different can be compensated by multiplying $D^{(i,\bullet)}_{p+q}/D^{(i,\bullet)}_{p+q}$, where
\bea
D^{(i,\bullet)}_{p+q}=s_{iA_1\cdots A_p B_1\cdots B_q}=\Big(k_i+\sum_{\a=1}^p\,k_{A_\a}+\sum_{\b=1}^q\,k_{B_\b}\Big)^2=\Big(\sum_{m=j+1}^{j-1}\,k_m\Big)^2=k_j^2\,.
\eea
The above manipulation yields the r.h.s. of Fig.\ref{Fig5}, namely,
\bea
\sum_{\shuffle(p,q)}\,\prod_{t=1}^{p+q-1}\,{1\over D_t^{(i,\bullet)}}\,&\xrightarrow[]{\eref{kine-condi-0-scalar}}&\,\Big(\prod_{\a=1}^p\,{1\over s_{iA_1\cdots A_{\a}}}\Big)\,
\Big(\prod_{\b=1}^q\,{1\over s_{iB_1\cdots B_{\b}}}\Big)\,k_j^2\,.~~\label{0-block}
\eea
Obviously, $k_j^2$ in the above expression will never be canceled by any $1/s_{iA_1\cdots A_{\a}}$ or $1/s_{iB_1\cdots B_{\b}}$. Thus, the on-shell condition $k_j^2=0$ implies the vanishing of \eref{0-block}. The above phenomenon is valid for any divisions of $\pmb A$ and $\pmb B$; therefore, after summing over divisions, we obtain the hidden zero,
\bea
{\cal A}_n^{{\rm Tr}(\phi^3)}(1,\cdots,n)\,&\xrightarrow[]{\eref{kine-condi-0-scalar}}&\,0\,.~~\label{zero-phi3}
\eea
\begin{figure}
  \centering
   \includegraphics[width=16cm]{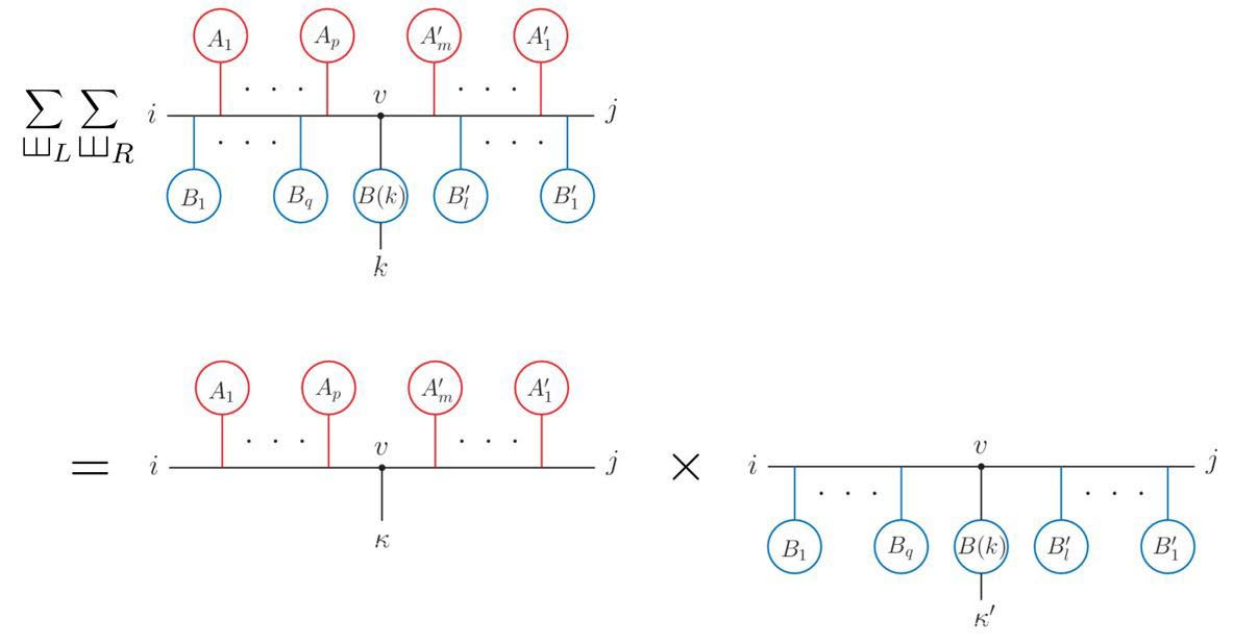} \\
  \caption{Diagrammatical interpretation for $2$-split. $\shuffle_L$ and $\shuffle_{R}$ in this graph correspond to $\shuffle_{(p,q)}$ and $\shuffle_{m,l}$ in \eref{two shuffles}, respectively. Although not shown explicitly, the first row of the figure actually contains all shuffles of $\{A_1,\cdots,A_p\}$ and $\{B_1,\cdots,B_q\}$ along the line $L_{(i,v)}$, as well as all shuffles of $\{A'_1,\cdots,A'_m\}$ and $\{B'_1,\cdots,B'_l\}$ along the line $L_{j,v}$.}\label{Fig6}
\end{figure}
%

\subsubsection{$2$-split}

The kinematic condition for $2$-split near the above zero can be achieved by slightly perturbing the zero kinematics \eref{kine-condi-0-scalar} as follows,
\bea
k_a\cdot k_b=0\,,~~~~{\rm for}~\forall\,a\in\pmb A\,,\,b\in\pmb B\setminus k\,.~~\label{kine-condi-2split-scalar}
\eea
Note that one can also remove a special element $k$ from $\pmb A$. Without loss of generality, in this paper we choose $k\in\pmb B$.
In each connected Feynman diagram, one can always find a vertex $v$ where three lines $L_{(i,v)}$, $L_{(j,v)}$ and $L_{(k,v)}$ meet. Since this is a common feature valid for any diagram, we can apply the SFASL to lines $L_{(i,v)}$ and $L_{(j,v)}$, obtaining
\bea
\sum_{\shuffle(p,q)}\,\prod_{t=1}^{p+q}\,{1\over D_t^{(i,v)}}\,&\xrightarrow[]{\eref{kine-condi-2split-scalar}}&\,\Big(\prod_{\a=1}^p\,{1\over s_{iA_1\cdots A_{\a}}}\Big)\,
\Big(\prod_{\b=1}^q\,{1\over s_{iB_1\cdots B_{\b}}}\Big)\,,\nn
\sum_{\shuffle(m,l)}\,\prod_{t=1}^{m+l}\,{1\over D_t^{(j,v)}}\,&\xrightarrow[]{\eref{kine-condi-2split-scalar}}&\,\Big(\prod_{\a=1}^m\,{1\over s_{iA'_1\cdots A'_{\a}}}\Big)\,
\Big(\prod_{\b=1}^l\,{1\over s_{iB'_1\cdots B'_{\b}}}\Big)\,.~~\label{two shuffles}
\eea
The divisions of $\pmb A$ and $\pmb B$ corresponding to \eref{two shuffles} are given by
\bea
\pmb A=\{A_1,\cdots,A_p,A'_m,\cdots,A'_1\}\,,~~~~\pmb B=\{B'_1,\cdots,B'_l,B(k),B_q,\cdots,B_1\}\,,~~\label{division-2split-tree}
\eea
as illustrated in Fig.\ref{Fig6}. The SFASL in \eref{two shuffles} leads to
\bea
&&\Big(\sum_{\shuffle(p,q)}\,\prod_{t=1}^{p+q}\,{1\over D_t^{(i,v)}}\Big)\,\Big(\sum_{\shuffle(m,l)}\,\prod_{t=1}^{m+l}\,{1\over D_t^{(j,v)}}\Big)\,
f^{{\rm Tr}(\phi^3)}(R)\,\xrightarrow[]{\eref{kine-condi-2split-scalar}}\nn
&&\Big[\Big(\prod_{\a=1}^p\,{1\over s_{iA_1\cdots A_{\a}}}\Big)\,\Big(\prod_{\a=1}^m\,{1\over s_{iA'_1\cdots A'_{\a}}}\Big)\,f^{{\rm Tr}(\phi^3)}_A(R)\,\Big]\,\times\,\Big[\Big(\prod_{\b=1}^q\,{1\over s_{iB_1\cdots B_{\b}}}\Big)\,\Big(\prod_{\b=1}^l\,{1\over s_{iB'_1\cdots B'_{\b}}}\Big)\,f^{{\rm Tr}(\phi^3)}_B(R)\,\Big]\,,~~\label{meaning-fig6}
\eea
where $f^{{\rm Tr}(\phi^3)}(R)$, $f^{{\rm Tr}(\phi^3)}_A(R)$ and $f^{{\rm Tr}(\phi^3)}_B(R)$ are contributions from remaining parts of diagrams, and each of them behaves as a product of contributions from blocks exhibited in Fig.\ref{Fig6}, namely:
\bea
f^{{\rm Tr}(\phi^3)}(R)&=&\Big(\prod_{\a=1}^p\,{\cal J}^{{\rm Tr}(\phi^3)}_{A_\a}\,{1\over s_{A_\a}}\Big)\,\Big(\prod_{\a=1}^m\,{\cal J}^{{\rm Tr}(\phi^3)}_{A'_\a}\,{1\over s_{A'_\a}}\Big)\,
\Big(\prod_{\b=1}^q\,{\cal J}^{{\rm Tr}(\phi^3)}_{B_\b}\,{1\over s_{B_\b}}\Big)\,\Big(\prod_{\b=1}^l\,{\cal J}^{{\rm Tr}(\phi^3)}_{B'_\b}\,{1\over s_{B'_\b}}\Big)\nn
&&\Big({\cal J}^{{\rm Tr}(\phi^3)}_{B(k)}\,{1\over s_{B(k)}}\Big)\,,\nn
f^{{\rm Tr}(\phi^3)}_A(R)&=&\Big(\prod_{\a=1}^p\,{\cal J}^{{\rm Tr}(\phi^3)}_{A_\a}\,{1\over s_{A_\a}}\Big)\,\Big(\prod_{\a=1}^m\,{\cal J}^{{\rm Tr}(\phi^3)}_{A'_\a}\,{1\over s_{A'_\a}}\Big)\,,\nn
f^{{\rm Tr}(\phi^3)}_B(R)&=&
\Big(\prod_{\b=1}^q\,{\cal J}^{{\rm Tr}(\phi^3)}_{B_\b}\,{1\over s_{B_\b}}\Big)\,\Big(\prod_{\b=1}^l\,{\cal J}^{{\rm Tr}(\phi^3)}_{B'_\b}\,{1\over s_{B'_\b}}\Big)\,\Big({\cal J}^{{\rm Tr}(\phi^3)}_{B(k)}\,{1\over s_{B(k)}}\Big)\,,~~~\label{fR}
\eea
where each ${\cal J}^{{\rm Tr}(\phi^3)}$ is the BG current contributed by the corresponding subset.
These $f(R)$, $f^A(R)$ and $f^B(R)$ satisfy a simple but crucial factorization formula
\bea
f^{{\rm Tr}(\phi^3)}(R)\,=\,f^{{\rm Tr}(\phi^3)}_A(R)\,\times\,f^{{\rm Tr}(\phi^3)}_B(R)\,,~~\label{key-fR}
\eea
which guarantees the validity of \eref{meaning-fig6}.

The factorization structure obtained in \eref{meaning-fig6} holds for any divisions of $\pmb A$ and $\pmb B$. Summing over divisions, we get
\bea
{\cal A}_n^{{\rm Tr}(\phi^3)}(1,\cdots,n)&=&\sum_{{\rm div}\pmb A}\,\sum_{{\rm div}\pmb B}\,\Big(\sum_{\shuffle(p,q)}\,\prod_{t=1}^{p+q}\,{1\over D_t^{(i,v)}}\Big)\,\Big(\sum_{\shuffle(m,l)}\,\prod_{t=1}^{m+l}\,{1\over D_t^{(j,v)}}\Big)\,
f^{{\rm Tr}(\phi^3)}(R)\nn
&\xrightarrow[]{\eref{kine-condi-2split-scalar}}&\,\Big[\sum_{{\rm div}\pmb A}\Big(\prod_{\a=1}^p\,{1\over s_{iA_1\cdots A_{\a}}}\Big)\,\Big(\prod_{\a=1}^m\,{1\over s_{iA'_1\cdots A'_{\a}}}\Big)\,f^{{\rm Tr}(\phi^3)}_A(R)\,\Big]\nn
&&\times\,\Big[\sum_{{\rm div}\pmb B}\,\Big(\prod_{\b=1}^q\,{1\over s_{iB_1\cdots B_{\b}}}\Big)\,\Big(\prod_{\b=1}^l\,{1\over s_{iB'_1\cdots B'_{\b}}}\Big)\,f^{{\rm Tr}(\phi^3)}_B(R)\,\Big]\,.~~\label{2split-phi-detail}
\eea
This is precisely the $2$-split of tree ${\rm Tr}(\phi^3)$ amplitudes, and can be recast as
\bea
{\cal A}_n^{{\rm Tr}(\phi^3)}(1,\cdots,n)\,&\xrightarrow[]{\eref{kine-condi-2split-scalar}}&\,{\cal J}_{n_1}^{{\rm Tr}(\phi^3)}(i,\pmb A,j,\kappa)\,\times\,{\cal J}_{n+3-n_1}^{{\rm Tr}(\phi^3)}(j,\pmb B(\kappa'),i)\,,
~~\label{2split-tree}
\eea
where two currents are given as,
\bea
{\cal J}_{n_1}^{{\rm Tr}(\phi^3)}(i,\pmb A,j,\kappa)&=&\sum_{{\rm div}\pmb A}\Big(\prod_{\a=1}^p\,{1\over s_{iA_1\cdots A_{\a}}}\Big)\,\Big(\prod_{\a=1}^m\,{1\over s_{iA'_1\cdots A'_{\a}}}\Big)\,f^{{\rm Tr}(\phi^3)}_A(R)\,,\nn
{\cal J}_{n+3-n_1}^{{\rm Tr}(\phi^3)}(j,\pmb B(\kappa'),i)&=&\sum_{{\rm div}\pmb B}\,\Big(\prod_{\b=1}^q\,{1\over s_{iB_1\cdots B_{\b}}}\Big)\,\Big(\prod_{\b=1}^l\,{1\over s_{iB'_1\cdots B'_{\b}}}\Big)\,f^{{\rm Tr}(\phi^3)}_B(R)\,.
\eea
Two currents ${\cal J}_{n_1}^{{\rm Tr}(\phi^3)}$ and ${\cal J}_{n+3-n_1}^{{\rm Tr}(\phi^3)}$ carry the off-shell legs $\kappa$ and $\kappa'$, respectively. Momentum conservation indicates
\bea
k_\kappa=\sum_{\b=j+1}^{i-1}\,k_\b=k_{\pmb B}\,,~~~~k_{\kappa'}=k_k+\sum_{\a=i+1}^{j-1}\,k_\a=k_k+k_{\pmb A}\,.~~\label{k-kappa}
\eea
%

\subsection{Generalization of shuffle permutations}
\label{subsec-phi3-genera-shuffle}

In the ${\rm Tr}(\phi^3)$ model, since there are only cubic vertices and each vertex provides only a constant, the SFASL is particularly simple. To apply the similar idea to interpret the hidden zeros and $2$-split of the more complicated models of NLSM and YM, we need to make some generalizations to the SFASL.

\begin{figure}
  \centering
   \includegraphics[width=14cm]{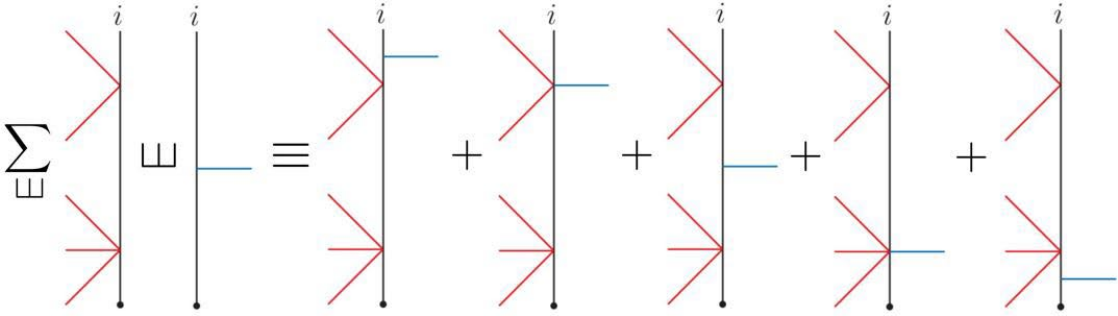} \\
  \caption{An example of generalized shuffle permutation.}\label{Fig7}
\end{figure}

In general, we shall allow vertices with valence higher than three. This means that, for a vertex on $L_{(i,\bullet)}$, the total number of $A$-lines and $B$-lines attached to this vertex can exceed one. Therefore, we introduce the sets of $A$-lines and the sets of $B$-lines, abbreviated as $A$-sets and $B$-sets. Each set is attached to an individual vertex. Then, we generalize the shuffle permutation of $A$-lines and $B$-lines to shuffle permutation of $A$-sets and $B$-sets. Take Fig.\ref{Fig7} as an example. In the figure, there are two $A$-sets and one $B$-set, containing two $A$-lines, three $A$-lines, and one $B$-line, respectively. The shuffle permutations of these two $A$-sets and one $B$-sets are shown on the r.h.s. of $\equiv$. Hereafter, we denote the $A$-set and $B$-set as $\{A\}$ and $\{B\}$, respectively, with potential subscripts and superscripts.

As can be seen from the shuffle permutations in Fig.\ref{Fig7}, we can classify the vertices on $L_{(i,\bullet)}$ into three types $V_{\{A\}}$, $V_{\{B\}}$ and $V_{\{A\}|\{B\}}$: those attached to an $A$-set only, those attached to a $B$-set only, and those attached to both an $A$-set and a $B$-set. The occurrence of mixed vertices $V_{\{A\}|\{B\}}$ constitutes the primary difference between the generalized shuffle permutation and the shuffle permutation represented by Fig.\ref{Fig2}. Each mixed vertex $V_{\{A\}|\{B\}}$ reduces the number of propagators by one. For example, the first three terms on the r.h.s. of $\equiv$ in Fig.\ref{Fig7} correspond to the shuffle permutations between an $A$-set and a $B$-set. The first and third diagrams each have three propagators, while the second diagram, which contains a mixed vertex, has only two propagators. This difference in the number of propagators implies that the mass dimensions of these vertices satisfy
\bea
[V_{\{A\}}]_{\rm mass}+[V_{\{B\}}]_{\rm mass}=[V_{\{A\}|\{B\}}]_{\rm mass}+2\,.~~\label{mass-dim-v}
\eea
In the following sections, we will show that the vertices of NLSM and YM all satisfy this relation.

\begin{figure}
  \centering
   \includegraphics[width=8cm]{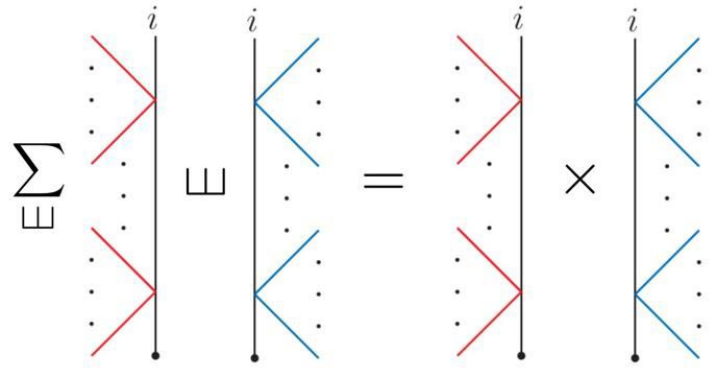} \\
  \caption{Generalized SFASL based on generalized shuffle permutation.}\label{Fig8}
\end{figure}

\begin{figure}
  \centering
   \includegraphics[width=7cm]{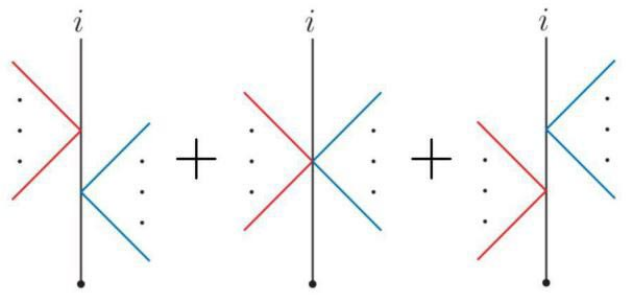} \\
  \caption{Cancelation when summing over generalized shuffle permutations: the middle diagram is canceled.}\label{cancel}
\end{figure}

Thus, for general forms of interactions, the SFASL is generalized as shown in Fig.\ref{Fig8}. It is worth pointing out that for a specific model, it is possible that various diagrams in Fig.\ref{Fig8} vanish, this is because the simultaneous existence of nonzero $V_{\{A\}}$, $V_{\{B\}}$ and $V_{\{A\}|\{B\}}$ is not necessary.
The factorization pattern on the r.h.s. of Fig.\ref{Fig8} does not receive any contribution from $V_{\{A\}|\{B\}}$-type vertices. It means these contributions are ultimately canceled.
In subsequent sections, we will see how such cancellations happen. The derivation process may be somewhat lengthy, but the cancellation mechanism is very simple. We provide a brief preview of this mechanism using Fig.\ref{cancel} as an example. In the first diagram of Fig.\ref{cancel}, starting from the external leg $i$ and moving along, vertex $V_{\{B\}}$ appears after vertex $V_{\{A\}}$. Thus,
$V_{\{B\}}$ can be divided into two parts\footnote{When no ambiguity is caused, we also use $V_{\{A\}}$/$V_{\{B\}}$/$V_{\{A\}|\{B\}}$ to denote the contribution of this vertex to the amplitude.}: one part is exactly the same as in the case where
$V_{\{A\}}$ is absent, which we call the part that commutes with $V_{\{A\}}$; the other part depends on
$V_{\{A\}}$, which we call the non-commuting part. Similarly, in the third diagram,
$V_{\{A\}}$ can also be separated into commuting and non-commuting parts. We will see that the non-commuting parts of the first and third diagrams cancel the second diagram contains the mixed vertex $V_{\{A\}|\{B\}}$, while the commuting parts of the two diagrams yield the SFASL. Regardless of the number of vertices on $L_{(i,\bullet)}$, the above mechanism holds recursively and is universal for both the NLSM and YM theories.

Of course, the generalized SFASL depicted in Fig.\ref{Fig8} is not a condition that tree-level amplitudes should satisfy. On the other hand, when the tree-level amplitudes of a model do satisfy the generalized SFASL, then this SFASL predicts that such amplitudes exhibit hidden zeros and $2$-split. In subsequent sections, we will show that all NLSM and YM amplitudes at tree-level satisfy the generalized SFASL in Fig.\ref{Fig8}. Moreover, the hidden zeros and $2$-splits of these amplitudes can be interpreted in terms of the generalized SFASL, as can be shown by a procedure closely analogous to that in subsection \ref{subsec-0and2split-phi3}.

\section{NLSM amplitudes}
\label{sec-NLSM}

In this section, we will demonstrate that the Feynman diagrams of NLSM satisfy the SFASL (generalized in the manner discussed in section \ref{subsec-phi3-genera-shuffle}), and from the SFASL, we will interpret the hidden zeros and $2$-split of tree NLSM amplitudes.

The $U(N)$ NLSM Lagrangian in its standard form, expressed via the Cayley parametrization, is given by
\bea
{\cal L}_{\rm NLSM}={1\over 8\lambda^2}\,{\rm Tr}(\partial_\mu {\rm U}^\dag\partial^\mu {\rm U})\,,~~\label{Lag-N}
\eea
where
\bea
{\rm U}=(\mathbb{I}+\lambda\Phi)\,(\mathbb{I}-\lambda\Phi)^{-1}\,,
\eea
with $\mathbb{I}$ denoting the identity matrix and $\Phi=\phi_IT^I$, where $T^I$ are the generators of $U(N)$.
The flavor-ordered Feynman rules for vertices are given as \cite{Kampf:2013vha},
\bea
V_{2n+1}&=&0\,,\nn
V_{2n+2}&=&{(-1)^n\over 2^{n+1}}\,\Big({1\over F}\Big)^{2n}\,\sum_{j=0}^n\,\sum_{i=1}^{2n+2}\,(k_i\cdot k_{i+2j+1})={(-1)^{n-1}\over 2^n}\,\Big({1\over F}\Big)^{2n}\,\Big(\sum_{i=0}^n\,k_{2i+1}\Big)^2\,.~~\label{cayley}
\eea
In \eref{cayley}, $V_{2n+2}$ has two equivalent forms. We refer to the form containing $(k_i\cdot k_{i+2j+1})$ as the $K\cdot K$-form, and the form containing $(\sum_{i=0}^n\,k_{2i+1})^2$ as the $K^2$-form.
In the rest of this section, we will ignore the decay constant $F$.
Since the mass dimension of each vertex is $2$, the relation \eref{mass-dim-v} is satisfied.

Based on the diagrammatic description in Fig.\ref{Fig8}, the SFASL for NLSM amplitudes is expressed as
\bea
\sum_{\shuffle(p,q)}\,\prod_{t=1}^{p+q-N_{A|B}}\,{V^{(i,\bullet)}_t\over D_t^{(i,\bullet)}}\,&\xrightarrow[]{\eref{kine-condi-shuffle}}&\,\Big(\prod_{\a=1}^p\,{V_{\{A\}_\a}\over s_{i\{A\}_1\cdots \{A\}_{\a}}}\Big)\,\times\,
\Big(\prod_{\b=1}^q\,{V_{\{B\}_\b}\over s_{i\{B\}_1\cdots \{B\}_{\b}}}\Big)\,,~~\label{fac-propa+v-gen}
\eea
which can be understood as follows.
The $A$-sets and $B$-sets are given by $\{A\}_1,\cdots,\{A\}_p$ and $\{B\}_1,\cdots,\{B\}_q$. Each $\{A\}_\a$ or $\{B\}_\b$
is attached to a vertex on $L_{(i,\bullet)}$, and each vertex launches a propagator along $L_{(i,\bullet)}$. Therefore, on the l.h.s of
\eref{fac-propa+v-gen}, each propagator $1/D_t^{(i,\bullet)}$ is accompanied with a vertex $V^{(i,\bullet)}_t$, both of them are along $L_{(i,\bullet)}$. Each vertex $V^{(i,\bullet)}_t$ can be of $V_{\{A\}}$-type, $V_{\{B\}}$-type or $V_{\{A\}|\{B\}}$-type. The number of $t$ is clearly $p+q-N_{A|B}$, where $N_{A|B}$ stands for the number of the mixed $V_{\{A\}|\{B\}}$-type vertices. The r.h.s of \eref{fac-propa+v-gen} represents the resulting factorization behavior.

In subsections \ref{subsec-NLSM-case1} and \ref{subsec-NLSM-case2}, we show the simplest SFASL, with only one $A$-set and only one $B$-set. As will be seen, the mixed vertex $V_{\{A\}|\{B\}}$ which hinder the SFASL, is canceled by certain terms from unmixed ones. In subsection \ref{subsec-NLSM-gen}, we give a recursive proof for the general SFASL. Then, in subsection \ref{subsec-NLSM-0and2split}, we use the SFASL to interpret the hidden zeros and $2$-split of tree NLSM amplitudes.

\subsection{Simplest shuffle permutation: first case}
\label{subsec-NLSM-case1}

We study the SFASL of NLSM by starting from the simplest shuffle permutation.
The simplest shuffle permutation is the one in which there is only one $A$-set and one $B$-set attached to $L_{(i,\bullet)}$. Under the Cayley parametrization, each nonzero vertex is even-point, thus the simplest shuffle permutations can be further classified as follows. First, both $n_A$ and $n_B$ are even, where $n_A$ is the number of $A$-lines contained in the $A$-set, and $n_B$ is the number of $B$-lines contained in the $B$-set. In this case, $V_{\{A\}}$, $V_{\{B\}}$, and $V_{\{A\}|\{B\}}$ can all exist. In the second case, both $n_A$ and $n_B$ are odd, where the mixed vertex $V_{\{A\}|\{B\}}$ exists, while $V_{\{A\}}$ and $V_{\{B\}}$ are forbidden. In the third case, one of $n_A$ and $n_B$ is even and the other is odd, in which no admissible graph exists. We will discuss the first case in this subsection and the second case in the next subsection.

\begin{figure}
  \centering
   \includegraphics[width=11cm]{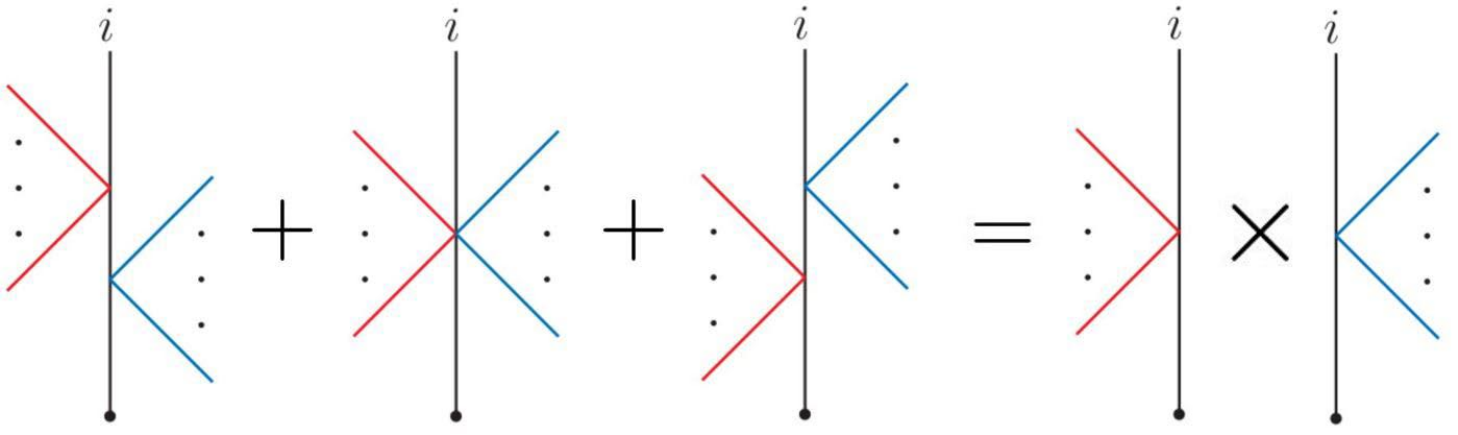} \\
  \caption{The first case of the simplest shuffle permutation.}\label{Fig9}
\end{figure}

For the first case, we denote $n_A$ and $n_B$ by $2m_A$ and $2m_B$, respectively. We will verify that the SFASL shown in Fig.\ref{Fig9} (expressed in \eref{fac-propa+v-gen} with $p=q=1$) holds when the kinematic condition \eref{kine-condi-shuffle} is satisfied.

Let us begin with the first diagram on the l.h.s. of Fig.\ref{Fig9}. By employing the $K\cdot K$-form of $V_{2n+2}$ in \eref{cayley}, one can evaluate the contribution from this diagram as,
\bea
F_1\,&\xrightarrow[]{\eref{kine-condi-shuffle}}&\,\Big[{(-1)^{m_A}\over 2^{m_A+1}}\,\Big({\cal V}_{\{A\}}-2k_i\cdot k_{i\{A\}}\Big)\Big]\,{1\over s_{i\{A\}}}\,\Big[{(-1)^{m_B}\over 2^{m_B+1}}\,\Big({\cal V}_{\{B\}}-2k_{i\{A\}}\cdot k_{i\{A\}\{B\}}\Big)\Big]\,{1\over s_{i\{A\}\{B\}}}\nn
&=&{(-1)^{m_A+m_B}\over 2^{m_A+m_B+2}}\,{1\over s_{i\{A\}}}\,{1\over s_{i\{A\}\{B\}}}\,\Big[\Big({\cal V}_{\{A\}}-2k_i\cdot k_{i\{A\}}\Big)\Big({\cal V}_{\{B\}}-2k_{i}\cdot k_{i\{B\}}\Big)\nn
&&~~~~~~~~~~~~~~~~~~~~~~~~~~~~~~~~~~~~-\Big({\cal V}_{\{A\}}-2k_i\cdot k_{i\{A\}}\Big)2k_{i\{A\}}^2\Big]\,.~~\label{NLSM-F1}
\eea
In the above, we denote by ${\cal V}_{\{A\}}$ the sum of all inner products involving $A$-line momenta at the
$V_{\{A\}}$ vertex, and by ${\cal V}_{\{B\}}$ the sum of all inner products involving $B$-line momenta at the $V_{\{B\}}$ vertex. Apart from
${\cal V}_{\{A\}}$, the
$V_{\{A\}}$ vertex also contains the inner product of two momenta
$k_i$ and $-k_{i\{A\}}$ propagate in the line $L_{(i,\bullet)}$. Similarly, in addition to
${\cal V}_{\{B\}}$, the
$V_{\{B\}}$ vertex also contains the inner product of two momenta
$k_{i\{A\}}$ and $-k_{i\{A\}\{B\}}$ propagate in the line $L_{(i,\bullet)}$. Here
\bea
k_{i\{A\}}=k_i+k_{\{A\}}\,,~~~~k_{i\{B\}}=k_i+k_{\{B\}}\,,~~~~k_{i\{A\}\{B\}}=k_i+k_{\{A\}}+k_{\{B\}}\,,
\eea
where $k_{\{A\}}$ and $k_{\{B\}}$ are total momenta of the $A$-set and the $B$-set, respectively.
For instance, suppose $\{B\}=\{b_1,b_2\}$, then the rule in \eref{cayley} yields
\bea
V_{\{B\}}={-1\over 2^2}\,\Big[\,{\cal V}_{\{B\}}-2k_{i\{A\}}\cdot k_{i\{A\}\{B\}}\,\Big]\,,
\eea
where
\bea
{\cal V}_{\{B\}}\,=\,2k_{i\{A\}}\cdot k_{b_1}+2k_{b_1}\cdot k_{b_2}-2k_{b_2}\cdot k_{i\{A\}\{B\}}\,.~~\label{example-VB}
\eea

In the second step of \eref{NLSM-F1}, we have used
\bea
k_{i\{A\}}\cdot k_{i\{A\}\{B\}}=k^2_{i\{A\}}+k_{i\{A\}}\cdot k_{\{B\}}\,&\xrightarrow[]{\eref{kine-condi-shuffle}}&\,k^2_{i\{A\}}+k_i\cdot k_{\{B\}}=k^2_{i\{A\}}+k_{i}\cdot k_{i\{B\}}\,,
\eea
due to the kinematic condition \eref{kine-condi-shuffle} and the on-shell condition $k_i^2=0$. The reason for this decomposition is to separate the vertex $V_{\{B\}}$ into two parts, such that the first part is completely insensitive to the momentum from $\{A\}$, as if the previous vertex $V_{\{A\}}$ did not exist. The above separation is guaranteed by the following observation: since all inner products in ${\cal V}_{\{B\}}$ involve $B$-line momenta, then the kinematic condition \eref{kine-condi-shuffle} implies that these $B$-line momenta annihilate the $A$-line momenta propagating through $L_{(i,\bullet)}$ within the inner products. For example, the kinematic condition \eref{kine-condi-shuffle} reduces the ${\cal V}_{\{B\}}$ in \eref{example-VB} to
\bea
{\cal V}_{\{B\}}\,&\xrightarrow[]{\eref{kine-condi-shuffle}}&\,2k_{i}\cdot k_{b_1}+2k_{b_1}\cdot k_{b_2}-2k_{b_2}\cdot k_{i\{B\}}\,,~~\label{example-VB-2}
\eea
with $k_{\{A\}}$ removed, then ${\cal V}_{\{B\}}-2k_i\cdot k_{i\{B\}}$ is manifestly independent of $V_{\{A\}}$. In this sense, if we temporarily disregard the propagators and consider only the numerators, the first part of $V_{\{B\}}$ commutes with the previous vertex $V_{\{A\}}$. From now on, when swapping two vertices on
$L_{(i,\bullet)}$, if the numerator generated by the two vertices together remains unchanged, we say that the two vertices are commuting.

Through the similar process, the contribution from the third diagram on the l.h.s. of Fig.\ref{Fig9} can be computed as
\bea
F_3\,&\xrightarrow[]{\eref{kine-condi-shuffle}}&\,\Big[{(-1)^{m_B}\over 2^{m_B+1}}\,\Big({\cal V}_{\{B\}}-2k_i\cdot k_{i\{B\}}\Big)\Big]\,{1\over s_{i\{B\}}}\,\Big[{(-1)^{m_A}\over 2^{m_A+1}}\,\Big({\cal V}_{\{A\}}-2k_{i\{B\}}\cdot k_{i\{A\}\{B\}}\Big)\Big]\,{1\over s_{i\{A\}\{B\}}}\nn
&=&{(-1)^{m_A+m_B}\over 2^{m_A+m_B+2}}\,{1\over s_{i\{B\}}}\,{1\over s_{i\{A\}\{B\}}}\,\Big[\Big({\cal V}_{\{B\}}-2k_i\cdot k_{i\{B\}}\Big)\Big({\cal V}_{\{A\}}-2k_{i}\cdot k_{i\{A\}}\Big)\nn
&&~~~~~~~~~~~~~~~~~~~~~~~~~~~~~~~~~~~~-\Big({\cal V}_{\{B\}}-2k_i\cdot k_{i\{B\}}\Big)2k_{i\{B\}}^2\Big]\,.~~\label{NLSM-F3}
\eea
Similar to the case in $F_1$, the term $V_{\{A\}}$ is divided into a part that commutes with $V_{\{B\}}$, and a part that does not. It should be emphasized that ${\cal V}_{\{A\}}$ and ${\cal V}_{\{B\}}$ in the above $F_3$ are exactly identical to those in $F_1$, due to the reduction for $V_{\{B\}}$ exemplified in \eref{example-VB-2} and the similar reduction for $V_{\{A\}}$ in $F_3$.

Combining $F_1$ and $F_3$ in \eref{NLSM-F1} and \eref{NLSM-F3} together, we obtain
\bea
F_1+F_3&=&P_1+P_2\,,
\eea
where
\bea
P_1&=&{(-1)^{m_A+m_B}\over 2^{m_A+m_B+2}}\,\Big({\cal V}_{\{B\}}-2k_i\cdot k_{i\{B\}}\Big)\Big({\cal V}_{\{A\}}-2k_{i}\cdot k_{i\{A\}}\Big)\,\Big[{1\over s_{i\{A\}}}\,{1\over s_{i\{A\}\{B\}}}+{1\over s_{i\{B\}}}\,{1\over s_{i\{A\}\{B\}}}\Big]\nn
P_2&=&-{(-1)^{m_A+m_B}\over 2^{m_A+m_B+1}}\,{1\over s_{i\{A\}\{B\}}}\,\Big[{\cal V}_{\{A\}}+{\cal V}_{\{B\}}-2k_i\cdot k_{i\{A\}}-2k_i\cdot k_{i\{B\}}\Big]\,.~~\label{P1P2}
\eea
When driving $P_2$, we have used $s_{i\{A\}}=k_{i\{A\}}^2$ and $s_{i\{B\}}=k_{i\{B\}}^2$.
Clearly, $P_1$ arises from the commuting parts. By utilizing the observation \eref{key-observation-general},
we find
\bea
P_1\,&\xrightarrow[]{\eref{kine-condi-shuffle}}&\,\Big[{(-1)^{m_A}\over 2^{m_A+1}}\,\Big({\cal V}_{\{A\}}-2k_i\cdot k_{i\{A\}}\Big)\,{1\over s_{i\{A\}}}\Big]\,\times\,\Big[{(-1)^{m_B}\over 2^{m_B+1}}\,\Big({\cal V}_{\{B\}}-2k_i\cdot k_{i\{B\}}\Big)\,{1\over s_{i\{B\}}}\Big]\nn
&=&{V_{\{A\}}\over s_{i\{A\}}}\,\times\,{V_{\{B\}}\over s_{i\{B\}}}\,,
\eea
which is precisely the factorization structure on the r.h.s. of Fig.\ref{Fig9} (r.h.s. of \eref{fac-propa+v-gen} with $p=q=1$). On the other hand, using
\bea
k_i\cdot k_{i\{A\}}+k_i\cdot k_{i\{B\}}=k_i\cdot k_{i\{A\}\{B\}}\,,
\eea
it can be seen immediately that $P_2$---which arises from the non-commuting parts---cancels the second diagram on the l.h.s. of Fig.\ref{Fig9}---a single $V_{\{A\}|\{B\}}$-type vertex.
Therefore, the SFASL shown in Fig.\ref{Fig9} is completely satisfied.

\subsection{Simplest shuffle permutation: second case}
\label{subsec-NLSM-case2}

In this subsection, we will show that when the kinematic condition \eref{kine-condi-shuffle} is satisfied, the second case, in which both $n_A$ and $n_B$ are odd, does not contribute to the amplitude at all.

This case corresponds to a single $V_{\{A\}|\{B\}}$-type vertex. We use the $K^2$-form of $V_{2n+2}$ in \eref{cayley} to deal with this vertex.
From \eref{cayley}, it is easy to see that when both $n_A$ and $n_B$ are odd, the two momenta provided by the line $L_{(i,\bullet)}$ either both appear in the $K^2$-form or both do not appear in the $K^2$-form. We choose the latter. The kinematic condition \eref{kine-condi-shuffle} then implies that this $V_{\{A\}|\{B\}}$-type vertex is divided into two parts, namely,
\bea
V_{\{A\}|\{B\}}\,&\xrightarrow[]{\eref{kine-condi-shuffle}}&\,K_A^2+K_B^2\,,~~\label{separate-V-NLSM}
\eea
where
\bea
&&K_A=k{{a}_1}+k_{{a}_3}+k_{{a}_5}+\cdots+k_{{a}_{n_A}}\,,~~~~~~~~K_B=k_{{b}_1}+k_{{b}_3}+k_{{b}_5}+\cdots+k_{{b}_{n_B}}\,,\nn
&&{\rm for}~~\{A\}=\{{a}_1,{a}_2,\cdots,{a}_{n_A}\}\,,~~~~~~\{B\}=\{{b}_1,{b}_2,\cdots,{b}_{n_B}\}\,,
\eea
due to the rule in \eref{cayley}.

\begin{figure}
  \centering
   \includegraphics[width=10cm]{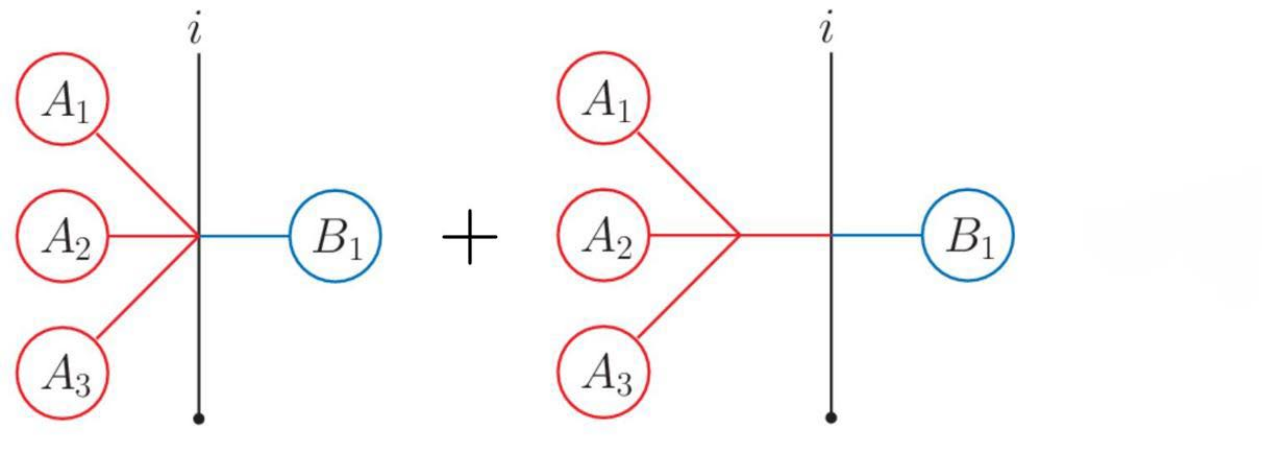} \\
  \caption{An example of the second case of the simplest shuffle permutation.}\label{Fig10}
\end{figure}

We first observe that if $n_A=1$ and this $A$-line is an on-shell external line, then $K_A^2$ vanishes. The same conclusion holds for $n_B$ and $K_B^2$. Thus, for the special case in which $n_A=n_B=1$, and both this $A$-line and this $B$-line are on-shell external legs, the contribution of this $V_{\{A\}|\{B\}}$ vertex vanishes. The above highly special conditions can be relaxed in two ways: (i) $n_A$ ($n_B$) is greater than $1$; (ii) $n_A$ ($n_B$) remains equal to $1$ but this $A$-line ($B$-line) is an internal line. In what follows, we demonstrate that the contributions from these two cases cancel each other.

To see how this cancellation arises, let us first consider
the simple example shown in Fig.\ref{Fig10}. From the standpoint of $L_{(i,\bullet)}$, the left diagram in Fig.\ref{Fig10} corresponds to $n_A=3$, whereas the right diagram corresponds to $n_A=1$, with the associated $A$-line being an internal line.
By using the $K^2$-form of $V_{2n+2}$ in \eref{cayley}, the left diagram of Fig.\ref{Fig10} can be evaluated as
\bea
F_L&=&\Big[\,{-1\over 2^2}\,\Big(k_{A_1}+k_{A_3}+k_{B_1}\Big)^2\,\Big]\,{1\over s_{iA_1A_2A_3B_1}}\nn
&\xrightarrow[]{\eref{kine-condi-shuffle}}&\,{-1\over 2^2}\,\Big(k_{A_1A_3}^2+k^2_{B_1}\Big)\,{1\over s_{iA_1A_2A_3B_1}}\,.~~\label{Fig10-L}
\eea
Meanwhile, the right diagram can be evaluated as
\bea
F_R&=&\Big[\,{1\over 2}\Big(k_{A_1}+k_{A_3}\Big)^2\,\Big]\,{1\over s_{A_1A_2A_3}}\,\Big[\,{1\over 2}\Big(k_{A_1A_2A_3}+k_{B_1}\Big)^2\,\Big]\,{1\over s_{iA_1A_2A_3B_1}}\nn
&\xrightarrow[]{\eref{kine-condi-shuffle}}&\,{1\over 2^2}\,k_{A_1A_3}^2\,{1\over s_{A_1A_2A_3}}\,\Big(k^2_{A_1A_2A_3}+k^2_{B_1}\Big)\,{1\over s_{iA_1A_2A_3B_1}}\nn
&=&{1\over 2^2}\,k_{A_1A_3}^2\,{1\over s_{iA_1A_2A_3B_1}}\,+\,{1\over 2^2}\,k_{A_1A_3}^2\,{1\over s_{A_1A_2A_3}}\,k_{B_1}^2\,{1\over s_{iA_1A_2A_3B_1}}\,.
\eea
As can be seen, in the summation over Feynman diagrams, the first term of $F_R$ cancels the first term of $F_L$.

\begin{figure}
  \centering
   \includegraphics[width=14cm]{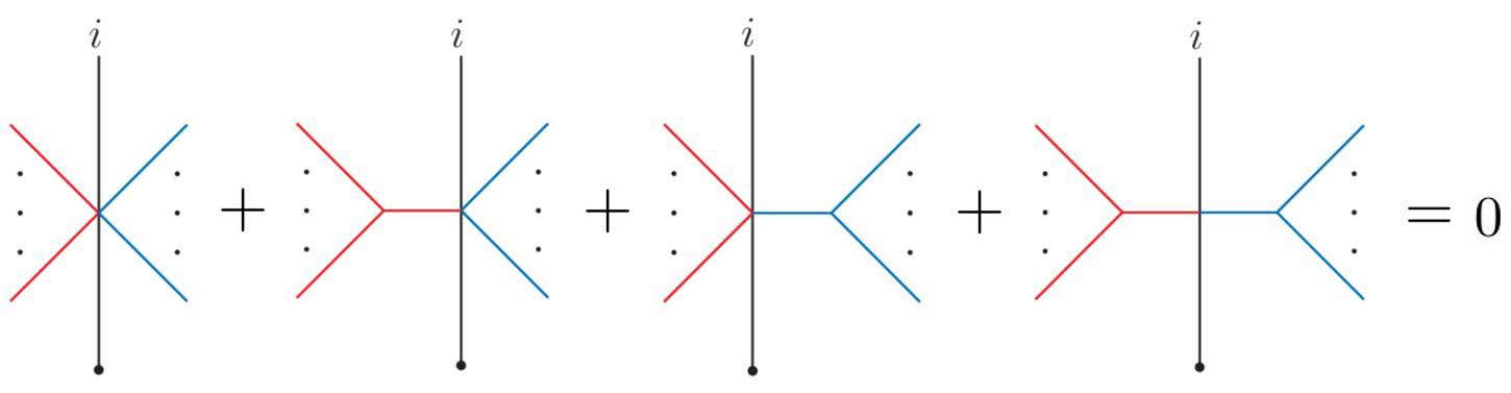} \\
  \caption{Cancelation in the second case.}\label{Fig11}
\end{figure}

The above cancellation mechanism extends directly to the general case. Provided that $n_A$ is an odd integer greater than $1$, in the summation over Feynman diagrams, there always exists another diagram in which the $A$-set, instead of being directly attached to $L_{(i,\bullet)}$, is first connected to an internal line, which is in turn connected to $L_{(i,\bullet)}$. The same circumstance holds for the $B$-set. Therefore, we must sum over the four diagrams shown on the l.h.s. of Fig.\ref{Fig11}; the cancellation occurs precisely within this summation.

We now examine that the four diagrams in Fig.\ref{Fig11} cancel each other. By employing the $K^2$-form of $V_{2n+2}$ in \eref{cayley}, as well as the observation \eref{separate-V-NLSM}, these diagrams can be evaluated in turn:
\bea
F_1&\xrightarrow[]{\eref{kine-condi-shuffle}}&\,\Big[\,{(-1)^{(n_A+n_B-2)/2}\over 2^{(n_A+n_B)/2}}\,\Big(K_A^2+K_B^2\Big)\,\Big]\,{1\over s_{i\{A\}\{B\}}}\,,\nn
F_2&\xrightarrow[]{\eref{kine-condi-shuffle}}&\,\Big[\,{(-1)^{(n_A-3)/2}\over 2^{(n_A-1)/2}}\,K_A^2\,\Big]\,{1\over s_{\{A\}}}\,\Big[\,{(-1)^{(n_B-1)/2}\over 2^{(n_B+1)/2}}\,\Big(k_{\{A\}}^2+K_B^2\Big)\,\Big]\,{1\over s_{i\{A\}\{B\}}}\,,\nn
F_3&\xrightarrow[]{\eref{kine-condi-shuffle}}&\,\Big[\,{(-1)^{(n_B-3)/2}\over 2^{(n_B-1)/2}}\,K_B^2\,\Big]\,{1\over s_{\{B\}}}\,\Big[\,{(-1)^{(n_A-1)/2}\over 2^{(n_A+1)/2}}\,\Big(k_{\{B\}}^2+K_A^2\Big)\,\Big]\,{1\over s_{i\{A\}\{B\}}}\,,\nn
F_4&\xrightarrow[]{\eref{kine-condi-shuffle}}&\,\Big[\,{(-1)^{(n_A-3)/2}\over 2^{(n_A-1)/2}}\,K_A^2\,\Big]\,{1\over s_{\{A\}}}\,
\Big[\,{(-1)^{(n_B-3)/2}\over 2^{(n_B-1)/2}}\,K_B^2\,\Big]\,{1\over s_{\{B\}}}\,\Big[{1\over2}\,\Big(k_{\{A\}}^2+k_{\{B\}}^2\Big)\Big]\,{1\over s_{i\{A\}\{B\}}}\,.
\eea
Putting them together, we get
\bea
F_1+F_2+F_3+F_4=0\,.
\eea
Thus we conclude that, under the kinematic condition \eref{kine-condi-shuffle}, the contributions of vertices with both $n_A$ and $n_B$ odd are canceled in the summation over Feynman diagrams.

\subsection{General SFASL: recursive proof}
\label{subsec-NLSM-gen}

Now we present a general proof of the SFASL expressed in \eref{fac-propa+v-gen}, for the NLSM diagrams. First, we note that our treatment in the previous subsection for the case where both $n_A$ and $n_B$ are odd is generic; under the kinematic constraint \eref{kine-condi-shuffle}, this case never contributes to the amplitude. Therefore, we only need to consider the case where both $n_A$ and $n_B$ are even. The proof will be carried out in a recursive manner.

In subsection \ref{subsec-NLSM-case1}, we showed that the SFASL in \eref{fac-propa+v-gen} holds for $p=q=1$. Moreover, when $p=0$ or $q=0$, the SFASL in \eref{fac-propa+v-gen} trivially holds, since there are no shuffle permutations in these cases. Now we assume that \eref{fac-propa+v-gen} holds for $(p,q-1)$, $(p-1,q)$, and $(p-1,q-1)$. To apply this assumption, we separate the l.h.s. of \eref{fac-propa+v-gen} into three parts,
\bea
\sum_{\shuffle(p,q)}\,\prod_{t=1}^{p+q-N_{A|B}}\,{V^{(i,\bullet)}_t\over D_t^{(i,\bullet)}}\,&=&\,\Big(\,\sum_{\shuffle(p,q-1)}\,\prod_{t=1}^{p+q-1-N_{A|B}}\,{V^{(i,\bullet)}_t\over D_t^{(i,\bullet)}}\,\Big)\,{V_{\{B\}_q}\over s_{i\underline{\{A\}}\{A\}_p\underline{\{B\}}\{B\}_q}}\nn
&&+\Big(\,\sum_{\shuffle(p-1,q)}\,\prod_{t=1}^{p+q-1-N'_{A|B}}\,{V^{(i,\bullet)}_t\over D_t^{(i,\bullet)}}\,\Big)\,{V_{\{A\}_p}\over s_{i\underline{\{A\}}\{A\}_p\underline{\{B\}}\{B\}_q}}\nn
&&+\Big(\,\sum_{\shuffle(p-1,q-1)}\,\prod_{t=1}^{p+q-2-N''_{A|B}}\,{V^{(i,\bullet)}_t\over D_t^{(i,\bullet)}}\,\Big)\,{V_{\{A\}_p|\{B\}_q}\over s_{i\underline{\{A\}}\{A\}_p\underline{\{B\}}\{B\}_q}}\,,~~~\label{lhs-SFASL-NLSM}
\eea
where we have defined $\underline{\{A\}}\equiv\{A\}_1\cdots\{A\}_{p-1}$ and $\underline{\{B\}}\equiv\{B\}_1\cdots\{B\}_{q-1}$ to simplify the expression. In \eref{lhs-SFASL-NLSM},
three parts correspond to three cases $(p,q-1)$, $(p-1,q)$ and $(p-1,q-1)$, respectively. Using the SFASL, as well as the $K\cdot K$-form of $V_{2n+2}$ in \eref{cayley}, we can evaluate the first part as,
\bea
F_{(p,q-1)}&=&\Big(\,\sum_{\shuffle(p,q-1)}\,\prod_{t=1}^{p+q-1-N_{A|B}}\,{V^{(i,\bullet)}_t\over D_t^{(i,\bullet)}}\,\Big)\,{V_{\{B\}_q}\over s_{i\underline{\{A\}}\{A\}_p\underline{\{B\}}\{B\}_q}}\nn
&\xrightarrow[]{\eref{kine-condi-shuffle}}&\,\Big[\Big(\prod_{\a=1}^p\,{V_{\{A\}_\a}\over s_{i\{A\}_1\cdots \{A\}_{\a}}}\Big)\,
\Big(\prod_{\b=1}^{q-1}\,{V_{\{B\}_\b}\over s_{i\{B\}_1\cdots \{B\}_{\b}}}\Big)\Big]\,{V_{\{B\}_q}\over s_{i\underline{\{A\}}\{A\}_p\underline{\{B\}}\{B\}_q}}\nn
&=&\Big[\Big(\prod_{\a=1}^{p-1}\,{V_{\{A\}_\a}\over s_{i\{A\}_1\cdots \{A\}_{\a}}}\Big)\,
\Big(\prod_{\b=1}^{q-1}\,{V_{\{B\}_\b}\over s_{i\{B\}_1\cdots \{B\}_{\b}}}\Big)\Big]\,{V_{\{A\}_p}\over s_{i\underline{\{A\}}\{A\}_p}}\,{V_{\{B\}_q}\over s_{i\underline{\{A\}}\{A\}_p\underline{\{B\}}\{B\}_q}}\nn
&=&C_{(p-1,q-1)}\,{(-1)^{m_{A_p}}\over2^{m_{A_p}+1}}\,{{\cal V}_{\{A\}_p}-2k_{i\underline{\{A\}}}\cdot k_{i\underline{\{A\}}\{A\}_p}\over s_{i\underline{\{A\}}\{A\}_p}}\nn
&&~~~~~~~~~~~~~{(-1)^{m_{B_q}}\over2^{m_{B_q}+1}}\,{{\cal V}_{\{B\}_q}-2k_{i\underline{\{A\}}\{A\}_p\underline{\{B\}}}\cdot k_{i\underline{\{A\}}\{A\}_p\underline{\{B\}}\{B\}_q}\over s_{i\underline{\{A\}}\{A\}_p\underline{\{B\}}\{B\}_q}}\nn
&=&C_{(p-1,q-1)}\,{(-1)^{m_{A_p}+m_{B_q}}\over2^{m_{A_p}+m_{B_q}+2}}\,{{\cal V}_{\{A\}_p}-2k_{i\underline{\{A\}}}\cdot k_{i\underline{\{A\}}\{A\}_p}\over s_{i\underline{\{A\}}\{A\}_p}}\,{{\cal V}_{\{B\}_q}-2k_{i\underline{\{B\}}}\cdot k_{i\underline{\{B\}}\{B\}_q}\over s_{i\underline{\{A\}}\{A\}_p\underline{\{B\}}\{B\}_q}}\nn
&&+C_{(p-1,q-1)}\,{{\cal V}_{\{A\}_p}-2k_{i\underline{\{A\}}}\cdot k_{i\underline{\{A\}}\{A\}_p}\over s_{i\underline{\{A\}}\{A\}_p}}\,{-2k^2_{i\underline{\{A\}}\{A\}_p}\over s_{i\underline{\{A\}}\{A\}_p\underline{\{B\}}\{B\}_q}}\,,~~\label{F1-NLSM-gen}
\eea
where the definitions of ${\cal V}_{\{A\}_p}$, ${\cal V}_{\{B\}_q}$, $m_{A_p}$ and $m_{B_q}$ are the same as those in subsection \ref{subsec-NLSM-case1}.
The factor $C_{(p-1,q-1)}$ is introduced for simplicity,
\bea
C_{(p-1,q-1)}=\Big(\prod_{\a=1}^{p-1}\,{V_{\{A\}_\a}\over s_{i\{A\}_1\cdots \{A\}_{\a}}}\Big)\,
\Big(\prod_{\b=1}^{q-1}\,{V_{\{B\}_\b}\over s_{i\{B\}_1\cdots \{B\}_{\b}}}\Big)\,.
\eea
Notice that $V_{\{A\}_p}$ in the above is ${\cal V}_{\{A\}_p}-2k_{i\underline{\{A\}}}\cdot k_{i\underline{\{A\}}\{A\}_p}$,  this is because it is taken from the factorized form in the second line. In the last step, we have used
\bea
k_{i\underline{\{A\}}\{A\}_p\underline{\{B\}}}\cdot k_{i\underline{\{A\}}\{A\}_p\underline{\{B\}}\{B\}_q}
\,\xrightarrow[]{\eref{kine-condi-shuffle}}\,k_{i\underline{\{B\}}}\cdot k_{i\underline{\{B\}}\{B\}_q}+k^2_{i\underline{\{A\}}\{A\}_p}\,.
\eea
As in subsection \ref{subsec-NLSM-case1}, this decomposition aims to separate $V_{\{B\}_q}$ into a part that is completely independent of the momentum from $A$-lines and commutes with $V_{\{A\}_p}$, and a part that depends on the momentum from $A$-lines and does not commute with $V_{\{A\}_p}$. For the convenience of the following discussion, we say that the entire $F_{(p,q-1)}$ is divided into a commuting part and an non-commuting part.

The second part in \eref{lhs-SFASL-NLSM} can be evaluated similarly,
\bea
F_{(p-1,q)}&=&\Big(\,\sum_{\shuffle(p-1,q)}\,\prod_{t=1}^{p+q-1-N'_{A|B}}\,{V^{(i,\bullet)}_t\over D_t^{(i,\bullet)}}\,\Big)\,{V_{\{A\}_p}\over s_{i\underline{\{A\}}\{A\}_p\underline{\{B\}}\{B\}_q}}\nn
&\xrightarrow[]{\eref{kine-condi-shuffle}}&\,\Big[\Big(\prod_{\a=1}^{p-1}\,{V_{\{A\}_\a}\over s_{i\{A\}_1\cdots \{A\}_{\a}}}\Big)\,
\Big(\prod_{\b=1}^{q-1}\,{V_{\{B\}_\b}\over s_{i\{B\}_1\cdots \{B\}_{\b}}}\Big)\Big]\,{V_{\{B\}_q}\over s_{i\underline{\{B\}}\{B_q\}}}\,{V_{\{A\}_p}\over s_{i\underline{\{A\}}\{A\}_p\underline{\{B\}}\{B\}_q}}\nn
&=&C_{(p-1,q-1)}\,{(-1)^{m_{A_p}+m_{B_q}}\over2^{m_{A_p}+m_{B_q}+2}}\,{{\cal V}_{\{B\}_q}-2k_{i\underline{\{B\}}}\cdot k_{i\underline{\{B\}}\{B\}_q}\over s_{i\underline{\{B\}}\{B_q\}}}\,{{\cal V}_{\{A\}_p}-2k_{i\underline{\{A\}}}\cdot k_{i\underline{\{A\}}\{A\}_p}\over s_{i\underline{\{A\}}\{A\}_p\underline{\{B\}}\{B\}_q}}\nn
&&+C_{(p-1,q-1)}\,{(-1)^{m_{A_p}+m_{B_q}}\over2^{m_{A_p}+m_{B_q}+2}}\,{{\cal V}_{\{B\}_q}-2k_{i\underline{\{B\}}}\cdot k_{i\underline{\{B\}}\{B\}_q}\over s_{i\underline{\{B\}}\{B_q\}}}\,{-2k^2_{i\underline{\{B\}}\{B\}_q}\over s_{i\underline{\{A\}}\{A\}_p\underline{\{B\}}\{B\}_q}}\,.
\eea
Again, $F_{(p-1,q)}$ is divided into a commuting part and an non-commuting part.
Meanwhile, the third part can be computed as
\bea
F_{(p-1,q-1)}&=&\Big(\,\sum_{\shuffle(p-1,q-1)}\,\prod_{t=1}^{p+q-2-N'_{A|B}}\,{V^{(i,\bullet)}_t\over D_t^{(i,\bullet)}}\,\Big)\,{V_{\{A\}_p|\{B\}_q}\over s_{i\underline{\{A\}}\{A\}_p\underline{\{B\}}\{B\}_q}}\nn
&\xrightarrow[]{\eref{kine-condi-shuffle}}&\,\Big[\Big(\prod_{\a=1}^{p-1}\,{V_{\{A\}_\a}\over s_{i\{A\}_1\cdots \{A\}_{\a}}}\Big)\,
\Big(\prod_{\b=1}^{q-1}\,{V_{\{B\}_\b}\over s_{i\{B\}_1\cdots \{B\}_{\b}}}\Big)\Big]\,{V_{\{A\}_p|\{B\}_q}\over s_{i\underline{\{A\}}\{A\}_p\underline{\{B\}}\{B\}_q}}\nn
&=&C_{(p-1,q-1)}\,{(-1)^{m_{A_p}+m_{B_q}}\over2^{m_{A_p}+m_{B_q}+1}}\,{{\cal V}_{\{A\}_p}+{\cal V}_{\{B\}_q}-2k_{i\underline{\{A\}}\underline{\{B\}}}\cdot k_{i\underline{\{A\}}\{A\}_p\underline{\{B\}}\{B\}_q}\over s_{i\underline{\{A\}}\{A\}_p\underline{\{B\}}\{B\}_q}}\,.
\eea
Adding $F_{(p,q-1)}$, $F_{(p-1,q)}$ and $F_{(p-1,q-1)}$ together, and using $s_{i\underline{\{A\}}\{A_p\}}=k^2_{i\underline{\{A\}}\{A_p\}}$,
$s_{i\underline{\{B\}}\{B_q\}}=k^2_{i\underline{\{B\}}\{B_q\}}$,
as well as the observation
\bea
k_{i\underline{\{A\}}\underline{\{B\}}}\cdot k_{i\underline{\{A\}}\{A\}_p\underline{\{B\}}\{B\}_q}
&\xrightarrow[]{\eref{kine-condi-shuffle}}& k_{i\underline{\{A\}}}\cdot k_{i\underline{\{A\}}\{A\}_p}+k_{i\underline{\{B\}}}\cdot k_{i\underline{\{B\}}\{B\}_q}\,,
\eea
we see that $F_{(p-1,q-1)}$ is canceled by non-commuting parts of $F_{(p,q-1)}$ and $F_{(p-1,q)}$. The final result comes from the commuting parts of $F_{(p,q-1)}$ and $F_{(p-1,q)}$,
\bea
&&F_{(p,q-1)}+F_{(p-1,q)}+F_{(p-1,q-1)}\nn
&=&C_{(p-1,q-1)}\,{(-1)^{m_{A_p}+m_{B_q}}\over2^{m_{A_p}+m_{B_q}+2}}\,\Big[\,\big({\cal V}_{\{A\}_p}-2k_{i\underline{\{A\}}}\cdot k_{i\underline{\{A\}}\{A\}_p}\big)\,\big({\cal V}_{\{B\}_q}-2k_{i\underline{\{B\}}}\cdot k_{i\underline{\{B\}}\{B\}_q}\big)\,\Big]\nn
&&\Big(\,{1\over s_{i\underline{\{A\}}\{A\}_p}}+{1\over s_{i\underline{\{B\}}\{B\}_q}}\,\Big)\,{1\over s_{i\underline{\{A\}}\{A\}_p\underline{\{B\}}\{B\}_q}}\nn
\,&\xrightarrow[]{\eref{kine-condi-shuffle}}&\,C_{(p-1,q-1)}\,\Big({(-1)^{m_{A_p}}\over2^{m_{A_p}+1}}\,{{\cal V}_{\{A\}_p}-2k_{i\underline{\{A\}}}\cdot k_{i\underline{\{A\}}\{A\}_p}\over s_{i\underline{\{A\}}\{A\}_p}}\Big)\,\Big({(-1)^{m_{B_q}}\over2^{m_{B_q}+1}}\,{{\cal V}_{\{B\}_q}-2k_{i\underline{\{B\}}}\cdot k_{i\underline{\{B\}}\{B\}_q}\over s_{i\underline{\{B\}}\{B\}_q}}\Big)\nn
&=&\Big(\prod_{\a=1}^{p}\,{V_{\{A\}_\a}\over s_{i\{A\}_1\cdots \{A\}_{\a}}}\Big)\,\times\,
\Big(\prod_{\b=1}^{q}\,{V_{\{B\}_\b}\over s_{i\{B\}_1\cdots \{B\}_{\b}}}\Big)\,,~~\label{SFASL-NLSM-proof}
\eea
where we have used the observation \eref{key-observation-general}.
The factorization formula in the last line of \eref{SFASL-NLSM-proof} completes the recursive proof. As can be seen, apart from the factor $C_{(p-1,q-1)}$, the above process is exactly the same as that in subsection \ref{subsec-NLSM-case1}. That is, the SFASL can be achieved stepwise along $L_{(i,\bullet)}$ via an unique mechanism.

\subsection{From SFASL to hidden zero and $2$-split}
\label{subsec-NLSM-0and2split}

In this subsection, we will interpret the hidden zeros and $2$-split of tree NLSM amplitudes, via the SFASL exhibited by the Feynman diagrams of NLSM.

\subsubsection{Hidden zero}

The kinematic condition for the hidden zeros of NLSM amplitudes is also given by \eref{kine-condi-0-scalar}, and the interpretation of the hidden zeros is in complete agreement with the ${\rm Tr}(\phi^3)$ case. As in the ${\rm Tr}(\phi^3)$ case, we can divide $\pmb A$ and
$\pmb B$ separately into a series of subsets, each of which will contribute the corresponding BG current of the NLSM. Similar to Fig.\ref{Fig5}, for a given division, the summation over Feynman diagrams involves a summation over shuffle permutations. Applying the SFASL to the line $L_{(i,j)}$ (multiplied by $D^{(i,\bullet)}_{p+q}/D^{(i,\bullet)}_{p+q}$ to compensate for the difference between internal and external lines), we find that, in the NLSM case, \eref{0-block} is replaced by
\bea
\Big[\sum_{\shuffle(p,q)}\,\prod_{t=1}^{p+q-N_{A|B}}\,{V^{(i,\bullet)}_t\over D_t^{(i,\bullet)}}\Big]\,s_{i\{A\}_1\cdots\{A\}_p\{B\}_1\cdots\{B\}_q}\,&\xrightarrow[]{\eref{kine-condi-0-scalar}}&\,\Big(\prod_{\a=1}^p\,{V_{\{A\}_\a}\over s_{i\{A\}_1\cdots \{A\}_{\a}}}\Big)\,
\Big(\prod_{\b=1}^q\,{V_{\{B\}_\b}\over s_{i\{B\}_1\cdots \{B\}_{\b}}}\Big)\,k_j^2\,.~~\label{0-NLSM}
\eea
Note that $p$ and $q$ in the above denote the numbers of $A$-sets and $B$-sets, respectively, rather than the numbers of subsets in the given division.
Although the expression becomes more complicated, the factor $k_j^2$ remains unchanged. Again, the on-shell condition $k_j^2=0$ indicates the vanishing of \eref{0-NLSM}. This phenomenon holds for any divisions of $\pmb A$ and $\pmb B$, thus we arrive at the following hidden zero
\bea
{\cal A}_{2n}^{\rm NLSM}(1,\cdots,2n)\,&\xrightarrow[]{\eref{kine-condi-0-scalar}}&\,0\,.~~\label{zero-NLSM}
\eea
%

\subsubsection{$2$-split: first case}

The kinematic condition for the $2$-split remains given by \eref{kine-condi-2split-scalar}. To reproduce the $2$-split of NLSM amplitudes, we first write down the analogue of the first line of \eref{2split-phi-detail}
\bea
{\cal A}_{2n}^{\rm NLSM}(1,\cdots,2n)&=&\sum_{{\rm div}\pmb A}\,\sum_{{\rm div}\pmb B}\,\sum_{{\cal P}_{\pmb A}}\,\sum_{{\cal P}_{\pmb B}}\,\Big(\sum_{\shuffle(p,q)}\,\prod_{t=1}^{p+q-N_{A|B}}\,{V_t^{(i,v)}\over D_t^{(i,v)}}\Big)\,\Big(\sum_{\shuffle(m,l)}\,\prod_{t=1}^{m+l-N'_{A|B}}\,{V_t^{(j,v)}\over D_t^{(j,v)}}\Big)\,V_v\,
f^{\rm NLSM}(R)\,.\nn~~\label{NLSM-forsplit}
\eea
The above expression should be understood as follows. We divide $\pmb A$ into $r$ subsets and $\pmb B$ into $h$ subsets,
namely
\bea
\pmb A=\{A_1,\cdots,A_r\}\,,~~~~~~~~\pmb B=\{B_1,\cdots,B(k),\cdots,B_h\}\,,~~\label{div-AB-NLSM}
\eea
with the external leg $k$ belonging to $B(k)$. Each subset (except $B(k)$) will generate a corresponding BG current, which is attached to $L_{(i,j)}$ via either an $A$-line or a $B$-line. The subset $B(k)$ gives rise to a BG current attached to $v$, where $v$ remains the vertex at which $L_{(i,v)}$, $L_{(j,v)}$, and $L_{(k,v)}$ meet. We refer to each specific way of attaching the subsets of $\pmb A$ to $L_{(i,j)}$ as an $\pmb A$-side partition, denoted as ${\cal P}_{\pmb A}$. We further require that, for a given
${\cal P}_{\pmb A}$, it is determined which subsets of $\pmb A$ are attached to $L_{(i,v)}$, which to $L_{(j,v)}$, and which to $v$. The definition of ${\cal P}_{\pmb B}$ is completely analogous. Once ${\cal P}_{\pmb A}$ and ${\cal P}_{\pmb B}$ are given, all
$A$-sets and $B$-sets attached onto $L_{(i,v)}$ and $L_{(j,v)}$ are determined accordingly. Summing over Feynman diagrams under a given partition and given
${\cal P}_{\pmb A}$ and ${\cal P}_{\pmb B}$ yields, in addition to the BG currents corresponding to subsets of $\pmb A$ and $\pmb B$, a summation over shuffle permutations among the $A$-sets and $B$-sets on $L_{(i,v)}$ and $L_{(j,v)}$. This leads to the amplitude being expressible in the formula \eref{NLSM-forsplit}. In \eref{NLSM-forsplit}, $V_v$ is the contribution from the vertex $v$. The summation over Feynman diagrams is separated into the summation over divisions, the summation over partitions ${\cal P}_{\pmb A}$ and ${\cal P}_{\pmb B}$, as well as the summation over shuffle permutations.

Applying the SFASL to \eref{NLSM-forsplit} yields the factorization structures
\bea
\sum_{\shuffle(p,q)}\,\prod_{t=1}^{p+q-N_{A|B}}\,{V_t^{(i,v)}\over D_t^{(i,v)}}\,&\xrightarrow[]{\eref{kine-condi-2split-scalar}}&\,\Big(\prod_{\a=1}^p\,{V_{\{A\}_\a}\over s_{i\{A\}_1\cdots \{A\}_{\a}}}\Big)\,\times\,
\Big(\prod_{\b=1}^q\,{V_{\{B\}_\b}\over s_{i\{B\}_1\cdots \{B\}_{\b}}}\Big)\,,\nn
\sum_{\shuffle(m,l)}\,\prod_{t=1}^{m+l-N'_{A|B}}\,{V_t^{(i,v)}\over D_t^{(i,v)}}\,&\xrightarrow[]{\eref{kine-condi-2split-scalar}}&\,\Big(\prod_{\a=1}^m\,{V_{\{A\}'_\a}\over s_{i\{A\}'_1\cdots \{A\}'_{\a}}}\Big)\,\times\,
\Big(\prod_{\b=1}^l\,{V_{\{B\}'_\b}\over s_{i\{B\}'_1\cdots \{B\}'_{\b}}}\Big)\,.~~\label{fac-alongLR-NLSM}
\eea
Meanwhile, based on the divisions in \eref{div-AB-NLSM}, we know that $f^{\rm NLSM}(R)$ takes the form
\bea
f^{\rm NLSM}(R)&=&\Big(\prod_{\a=1}^r\,{\cal J}^{\rm NLSM}_{A_\a}\,{1\over s_{A_\a}}\Big)\,\Big(\prod_{\b=1}^h\,{\cal J}^{\rm NLSM}_{B_\b}\,{1\over s_{B_\b}}\Big)\,,~~\label{f(R)-NLSM}
\eea
which automatically factorizes as
\bea
f^{\rm NLSM}(R)=f^{\rm NLSM}_A(R)\,\times\,f^{\rm NLSM}_B(R)\,,~~\label{fac-f-NLSM}
\eea
where
\bea
f^{\rm NLSM}_A(R)=\prod_{\a=1}^r\,{\cal J}^{\rm NLSM}_{A_\a}\,{1\over s_{A_\a}}\,,~~~~~~~~f^{\rm NLSM}_B(R)=\prod_{\b=1}^h\,{\cal J}^{\rm NLSM}_{B_\b}\,{1\over s_{B_\b}}\,.
\eea
All these factorizations are in one-to-one correspondence with the ${\rm Tr}(\phi^3)$ case, as illustrated in Fig.\ref{Fig6}, where cubic vertices are replaced by general vertices carrying $A$-sets and $B$-sets.

To reproduce the complete $2$-split behavior of the NLSM amplitudes, we next need to analyze the behavior of the term $V_v$ coming from vertex $v$. This is a new situation, since in the ${\rm Tr}(\phi^3)$ case $V_v$ is a trivial constant $1$.

\begin{figure}
  \centering
   \includegraphics[width=16cm]{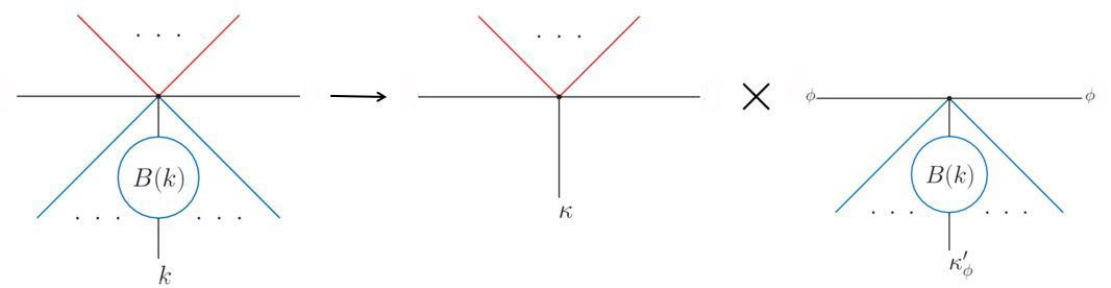} \\
  \caption{Factorization of $V_v$ when $n_A$ is odd and $n_B$ is even. The vertex $v$ is represented by $\bullet$.}\label{Fig12}
\end{figure}

Suppose that in addition to three lines $L_{(i,v)}$, $L_{(j,v)}$ and $L_{(k,v)}$, there are
$n_A$ $A$-lines and $n_B$ $B$-lines attached to $v$. Since the NLSM theory contains only even-point vertices, there are two possibilities. The first is that $n_A$ is odd and $n_B$ is even; the second is that $n_A$ is even and $n_B$ is odd. We begin with the first case. Using the $K^2$-form of $V_{2n+2}$ in \eref{cayley}, one can see that when $n_A$ is odd and
$n_B$ is even, the contribution of this vertex takes the form
\bea
V_v={(-1)^{(n_A+n_B-1)/2}\over 2^{(n_A+n_B+1)/2}}\,(K'_A+K'_B+k_{B(k)})^2\,,~~~~{\rm or}~~~~V_v={(-1)^{(n_A+n_B-1)/2}\over 2^{(n_A+n_B+1)/2}}\,(K'_A+K'_B)^2\,,~~\label{2Vv-odd}
\eea
where $K'_A$ is the sum of the momenta of some (not all) of the $A$-lines written down according to the rule in \eref{cayley}, and
$K'_B$ is the sum of the momenta of some of the $B$-lines by the same rule. As the simplest example, suppose
$v$ is a quartic vertex, and in addition to $L_{(i,v)}$, $L_{(j,v)}$, and $L_{(k,v)}$, there is a subset $A_0$ of
$\pmb A$ attached to $v$ via an $A$-line. In this case, we have
$V_v=(k_{A_0}+k_{B(k)})^2$,
which exactly corresponds to the first case in \eref{2Vv-odd}, with $K'_A=k_{A_0}$, $K'_B=0$.

Repeating the discussion in subsection \ref{subsec-NLSM-case2}, we see that the second case in \eref{2Vv-odd} cancels out in the summation over Feynman diagrams, via the same mechanism. Therefore, we only need to consider the first case in \eref{2Vv-odd}. Under the kinematic constraint \eref{kine-condi-2split-scalar}, the first case satisfies
\bea
V_v\,\xrightarrow[]{\eref{kine-condi-2split-scalar}}\,{(-1)^{(n_A+n_B-1)/2}\over 2^{(n_A+n_B+1)/2}}\,(K'_A+k_k)^2+{(-1)^{(n_A+n_B-1)/2}\over 2^{(n_A+n_B+1)/2}}\,(K'_B+k_{B(k)})^2\,.
\eea
The $(K'_B+k_{B(k)})^2$-part cancels out in the summation over Feynman diagrams, because $\{B\}_v\cup L_{(k,v)}$ contains an odd number of lines, so the cancellation mechanism discussed in subsection \ref{subsec-NLSM-case2} applies perfectly to this case. Here $\{B\}_v$ denotes the set of $B$-lines coupled to $v$. The
$(K'_A+k_k)^2$-part, on the other hand, exactly gives the contribution of an
$(n_A+3)$-point NLSM vertex. We thus obtain the following effective factorization formula
\bea
V_v\,\xrightarrow[]{\eref{kine-condi-2split-scalar}}\,V^{\rm NLSM}_{n_A+3}\,\times\,V^{{\rm NLSM}\oplus{\rm Tr}(\phi^3)}_{n_B+3}\,,~~\label{fac1-v-NLSM}
\eea
where
\bea
V^{\rm NLSM}_{n_A+3}={(-1)^{(n_A-1)/2}\over 2^{(n_A+1)/2}}\,(K'_A+k_k)^2\,,~~~~V^{{\rm NLSM}\oplus{\rm Tr}(\phi^3)}_{n_B+3}={(-1)^{n_B/2}\over 2^{n_B/2}}\,.
\eea
as illustrated in Fig.\ref{Fig12}.
In \eref{fac1-v-NLSM}, $V^{\rm NLSM}_{n_A+3}$ is a pure NLSM vertex, while $V^{{\rm NLSM}\oplus{\rm Tr}(\phi^3)}_{n_B+3}$ is an NLSM$\oplus{\rm Tr}(\phi^3)$ vertex describing three ${\rm Tr}(\phi^3)$-scalars coupled to $n_B$ pions \cite{Low:2017mlh,Yin:2018hht,Low:2018acv}.

Plugging \eref{fac1-v-NLSM} into \eref{NLSM-forsplit} and using \eref{fac-alongLR-NLSM}, \eref{fac-f-NLSM}, we ultimately get the first version of the $2$-split of NLSM amplitudes,
\bea
{\cal A}^{\rm NLSM}_{2n}(1,\cdots,2n)\,\xrightarrow[]{\eref{kine-condi-2split-scalar}}\,{\cal J}^{\rm NLSM}_{n_1}(i,\pmb A,j,\kappa)\,\times\,{\cal J}^{{\rm NLSM}\oplus{\rm Tr}(\phi^3)}_{2n+3-n_1}(j_\phi,\pmb B(\kappa'_\phi),i_\phi)\,,~~\label{2split-NLSM-v1}
\eea
where
\bea
{\cal J}^{\rm NLSM}_{n_1}(i,\pmb A,j,\kappa)&=&\sum_{{\rm div}\pmb A}\,\sum_{{\cal P}_{\pmb A}}\,\Big(\prod_{\a=1}^p\,{V_{\{A\}_\a}\over s_{i\{A\}_1\cdots \{A\}_{\a}}}\Big)\,\Big(\prod_{\a=1}^m\,{V_{\{A\}_\a}\over s_{i\{A\}_1\cdots \{A\}_{\a}}}\Big)\,V^{\rm NLSM}_{n_A+3}\,f^{\rm NLSM}_A(R)\,,\nn
{\cal J}^{{\rm NLSM}\oplus{\rm Tr}(\phi^3)}_{2n+3-n_1}(j_\phi,\pmb B(\kappa'_\phi),i_\phi)&=&\sum_{{\rm div}\pmb B}\,\sum_{{\cal P}_{\pmb B}}\,\Big(\prod_{\b=1}^q\,{V_{\{B\}_\b}\over s_{i\{B\}_1\cdots \{B\}_{\b}}}\Big)\,\Big(\prod_{\b=1}^l\,{V_{\{B\}_\b}\over s_{i\{B\}_1\cdots \{B\}_{\b}}}\Big)\,V^{{\rm NLSM}\oplus{\rm Tr}(\phi^3)}_{n_B+3}\,f^{\rm NLSM}_B(R)\,.\nn
\eea
From the facts that the NLSM has only even-point vertices and that
$n_A$ is odd, it is easy to see that
$n_1$ is even and $2n+3-n_1$ is odd.

\subsubsection{$2$-split: second case}

%
\begin{figure}
  \centering
   \includegraphics[width=16cm]{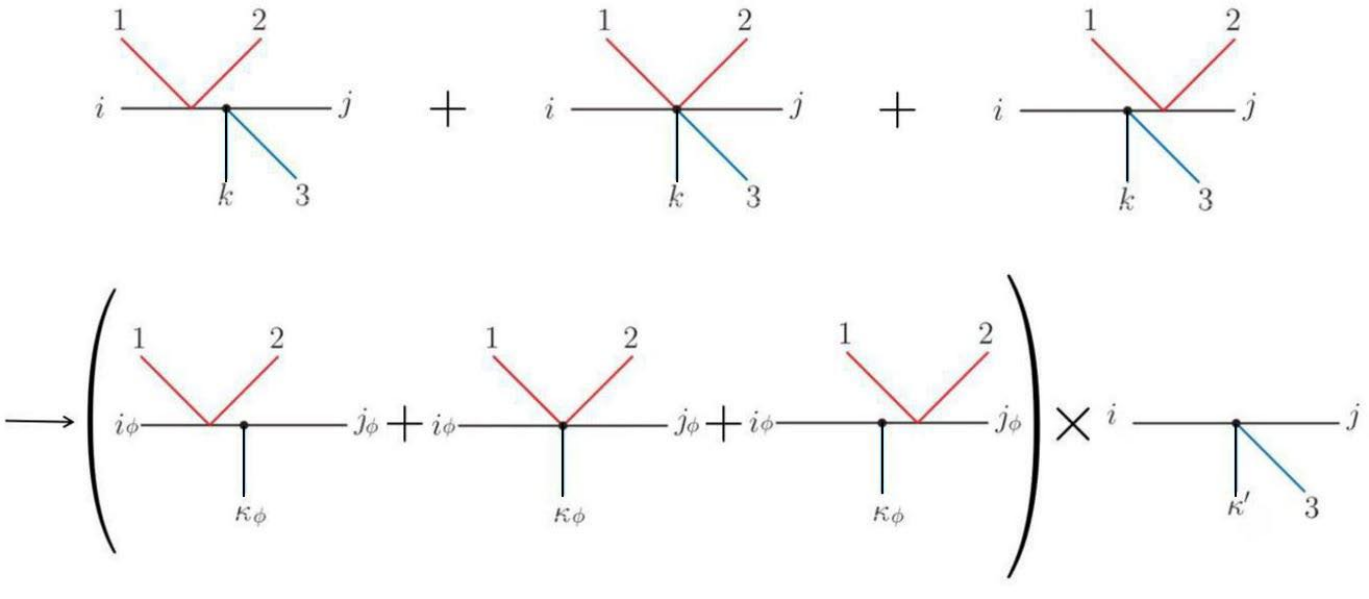} \\
  \caption{A simple example of the cancelation in the second case of $2$-split.}\label{Fig13}
\end{figure}

We now turn to the second case, where $n_A$ is even and $n_B$ is odd. In this case, we will encounter a new cancellation mechanism. To see this cancellation, let us first consider the example shown in Fig.\ref{Fig13}. This $6$-point amplitude has four Feynman diagrams. One of them vanishes under the kinematic condition \eref{kine-condi-2split-scalar} (with $\pmb A=\{1,2\}$, $\pmb B=\{3,k\}$), and the remaining three are displayed in the first line of Fig.\ref{Fig13}. We analyze the behavior of vertex $v$ using the $K^2$-form in \eref{cayley}. If we stipulate that the external momentum $k_k$ does not appear in the chosen $K^2$-form, then the contributions of the three diagrams can be calculated sequentially as
\bea
F_1&=&\Big[\,{1\over2}\,\big(k_i+k_2\big)^2\,\Big]\,{1\over s_{i12}}\,\Big[\,{1\over 2}\,\big(k_{i12}+k_3\big)^2\,\Big]\,,\nn
F_2&=&{-1\over 2^2}\,\big(k_i+k_2+k_3\big)^2\,,\nn
F_3&=&\Big[\,{1\over2}\,\big(k_i+k_3\big)^2\,\Big]\,{1\over s_{12j}}\,V'_A\,,
\eea
where $V'_A$ denotes the contribution from the right vertex in the third diagram. Using
\bea
k^2_{i123}\,\xrightarrow[]{\eref{kine-condi-2split-scalar}}\,k^2_{i12}+k^2_{i3}\,,~~~~~~~~k^2_{i23}\,\xrightarrow[]{\eref{kine-condi-2split-scalar}}\,
k^2_{i2}+k^2_{i3}\,,
\eea
we see that
\bea
F_1+F_2\,&\xrightarrow[]{\eref{kine-condi-2split-scalar}}&\,{1\over 4}\,{1\over s_{i12}}\,k^2_{i2}\,\big(k^2_{i12}+k^2_{i3}\big)
-{1\over4}\,\big(k^2_{i2}+k^2_{i3}\big)\nn
&=&{1\over 4}\,{1\over s_{i12}}\,k^2_{i2}\,k^2_{i3}-{1\over4}\,k^2_{i3}\,,
\eea
where $k_{i2}^2/4$ in the first and second terms cancel each other.
Therefore,
\bea
F_1+F_2+F_3\,&\xrightarrow[]{\eref{kine-condi-2split-scalar}}&\,\Big({1\over2}\,\big(k_i+k_2\big)^2\,{1\over s_{i12}}-{1\over2}+{1\over s_{12j}}\,V'_A\Big)\,\times\,{1\over2}\,\big(k_i+k_3\big)^2\nn
&=&{\cal J}^{{\rm NLSM}\oplus{\rm Tr}(\phi^3)}_5(i_\phi,1,2,j_\phi,\kappa_\phi)\,\times\,{\cal J}^{\rm NLSM}_4(j,3,\kappa,i)\,,
\eea
as illustrated in Fig.\ref{Fig13}.

In the first line of Fig.\ref{Fig13}, the first and third diagrams appear symmetric. However, the vertex
$v$ in the third diagram already takes its final form from the very beginning, whereas the vertex
$v$ in the first diagram does not. This is because our attempt to choose the
$K^2$-form such that the external momentum $k_k$ does not appear breaks the symmetry.

\begin{figure}
  \centering
   \includegraphics[width=16cm]{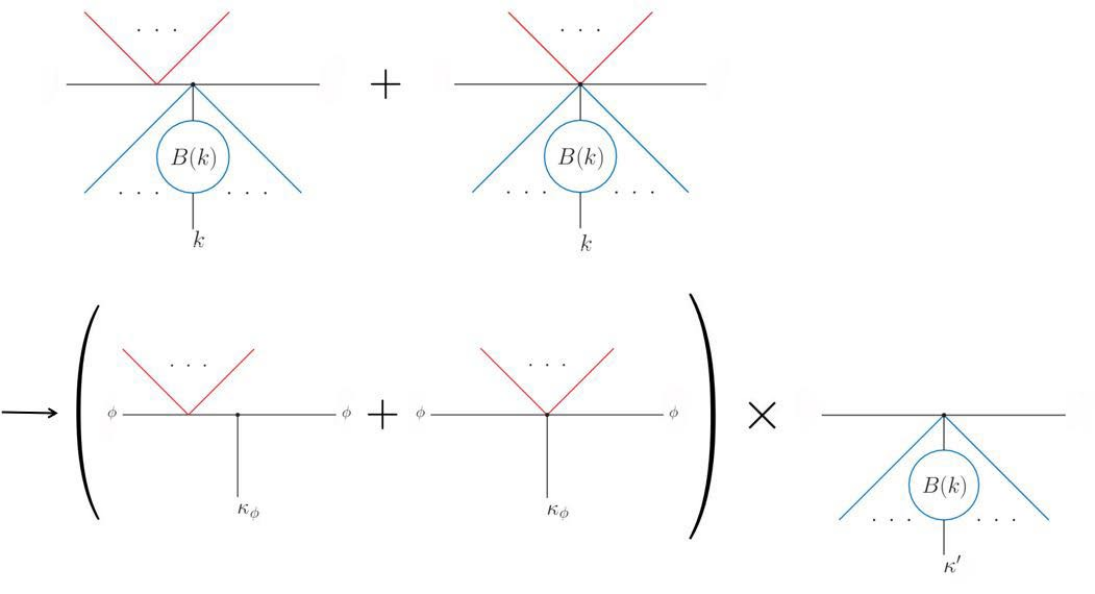} \\
  \caption{Factorization of $V_v$ when $n_A$ is even and $n_B$ is odd. The vertex $v$ is represented by $\bullet$.}\label{Fig14}
\end{figure}

\begin{figure}
  \centering
   \includegraphics[width=4.3cm]{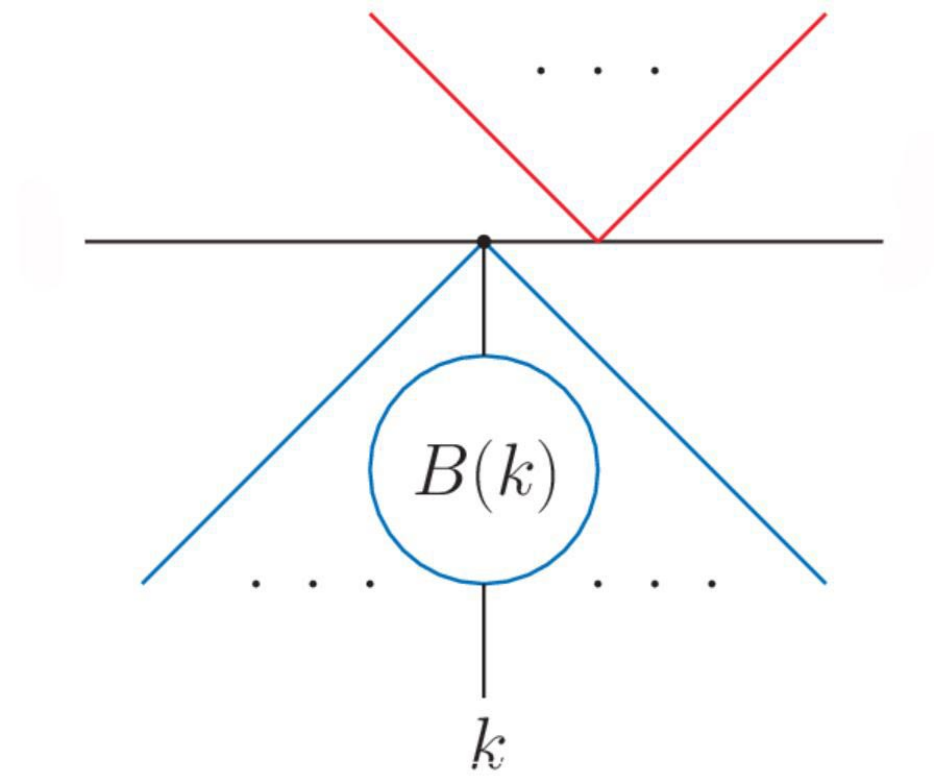} \\
  \caption{The diagram which does not participate the cancelation.}\label{Fig15}
\end{figure}

The cancellation between
$F_1$ and $F_2$ in this example can be generalized to the generic case. In the generic case, when summing over ${\cal P}_{\pmb A}$, the counterparts of the diagrams in Fig.\ref{Fig13} also appear (the diagrams in the first line of Fig.\ref{Fig14}, and the diagram in Fig.\ref{Fig15}). The vertex $v$ receives momenta from $L_{(i,v)}$ and $L_{(j,v)}$. When $n_A$ is even, one of these two momenta will appear in the $K^2$-form. If we still choose the
$K^2$-form such that the external momentum $k_k$ does not appear, then which of the two momenta appears in the
$K^2$-form will be determined accordingly. Without loss of generality, we assume that the momentum from $L_{(i,v)}$ appears in the
$K^2$-form. Then, the cancellation will occur between the two diagrams in the first line of Fig.\ref{Fig14}.

The computation of the two diagrams in the first line of Fig.\ref{Fig14} is straightforward:
\bea
F_1&=&\Big[\,{(-1)^{(n_A-2)/2}\over2^{n_A/2}}\,\big(k_{i\underline{\{A\}}}+K''_A\big)^2\,\Big]\,{1\over s_{i\underline{\{A\}}\{A\}_s}}\,\Big[\,{(-1)^{(n_B-1)/2}\over 2^{(n_B+1)/2}}\,\big(k_{i\underline{\{A\}}\{A\}_s\underline{\{B\}}}+K''_B\big)^2\,\Big]\,{\cal R}\,,\nn
F_2&=&{(-1)^{(n_A+n_B-1)/2}\over 2^{(n_A+n_B+1)/2}}\,\big(k_{i\underline{\{A\}}\underline{\{B\}}}+K''_A+K''_B\big)^2\,{\cal R}\,,~~\label{44}
\eea
where the $A$-set shown in Fig.\ref{Fig14} is denoted as $\{A\}_s$, and $\underline{\{A\}}$ stands for the series of $A$-sets on the l.h.s. of $\{A\}_s$, $\underline{\{B\}}$ stands for the series of $B$-sets on the l.h.s. of $v$. According to the $K^2$-form in \eref{cayley}, $K''_A$ is the sum of momenta carried by a proper subset of $A$-lines, while $K''_B$ is the sum of momenta carried by a proper subset of $B$-lines. Notice that the first vertex and the propagator in $F_1$ are expressed as in the factorized form \eref{fac-alongLR-NLSM}. That is, they are written down after applying the SFASL. Clearly, $K''_A$ and $K''_B$ in $F_1$ and $F_2$ are the same. Finally, ${\cal R}$ denotes the contribution from the remaining parts of diagrams, which is exactly the same for $F_1$ and $F_2$.

Using
\bea
\big(k_{i\underline{\{A\}}\{A\}_s\underline{\{B\}}}+K''_B\big)^2\,&\xrightarrow[]{\eref{kine-condi-2split-scalar}}&\,k^2_{i\underline{\{A\}}\{A\}_s}
+\big(k_{i\underline{\{B\}}}+K''_B\big)^2\,,\nn
\big(k_{i\underline{\{A\}}\underline{\{B\}}}+K''_A+K''_B\big)^2\,&\xrightarrow[]{\eref{kine-condi-2split-scalar}}&\,\big(k_{i\underline{\{A\}}}+K''_A\big)^2
+\big(k_{i\underline{\{B\}}}+K''_B\big)^2\,,
\eea
which are based on the observation \eref{key-observation-general},
we find
\bea
F_1+F_2\,&\xrightarrow[]{\eref{kine-condi-2split-scalar}}&\,{\cal R}\,
\Big[\,\Big({(-1)^{(n_A-2)/2}\over2^{n_A/2}}\,\big(k_{i\underline{\{A\}}}+K''_A\big)^2\Big)\,{1\over s_{i\underline{\{A\}}\{A\}_s}}+{(-1)^{n_A/2}\over 2^{n_A/2}}\Big]\nn
&&\times\,\Big[\,{(-1)^{(n_B-1)/2}\over 2^{(n_B+1)/2}}\,\big(k_{i\underline{\{B\}}}+K''_B\big)^2\,\Big]\nn
&=&{\cal R}\,\Big[\,V^{{\rm NLSM}\oplus{\rm Tr}(\phi^3)}_{n_A+2}{1\over s_{i\underline{\{A\}}\{A\}_s}}V^{{\rm Tr}(\phi^3)}_3+V^{{\rm NLSM}\oplus{\rm Tr}(\phi^3)}_{n_A+3}\,\Big]\,\times\,V^{\rm NLSM}_{n_B+3}\,,~~\label{F1F2-cancel}
\eea
as depicted in Fig.\ref{Fig14}. In the above, the cubic vertex $V^{{\rm Tr}(\phi^3)}_3$ is a trivial constant $1$. In the expression of ${\cal J}^{{\rm NLSM}\oplus{\rm Tr}(\phi^3)}_{n_1}$ in \eref{J-NLSM}, such a cubic vertex $V_3^{{\rm Tr}(\phi^3)}$
should be understood as the special case of $V^{{\rm NLSM}\oplus{\rm Tr}(\phi^3)}_{n_A+3}$, with $n_A=0$. The vertex $V^{{\rm NLSM}\oplus{\rm Tr}(\phi^3)}_{n_A+2}$ is equivalent to $V^{{\rm NLSM}}_{n_A+2}$, this is because of a well known fact: when only two ${\rm Tr}(\phi^3)$ scalars are involved, the NLSM$\oplus{\rm Tr}(\phi^3)$ vertex is equivalent to the NLSM vertex.

\begin{figure}
  \centering
   \includegraphics[width=10cm]{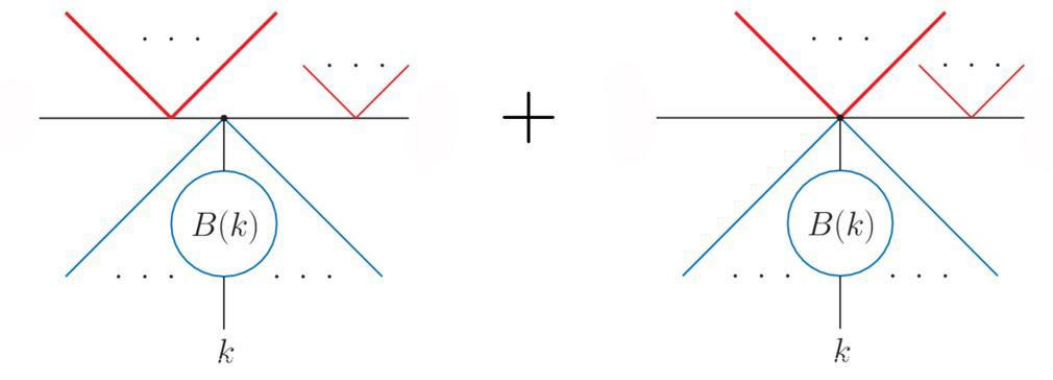} \\
  \caption{Treatment of Fig.\ref{Fig15}. The thin red line represents $\{A\}_s$, while the thick red line represents $\{A\}_{s-1}$.}\label{Fig16}
\end{figure}

We emphasize that Fig.\ref{Fig15} does not participate in the above cancellation process. For Fig.\ref{Fig15}, instead of concerning ourselves with the $A$-set $\{A\}_s$, we focus on the $A$-set $\{A\}_{s-1}$ on the left of $\{A\}_s$. In other words, we will consider the cancellation between the two diagrams in Fig.\ref{Fig16}, and the detailed process is exactly the same as that from \eref{44} to \eref{F1F2-cancel}.

By substituting the factorization formula \eref{F1F2-cancel} into \eref{NLSM-forsplit} and using \eref{fac-alongLR-NLSM}, \eref{fac-f-NLSM},
we obtain the second version of $2$-split,
\bea
{\cal A}^{\rm NLSM}_{2n}(1,\cdots,2n)\,\xrightarrow[]{\eref{kine-condi-2split-scalar}}\,{\cal J}^{{\rm NLSM}\oplus{\rm Tr}(\phi^3)}_{n_1}(i_\phi,\pmb A,j_\phi,\kappa_\phi)\,\times\,{\cal J}^{{\rm NLSM}}_{2n+3-n_1}(j,\pmb B(\kappa'),i)\,,~~\label{2split-NLSM-v2}
\eea
where
\bea
{\cal J}^{{\rm NLSM}\oplus{\rm Tr}(\phi^3)}_{n_1}(i_\phi,\pmb A,j_\phi,\kappa_\phi)&=&\sum_{{\rm div}\pmb A}\,\sum_{{\cal P}_{\pmb A}}\,\Big(\prod_{\a=1}^p\,{V_{\{A\}_\a}\over s_{i\{A\}_1\cdots \{A\}_{\a}}}\Big)\,\Big(\prod_{\a=1}^m\,{V_{\{A\}_\a}\over s_{i\{A\}_1\cdots \{A\}_{\a}}}\Big)\,V^{{\rm NLSM}\oplus{\rm Tr}(\phi^3)}_{n_A+3}\,f^{\rm NLSM}_A(R)\,,\nn
{\cal J}^{{\rm NLSM}}_{2n+3-n_1}(j,\pmb B(\kappa'),i)&=&\sum_{{\rm div}\pmb B}\,\sum_{{\cal P}_{\pmb B}}\,\Big(\prod_{\b=1}^q\,{V_{\{B\}_\b}\over s_{i\{B\}_1\cdots \{B\}_{\b}}}\Big)\,\Big(\prod_{\b=1}^l\,{V_{\{B\}_\b}\over s_{i\{B\}_1\cdots \{B\}_{\b}}}\Big)\,V^{\rm NLSM}_{n_B+3}\,f^{\rm NLSM}_B(R)\,.~~\label{J-NLSM}
\eea
In this case, $n_1$ is odd and $2n+3-n_1$ is even.

\section{YM amplitudes}
\label{sec-YM}

In this section, we turn to the SFASL of YM diagrams, as well as the corresponding interpretation of the hidden zeros and $2$-split of tree YM amplitudes.

\begin{figure}
  \centering
   \includegraphics[width=8cm]{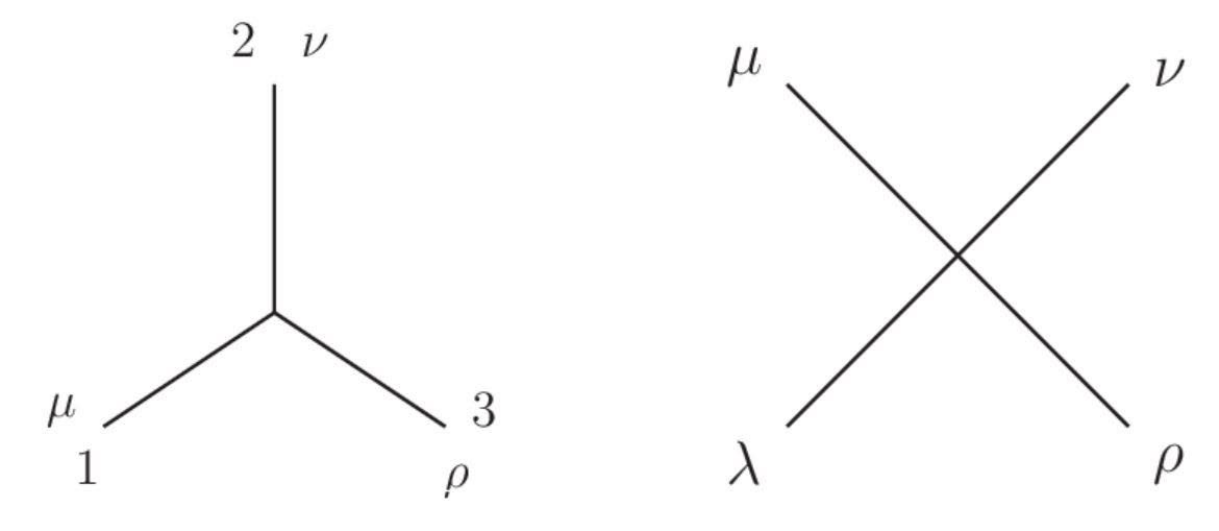} \\
  \caption{Lorentz indices of YM vertices correspond to Feynman rules in \eref{Feynmanrule-YM}.}\label{Index}
\end{figure}

Although it is well known, to avoid the ambiguity, we still list the color-ordered Feynman rules for YM.
According to the Lorentz indices labeled in Fig.\ref{Index}, the color-ordered Feynman rules for cubic and quartic vertices in the Lorentz-Feynman gauge are
\bea
V^{3g}_{\mu\nu\rho}&=&{i\over\sqrt{2}}\,\Big(\eta_{\mu\nu}\,(k_1-k_2)_\rho+\eta_{\nu\rho}\,(k_2-k_3)_\mu+\eta_{\rho\mu}\,(k_3-k_1)_\nu\Big)\,,\nn
V^{4g}_{\mu\nu\rho\lambda}&=&i\,\eta_{\mu\rho}\,\eta_{\nu\lambda}-{i\over2}\,(\eta_{\mu\nu}\,\eta_{\rho\lambda}+\eta_{\mu\lambda}\,\eta_{\nu\rho})\,.
~~\label{Feynmanrule-YM}
\eea
Meanwhile, the Feynman rule for the propagator is
\bea
\Delta^{\mu\nu}(k)=-i\,{\eta^{\mu\nu}\over k^2}\,.
\eea
In the subsequent discussion, we will omit these $i$ in the numerators.
Note that, according to the above convention of upper and lower indices, each BG current carries a lower index.
An crucial property of each BG current ${\cal J}_\a$ which will be used frequently in this section is
\bea
{\cal J}_o\cdot k_o=0\,,~~~~\label{BG-gauge}
\eea
due to the gauge invariance, where $o$ is the off-shell leg carried by this current \cite{Berends:1987me,Wu:2021exa}.

Form the Feynman rules in \eref{Feynmanrule-YM}, we see that the mass dimensions of cubic and quartic vertices are $1$ and $0$, respectively.
Therefore, the mass dimensions of YM vertices satisfy the requirement in \eref{mass-dim-v} (the case where $V_{\{A\}}$, $V_{\{B\}}$ and $V_{\{A\}|\{B\}}$ appear simultaneously in the shuffle permutation is shown in Fig.\ref{Fig20}).

In subsection \ref{subsec-subspace}, we introduce the idea of orthogonal subspaces, which is useful for generalizing the SFASL to the YM case where the external gluon $i$ carries a polarization vector. In subsections \ref{subsec-YM-case1}, \ref{subsec-YM-case2} and \ref{subsec-YM-case3}, we classify the simplest shuffle permutations into three cases, and show the corresponding SFASL in turn. In particular, in subsection \ref{subsec-YM-case3}, we will show that the mixed vertex which hinders the SFASL, is canceled by non-commuting parts in unmixed vertices, as in the NLSM case. Then, in subsection \ref{subsec-YM-gen}, we provide a recursive proof of the SFAS of YM diagrams. Finally, in subsection \ref{subsec-YM-0andsplit}, we interpret the hidden zeros and $2$-split of tree YM amplitudes through the SFASL.

\subsection{Orthogonal subspaces and generalization of SFASL}
\label{subsec-subspace}

For Feynman diagrams of YM, the SFASL expressed in \eref{fac-propa+v-gen} should be further generalized. Suppose that the particles propagating along the $A$-lines and $B$-lines are all gluons. Since gluons are vector particles, the role of the $A$-lines and $B$-lines is no longer merely to provide momenta for the vertices on $L_{(i,\bullet)}$. The BG currents attached to the $A$-lines or $B$-lines become contracted with the vertices on $L_{(i,\bullet)}$ via the Lorentz metric $\eta^{\mu\nu}$ carried by the propagators. Therefore, we first generalize the kinematic condition \eref{kine-condi-shuffle} from a constraint on momenta to a constraint on both momenta and BG currents,
\bea
k_{\hat{a}}\cdot k_{\hat{b}}=0\,,~~~~k_{\hat{a}}\cdot{\cal J}_{\hat{b}}=0\,,~~~~{\cal J}_{\hat{a}}\cdot k_{\hat{b}}=0\,,~~~~
{\cal J}_{\hat{a}}\cdot{\cal J}_{\hat{b}}=0\,, ~~\label{kine-condi-shuffle-YM}
\eea
where ${\cal J}_{\hat{a}}$ and ${\cal J}_{\hat{b}}$ are BG currents carried by the $A$-line $\hat{a}$ and $B$-line $\hat{b}$, respectively.

However, such a generalization is insufficient to realize the SFASL illustrated in Fig.\ref{Fig8}. The reason is that, since the external line $i$ is a gluon, it carries a polarization vector $\epsilon_i$. This immediately raises a question: in the factorization formula on the r.h.s. of Fig.\ref{Fig8}, which part should the polarization vector $\epsilon_i$ be assigned to? Obviously, assigning the $\epsilon_i$ to either part would break the symmetry and is therefore unjustified.

To address this difficulty, we observe that the kinematic condition \eref{kine-condi-shuffle-YM} can be understood as follows: the $A$-lines together with the blocks attached to them, and the $B$-lines together with the blocks attached to them, originate from two mutually orthogonal subspaces. That is, we decompose the $d$-dimensional spacetime into a $d_A$-dimensional subspace ${\cal S}_A$ and a $d_B$-dimensional subspace
${\cal S}_B$, satisfying $d_A+d_B=d$. Since the $d$-dimensional spacetime possesses only one time dimension, to define massless particles in each of the two subspaces, we should allow the momenta to take complex values. Of course, such a decomposition of the spacetime breaks manifest Lorentz invariance. However, since the amplitude is a function of various Lorentz invariants, for the hidden zeros and the $2$-split that we aim to study, the amplitude can only detect that certain Lorentz invariants vanish, but cannot detect why these Lorentz invariants become zero. Therefore, if we cause certain Lorentz invariants to vanish by decomposing the entire space into two mutually orthogonal subspaces, thereby obtaining a vanishing amplitude or a $2$-split behavior, the final conclusion can only state how the amplitude behaves when those Lorentz invariants are zero. Hence, the resulting statement remains Lorentz invariant.

Following the above idea, we can decompose the polarization vector into the two subspaces, i.e., $\epsilon_i=\epsilon_i^{{\cal S}_A}+\epsilon_i^{{\cal S}_B}$. Then, we can anticipate the SFASL illustrated in Fig.\ref{Fig8} as follows. When the polarization vector takes $\epsilon_i^{{\cal S}_A}$, the $A$-side part in the factorization formula acquires the polarization vector. For the $B$-side part, the polarization vector lies in the extra dimensions, so that the usual dimensional reduction idea implies that the particle propagating on $L_{(i,\bullet)}$ is a scalar. Conversely, when the polarization vector takes $\epsilon_i^{{\cal S}_B}$, the $B$-side part acquires the polarization vector. In this case, for the $A$-side part, the particle running along $L_{(i,\bullet)}$ is a scalar. For a generic polarization vector, the corresponding SFASL will contain both of the above contributions.

Therefore, for YM amplitudes, the SFASL takes the following extended form,
\bea
&&\sum_{\shuffle(p,q)}\,\Big(\epsilon_i\{\W{\cal J}_{\hat{a}}\}\{\W{\cal J}_{\hat{b}}\}\Big)^{\{\mu\}}\,\Big(\prod_{t=1}^{p+q-N_{A|B}}\,{V^{(i,\bullet)}_t\over D_t^{(i,\bullet)}}\Big)_{\{\mu\}\nu}\,\eta^{\nu\rho}\nn
&\xrightarrow[]{\eref{kine-condi-shuffle-YM}}&\,\Big[\,\Big(\epsilon_i^{{\cal S}_A}\{\W{\cal J}_{\hat{a}}\}\Big)^{\{\mu\}_1}\,\Big(\prod_{\a=1}^p\,{V_{\{A\}_\a}\over s_{i\{A\}_1\cdots \{A\}_{\a}}}\Big)_{\{\mu\}_1\nu}\,\eta^{\nu\rho}\,\Big]\,\times\,\Big[\,\Big(\{\W{\cal J}_{\hat{b}}\}\Big)^{\{\mu\}_2}\,
\Big(\prod_{\b=1}^q\,{V_{\{B\}_\b}\over s_{i\{B\}_1\cdots \{B\}_{\b}}}\Big)_{\{\mu\}_2}\,\Big]\nn
&&+\Big[\,\Big(\{\W{\cal J}_{\hat{a}}\}\Big)^{\{\mu\}_1}\,\Big(\prod_{\a=1}^p\,{V_{\{A\}_\a}\over s_{i\{A\}_1\cdots \{A\}_{\a}}}\Big)_{\{\mu\}_1}\,\Big]\,\times\,\Big[\,\Big(\epsilon_i^{{\cal S}_B}\{\W{\cal J}_{\hat{b}}\}\Big)^{\{\mu\}_2}\,
\Big(\prod_{\b=1}^q\,{V_{\{B\}_\b}\over s_{i\{B\}_1\cdots \{B\}_{\b}}}\Big)_{\{\mu\}_2\nu}\,\eta^{\nu\rho}\,\Big]\,.~~\label{fac-propa+v-YM}
\eea
In the above, $\{\mu\}$, $\{\mu_1\}$ and $\{\mu_2\}$ are sets of Lorentz indices. $\W{\cal J}_{\hat{a}}$ and $\W{\cal J}_{\hat{b}}$
are defined by combining BG currents with corresponding propagators, namely,
\bea
\W{\cal J}^\mu_{\hat{a}}={\cal J}_{\hat{a};\nu}\,{\eta^{\nu\mu}\over s_{\hat{a}}}\,,~~~~~~~~\W{\cal J}^\mu_{\hat{b}}={\cal J}_{\hat{b};\nu}\,{\eta^{\nu\mu}\over s_{\hat{b}}}\,,
~~\label{define-WJ}
\eea
where ${\cal J}_{\hat{a}}$ and ${\cal J}_{\hat{b}}$ are BG currents connected to $\hat{a}$ and $\hat{b}$, respectively.
The expression \eref{fac-propa+v-YM} is, of course, too abstract. We will explain the pattern of the contractions of indices in \eref{fac-propa+v-YM} in the rest of this subsection. In the examples provided in the subsequent subsections, the more concrete meaning of \eref{fac-propa+v-YM} will be seen.

In \eref{fac-propa+v-YM}, the set of upper indices $\{\mu\}$, provided by the collection of polarization vectors and BG currents, is contracted with the set of lower indices $\{\mu\}\cup\nu$, provided by the set of vertices. Note that except for the $\eta^{\nu\rho}$ carried by the last propagator, all upper indices carried by the other propagators on $L_{(i,\bullet)}$ have already been contracted with the lower indices of the vertices on $L_{(i,\bullet)}$, while the set $\{\mu\}$ of lower indices consists of those remaining after the contraction. After the contraction between the upper $\{\mu\}$ and lower $\{\mu\}$, the lower indices leave behind a single lower index
$\nu$. This index is raised by the metric $\eta^{\nu\rho}$ to an upper index
$\rho$. That is, the propagators on $L_{(i,\bullet)}$, along with the BG currents and the polarization vector $\epsilon_i$ connected to $L_{(i,\bullet)}$, ultimately forms a vector that carries an upper index $\rho$. This vector propagates through the vertex $\bullet$ and is eventually contracted at another vertex. This is the general strategy we adopt in the remainder of this section for contracting the Lorentz index of a given vertex with other vectors.

The above way of understanding the contractions is based on the following picture. In the preceding two sections, when proving the SFASL for the ${\rm Tr}(\phi^3)$ and NLSM diagrams using the recursive approach, we always started from the external leg
$i$ and progressively implemented factorization along $L_{(i,\bullet)}$, much like gradually unzipping a zipper. Following this pattern, for an
$n$-point vertex on $L_{(i,\bullet)}$, we adopt the following viewpoint:
$n-1$ vectors contract with $n-1$ indices of the vertex, thereby generating a new vector that propagates along
$L_{(i,\bullet)}$ and participates in contractions at other vertices. The only exception occurs when reproducing the $2$-split. At the vertex
$v$ where $L_{(i,v)}$, $L_{(j,v)}$, and $L_{(k,v)}$ meet, we take the viewpoint that all vectors contract with all indices of the vertex $v$.

\begin{figure}
  \centering
   \includegraphics[width=5cm]{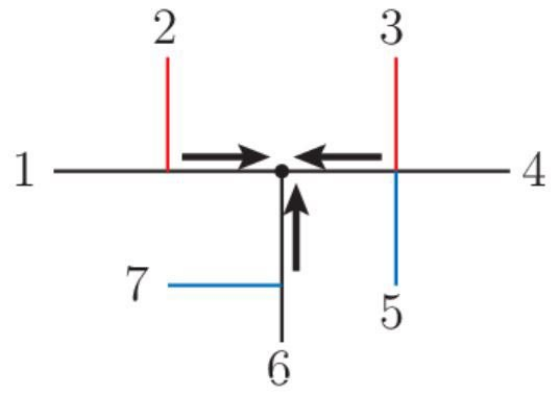} \\
  \caption{An example of the understanding of contractions of indices. The directions of currents are represented by arrows. The vertex $v$ is labeled by $\bullet$.}\label{Fig17}
\end{figure}

Fig.\ref{Fig17} provides an example of the above viewpoint. In Fig.\ref{Fig17}, the polarization vectors carried by external legs $1$ and $2$ are contracted at a cubic vertex, generating the first vector. The polarization vectors carried by external legs $3$, $4$, and $5$ are contracted at a quartic vertex, generating the second vector. The polarization vectors carried by external legs $6$ and $7$ are contracted at another cubic vertex, generating the third vector. The flow directions of the three vectors are all from the external-leg-side to the internal-line-side, as depicted by the arrows in Fig.\ref{Fig17}. According to these directions, the three vectors eventually meet at the vertex
$v$ labeled by $\bullet$, and are contracted with the indices of $v$.

\subsection{Simplest SFASL: first case}
\label{subsec-YM-case1}

Again, we study the SFASL by starting from the simplest shuffle permutation. For the YM diagrams, the simplest shuffle permutations can be classified into three cases, which will be studied in subsections \ref{subsec-YM-case1}, \ref{subsec-YM-case2} and \ref{subsec-YM-case3}, respectively.

\begin{figure}
  \centering
   \includegraphics[width=13cm]{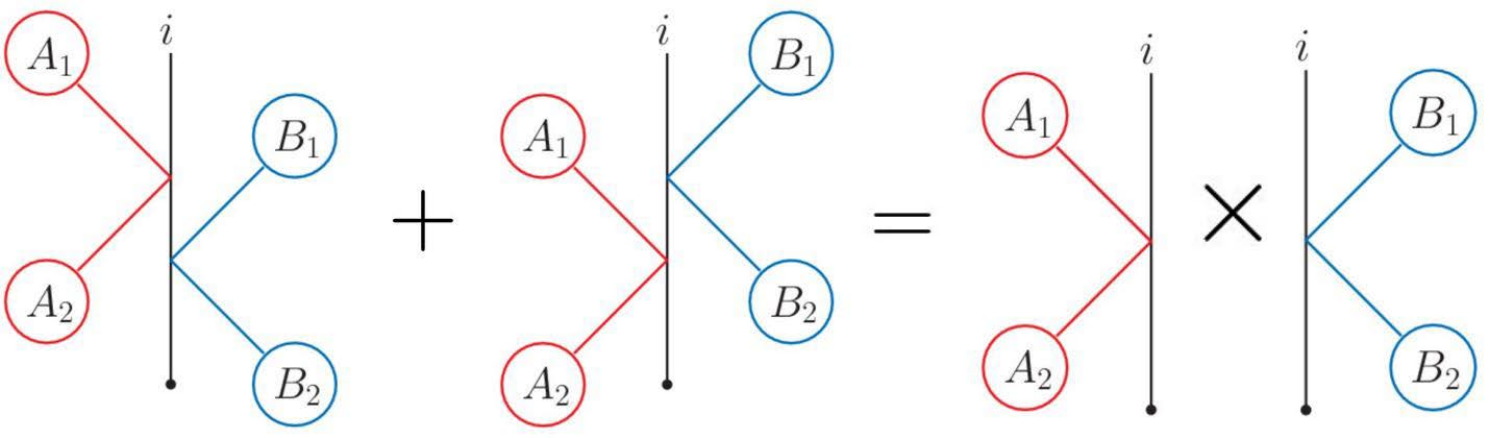} \\
  \caption{First case of the simplest shuffle permutation.}\label{Fig18}
\end{figure}

The first case is shown in Fig.\ref{Fig18}, where blocks $A_1$ and $A_2$ are connected to $L_{(i,\bullet)}$ via one quartic vertex, and blocks $B_1$ and $B_2$ are connected to $L_{(i,\bullet)}$ via another quartic vertex. The set of shuffle permutations consists of two elements. What we aim to show is the factorization behavior on the r.h.s. of Fig.\ref{Fig18}.

The contribution from the first diagram on the l.h.s. of Fig.\ref{Fig18} is
\bea
F_1^\rho=\Big[\,\big(\epsilon_i^{\mu_1}\W{\cal J}^{\mu_2}_{A_1}\W{\cal J}^{\mu_3}_{A_2}\,V^{4g}_{\mu_1\mu_2\mu_3\mu_4}\big)\,\eta^{\mu_4\nu_1}\,\W{\cal J}^{\nu_2}_{B_1}\W{\cal J}^{\nu_3}_{B_2}\,V^{4g}_{\nu_1\nu_2\nu_3\nu_4}\,\Big]\,\eta^{\nu_4\rho}\,{1\over s_{iA_1A_2}}\,{1\over s_{iA_1A_2B_1B_2}}\,,~~\label{YM-example1-F1}
\eea
where each $\W{\cal J}$ is defined in \eref{define-WJ}.
By employing the Feynman rules of $V^{4g}$ in \eref{Feynmanrule-YM}, we have
\bea
J^{\nu_1}_{i|A_1|A_2}&=&\epsilon_i^{\mu_1}\W{\cal J}^{\mu_2}_{A_1}\W{\cal J}^{\mu_3}_{A_2}\,V^{4g}_{\mu_1\mu_2\mu_3\mu_4}\,\eta^{\mu_4\nu_1}=(\epsilon_i\cdot \W{\cal J}_{A_2})\,\W{\cal J}^{\nu_1}_{A_1}-{1\over2}\,(\epsilon_i\cdot\W{\cal J}_{A_1})\,\W{\cal J}^{\nu_1}_{A_2}-{1\over2}\,(\W{\cal J}_{A_1}\cdot\W{\cal J}_{A_2})\,\epsilon_i^{\nu_1}\,.~~\label{YM-example1-v1}
\eea
Substituting this into \eref{YM-example1-F1} yields
\bea
F^\rho_1&=&\Big[\,(J_{i|A_1|A_2}\cdot \W{\cal J}_{B_2})\,\W{\cal J}^\rho_{B_1}-{1\over2}\,(J_{i|A_1|A_2}\cdot\W{\cal J}_{B_1})\,\W{\cal J}^\rho_{B_2}
-{1\over2}\,(\W{\cal J}_{B_1}\cdot\W{\cal J}_{B_1})\,J_{i|A_1|A_2}^\rho\,\Big]\,{1\over s_{iA_1A_2}}\,{1\over s_{iA_1A_2B_1B_2}}\nn
&\xrightarrow[]{\eref{kine-condi-shuffle-YM}}&\,{\cal V}^\rho\,{1\over s_{iA_1A_2}}\,{1\over s_{iA_1A_2B_1B_2}}\,,
\eea
where
\bea
{\cal V}^\rho&=&-{1\over2}\,(\W{\cal J}_{A_1}\cdot\W{\cal J}_{A_2})\,(\epsilon_i\cdot\W{\cal J}_{B_2})\,\W{\cal J}^\rho_{B_1}
+{1\over4}\,(\W{\cal J}_{A_1}\cdot\W{\cal J}_{A_2})\,(\epsilon_i\cdot\W{\cal J}_{B_1})\,\W{\cal J}^\rho_{B_2}\nn
&&\,-{1\over2}\,(\W{\cal J}_{B_1}\cdot\W{\cal J}_{B_2})\,(\epsilon_i\cdot\W{\cal J}_{A_2})\,\W{\cal J}^\rho_{A_1}
+{1\over4}\,(\W{\cal J}_{B_1}\cdot\W{\cal J}_{B_2})\,(\epsilon_i\cdot\W{\cal J}_{A_1})\,\W{\cal J}^\rho_{A_2}\nn
&&\,+{1\over4}\,(\W{\cal J}_{A_1}\cdot\W{\cal J}_{A_2})\,(\W{\cal J}_{B_1}\cdot\W{\cal J}_{B_2})\,\epsilon_i^\rho\,.
\eea

A similar process gives rise to
\bea
F^\rho_2\,\xrightarrow[]{\eref{kine-condi-shuffle-YM}}\,{\cal V}^\rho\,{1\over s_{iB_1B_2}}\,{1\over s_{iA_1A_2B_1B_2}}\,.
\eea
Therefore,
\bea
F^\rho_1+F^\rho_2\,\xrightarrow[]{\eref{kine-condi-shuffle-YM}}\,{\cal V}^\rho\,\Big({1\over s_{iA_1A_2}}+
{1\over s_{iB_1B_2}}\Big)\,{1\over s_{iA_1A_2B_1B_2}}&\xrightarrow[]{\eref{kine-condi-shuffle-YM}}&\,{\cal V}^\rho\,\Big({1\over s_{iA_1A_2}}\,\times\,{1\over s_{iB_1B_2}}\Big)\,,~~\label{F1F2-step1}
\eea
where the observation \eref{key-observation-general} has been used.

In \eref{F1F2-step1}, the propagator-part has already factorized as in the ${\rm Tr}(\phi^3)$ case, but the Lorentz vector ${\cal V}^\rho$ in the numerator does not factorize. The method to achieve the complete factorization, as discussed in subsection \ref{subsec-subspace}, is to decompose the polarization vector into orthogonal subspaces as $\epsilon_i=\epsilon_i^{{\cal S}_A}+\epsilon_i^{{\cal S}_B}$. After this decomposition, we obtain
\bea
F^\rho_1+F^\rho_2\,&\xrightarrow[]{\eref{kine-condi-shuffle-YM}}&\,\Big[\Big(\,
(\epsilon^{{\cal S}_A}_i\cdot\W{\cal J}_{A_2})\,\W{\cal J}^\rho_{A_1}
-{1\over2}\,(\epsilon^{{\cal S}_A}_i\cdot\W{\cal J}_{A_1})\,\W{\cal J}^\rho_{A_2}-{1\over2}\,(\W{\cal J}_{A_1}\cdot\W{\cal J}_{A_2})\,\epsilon^{{\cal S}_A;\rho}_i\,\Big)\,{1\over s_{iA_1A_2}}\Big]\nn
&&\times\,\Big[-{1\over2}\,(\W{\cal J}_{B_1}\cdot\W{\cal J}_{B_2})\,{1\over s_{iB_1B_2}}\Big]\nn
&&+\,\Big[-{1\over2}\,(\W{\cal J}_{A_1}\cdot\W{\cal J}_{A_2})\,{1\over s_{iA_1A_2}}\Big]\nn
&&\times\,\Big[\Big(\,(\epsilon^{{\cal S}_B}_i\cdot\W{\cal J}_{B_2})\,\W{\cal J}^\rho_{B_1}
-{1\over2}\,(\epsilon^{{\cal S}_B}_i\cdot\W{\cal J}_{B_1})\,\W{\cal J}^\rho_{B_2}-{1\over2}\,(\W{\cal J}_{B_1}\cdot\W{\cal J}_{B_2})\,\epsilon^{{\cal S}_B;\rho}_i\,\Big)\,{1\over s_{iB_1B_2}}\Big]\nn
&=&\Big[\epsilon_i^{{\cal S}_A;\mu_1}\W{\cal J}_{A_1}^{\mu_2}\W{\cal J}_{A_2}^{\mu_3}\,V^{4g}_{\mu_1\mu_2\mu_2\mu_4}\,\eta^{\mu_4\rho}\,{1\over s_{iA_1A_2}}\Big]\,\times\,\Big[\W{\cal J}^{\nu_1}_{B_1}\W{\cal J}^{\nu_2}_{B_2}\,V^{2g-2\phi}_{\nu_1\nu_2}\,{1\over s_{iB_1B_2}}\Big]\nn
&&+\Big[\W{\cal J}_{A_1}^{\mu_1}\W{\cal J}_{A_2}^{\mu_2}\,V^{2g-2\phi}_{\mu_1\mu_2}\,{1\over s_{iA_1A_2}}\Big]\,\times\,\Big[\epsilon_i^{{\cal S}_B;\nu_1}\W{\cal J}^{\nu_2}_{B_1}\W{\cal J}^{\nu_3}_{B_2}\,V^{4g}_{\nu_1\nu_2\nu_3\nu_4}\,\eta^{\nu_4\rho}\,{1\over s_{iB_1B_2}}\Big]\,,~~\label{result-example1-YM}
\eea
where $V^{2g-2\phi}$ is a YM$\oplus{\rm Tr}(\phi^3)$ vertex coupling two gluons and two scalars together.
Consequently, regardless of whether the polarization vector takes $\epsilon_i^{{\cal S}_A}$ or $\epsilon_i^{{\cal S}_B}$, $F^\rho_1+F^\rho_2$ always exhibits the factorization behavior shown on the r.h.s. of Fig.\ref{Fig18}.
When the polarization vector takes $\epsilon_i^{{\cal S}_A}$, in the $\pmb A$-side part in the factorization formula, the particle propagating in $L_{(i,\bullet)}$ behaves like a gluon, while in the $\pmb B$-side part, the particle propagating in $L_{(i,\bullet)}$ behaves like a ${\rm Tr}(\phi^3)$ scalar (as indicated by $V^{2g-2\phi}$). This picture is quite natural, due to the dimensional reduction interpretation discussed in subsection \ref{subsec-subspace}. When the polarization vector takes $\epsilon_i^{{\cal S}_B}$, the situation is exactly reversed.

Before ending this subsection, we briefly discuss why $F_1^\rho$ and $F_2^\rho$ share the same numerator ${\cal V}^\rho$. In the diagram in Fig.\ref{Fig18} corresponding to $F_1^\rho$, any contraction at the second vertex involves a current from the subspace ${\cal S}_B$, thereby annihilating the kinematic variables from ${\cal S}_A$. This implies that the second vertex is completely insensitive to the two currents from ${\cal S}_A$ that enter through the first vertex; all contractions occurring at the second vertex are identical to those in the absence of the first vertex. The same argument applies to the diagram corresponding to $F_2^\rho$. In this sense, the two vertices commute with each other. This is the reason why $F_1^\rho$ and $F_2^\rho$ share the same numerator.

\subsection{Simplest SFASL: second case}
\label{subsec-YM-case2}

\begin{figure}
  \centering
   \includegraphics[width=13cm]{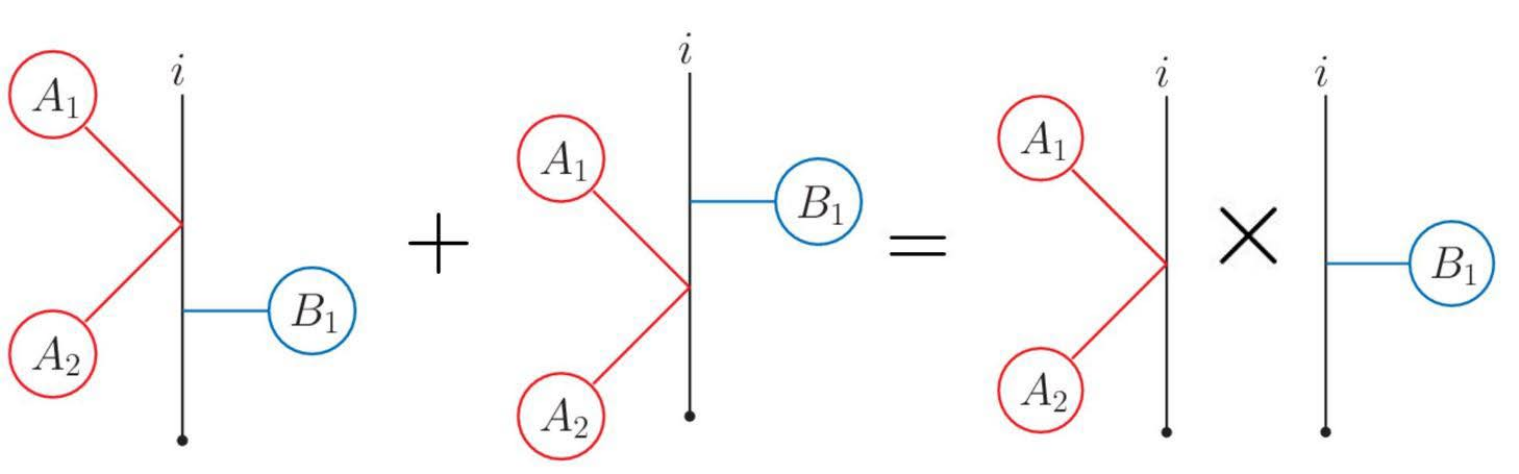} \\
  \caption{Second case of the simplest shuffle permutation.}\label{Fig19}
\end{figure}

The second case of the simplest SFASL in shown in Fig.\ref{Fig19}. Two blocks $A_1$ and $A_2$ are connected to $L_{(i,\bullet)}$ via a quartic vertex, and a block $B_1$ is connected to $L_{(i,\bullet)}$ via a cubic vertex. The set of shuffle permutations includes two elements. The purpose of this subsection is to show the factorization behavior on the r.h.s. of Fig.\ref{Fig19}.

The first diagram on the l.h.s. of Fig.\ref{Fig19} can be evaluated as
\bea
F_1^\rho=\Big[\,\big(\epsilon_i^{\mu_1}\W{\cal J}^{\mu_2}_{A_1}\W{\cal J}^{\mu_3}_{A_2}\,V^{4g}_{\mu_1\mu_2\mu_3\mu_4}\big)\,\eta^{\mu_4\nu_1}\,\W{\cal J}^{\nu_2}_{B_1}\,V^{3g}_{\nu_1\nu_2\nu_3}\,\Big]\,\eta^{\nu_3\rho}\,{1\over s_{iA_1A_2}}\,{1\over s_{iA_1A_2B_1}}\,.~~\label{YM-example2-F1}
\eea
In this case $J^{\nu_1}_{i|A_1|A_2}=\epsilon_i^{\mu_1}\W{\cal J}^{\mu_2}_{A_1}\W{\cal J}^{\mu_3}_{A_2}\,V^{4g}_{\mu_1\mu_2\mu_3\mu_4}\,\eta^{\mu_4\nu_1}$ is the same as that in \eref{YM-example1-v1}. Plugging this in to \eref{YM-example2-F1} and using the Feynman rules of $V^{3g}$ in \eref{Feynmanrule-YM}, we arrive at
\bea
F_1^\rho&=&{1\over\sqrt{2}}\,\Big[\,(J_{i|A_1|A_2}\cdot\W{\cal J}_{B_1})\,(k_{iA_1A_2}^\rho-k_{B_1}^\rho)+J_{i|A_1|A_2}\cdot(k_{B_1}-k_I)\,\W{\cal J}_{B_1}^\rho+\W{\cal J}_{B_1}\cdot(k_I-k_{iA_1A_2})\,J_{i|A_1|A_2}^\rho\,\Big]\nn
&&{1\over s_{iA_1A_2}}\,{1\over s_{iA_1A_2B_1}}\nn
&=&\sqrt{2}\,\Big[\,-(J_{i|A_1|A_2}\cdot\W{\cal J}_{B_1})\,k_{B_1}^\rho+J_{i|A_1|A_2}\cdot\Big(k_{B_1}+{1\over2}\,k_{iA_1A_2}\Big)\,\W{\cal J}_{B_1}^\rho-(k_{iA_1A_2}\cdot \W{\cal J}_{B_1})\,J_{i|A_1|A_2}^\rho\,\Big]\nn
&&{1\over s_{iA_1A_2}}\,{1\over s_{iA_1A_2B_1}}\,,~~\label{YM-example2-F1-key}
\eea
where $I$ is the internal line connecting $V^{3g}$ and $\bullet$, satisfying $k_I+k_{iA_1A_2}+k_{B_1}=0$.
In \eref{YM-example2-F1-key}, we have used the property \eref{BG-gauge} for BG currents, namely,
\bea
{\cal J}_{\overline{iA_1A_2B_1}}\cdot k_I=0\,,~~~~~~~~{\cal J}_{B_1}\cdot k_{B_1}=0\,,
\eea
where $\overline{iA_1A_2B_1}=\{1,\cdots,n\}\setminus\{i,A_1,A_2,B_1\}$.
Since $J^\mu_{i|A_1|A_2}$ is not a BG current, the component $J_{i|A_1|A_2}\cdot k_{iA_1A_2}$ does not vanish. However, after summing over all divisions of the set $i\cup A_1\cup A_2$, contributions containing this component will cancel, since the summation leads to the BG current ${\cal J}_{iA_1A_2}$, satisfying \eref{BG-gauge}, i.e., ${\cal J}_{iA_1A_2}\cdot k_{iA_1A_2}=0$. Consequently, when summing over all diagrams, this component can be removed. It is easy to see that, in \eref{YM-example2-F1-key}, $J_{i|A_1|A_2}\cdot k_{iA_1A_2}$ is the only term that can perceive the existence of another vertex $V^{4g}$. In other inner products in \eref{YM-example2-F1-key}, the current ${\cal J}_{B_1}$ from ${\cal S}_B$ annihilates the kinematic variables from ${\cal S}_A$. Since $J_{i|A_1|A_2}\cdot k_{iA_1A_2}$ can be eliminated in the summation over Feynman diagrams, the vertex $V^{3g}$ in $F_1^\rho$ effectively commutes with the vertex $V^{4g}$.
Consequently, after removing this un-effective $k_{iA_1A_2}\cdot J_{i|A_1|A_2}$, we get
\bea
F_1^\rho&\xrightarrow[]{\eref{kine-condi-shuffle-YM}}&\,{\cal V}^\rho_2\,{1\over s_{iA_1A_2}}\,{1\over s_{iA_1A_2B_1}}\,,
\eea
where ${\cal V}^{\rho}_2$ is given as,
\bea
{\cal V}^{\rho}_2&=&\sqrt{2}\,\Big[\,{1\over2}\,(\W{\cal J}_{A_1}\cdot\W{\cal J}_{A_2})\,(\epsilon_i\cdot\W{\cal J}_{B_1})\,k_{B_1}^\rho-{1\over2}\,(\W{\cal J}_{A_1}\cdot\W{\cal J}_{A_2})\,(\epsilon_i\cdot k_{B_1})\,\W{\cal J}_{B_1}^\rho\nn
&&-(k_i\cdot\W{\cal J}_{B_1})\,(\epsilon_i\cdot \W{\cal J}_{A_2})\,\W{\cal J}_{A_1}^\rho+{1\over2}\,(k_i\cdot\W{\cal J}_{B_1})\,(\epsilon_i\cdot \W{\cal J}_{A_1})\,\W{\cal J}_{A_2}^\rho+{1\over2}\,(k_i\cdot\W{\cal J}_{B_1})\,(\W{\cal J}_{A_1}\cdot \W{\cal J}_{A_2})\,\epsilon_i^\rho\,\Big]\,.
\eea
which does not contain any non-commuting component.

Then we move to the second diagram on the l.h.s. of Fig.\ref{Fig19}. The contribution from this diagram is
\bea
F_2^\rho=\Big[\,\big(\epsilon_i^{\mu_1}\W{\cal J}^{\mu_2}_{B_1}\,V^{3g}_{\mu_1\mu_2\mu_3}\big)\,\eta^{\mu_3\nu_1}\,\W{\cal J}^{\nu_2}_{A_1}\W{\cal J}^{\nu_3}_{A_2}\,V^{4g}_{\nu_1\nu_2\nu_3\nu_4}\,\Big]\,\eta^{\nu_4\rho}\,{1\over s_{iB_1}}\,{1\over s_{iA_1A_2B_1}}\,.~~\label{YM-example2-F2}
\eea
By utilizing the Feynman rule of $V^{3g}$, we obtain
\bea
J^{\nu_1}_{i|B_1}&=&\epsilon_i^{\mu_1}\W{\cal J}^{\mu_2}_{A_1}\,V^{3g}_{\mu_1\mu_2\mu_3}\,\eta^{\mu_3\nu_1}\nn
&=&{1\over\sqrt{2}}\,\Big[\,(\epsilon_i\cdot\W{\cal J}_{B_1})\,(k_i^{\nu_1}-k_{B_1}^{\nu_1})+\epsilon_i\cdot (k_{B_1}-k_{I'})\,\W{\cal J}_{B_1}^{\nu_1}
+\W{\cal J}_{B_1}\cdot(k_{I'}-k_i)\,\epsilon_i^{\nu_1}\,\Big]\nn
&=&\sqrt{2}\,\Big[\,-(\epsilon_i\cdot\W{\cal J}_{B_1})\,k_{B_1}^{\nu_1}+(\epsilon_i\cdot k_{B_1})\,\W{\cal J}_{B_1}^{\nu_1}
-(k_i\cdot\W{\cal J}_{B_1})\,\epsilon_i^{\nu_1}\,\Big]\,,~~\label{onshell}
\eea
where $I'$ is the internal line connecting $V^{3g}$ and $V^{4g}$, satisfying $k_{I'}+k_i+k_{B_1}=0$. In the above, we have used the property \eref{BG-gauge}, i.e.,
\bea
{\cal J}_{B_1}\cdot k_{B_1}=0\,,~~~~~~~~{\cal J}_{\overline{iB_1}}\cdot k_{I'}=0\,,
\eea
and the on-shell condition $\epsilon_i\cdot k_i=0$.
Plugging $J^{\nu_1}_{i|B_1}$ into \eref{YM-example2-F2} leads to
\bea
F_2^\rho&=&\Big[\,(J_{i|B_1}\cdot \W{\cal J}_{A_2})\,\W{\cal J}_{A_1}^\rho-{1\over2}\,(J_{i|B_1}\cdot \W{\cal J}_{A_1})\,\W{\cal J}_{A_2}^\rho
-{1\over2}\,(\W{\cal J}_{A_1}\cdot\W{\cal J}_{A_2})\,J_{i|B_1}^\rho\,\Big]\,{1\over s_{iB_1}}\,{1\over s_{iA_1A_2B_1}}\nn
&\xrightarrow[]{\eref{kine-condi-shuffle-YM}}&\,\,{\cal V}^\rho_2\,{1\over s_{iB_1}}\,{1\over s_{iA_1A_2B_1}}\,.~~\label{YM-example2-F2-final}
\eea
$F_2^\rho$ and $F_1^\rho$ share the same numerator ${\cal V}_2^\rho$. This is because, in the first line of \eref{YM-example2-F2-final}, the kinematic variables from the subspace ${\cal S}_B$ are annihilated by $\W{\cal J}_{A_1}$ or $\W{\cal J}_{A_2}$ in any inner product, leaving only the commuting part of the vertex $V^{4g}$. On the other hand, as mentioned earlier, the effective part of the vertex $V^{3g}$ in $F^\rho_1$ also commutes with another vertex. Consequently, the commutativity of $V^{3g}$ and $V^{4g}$ in $F_1^\rho$ and $F_2^\rho$ ensures the uniqueness of the numerator ${\cal V}_2^\rho$.

Therefore, the summation over shuffle permutations on the l.h.s. of Fig.\ref{Fig19} reads,
\bea
F_1^\rho+F_2^\rho\,&\xrightarrow[]{\eref{kine-condi-shuffle-YM}}&\,{\cal V}_2^\rho\,\Big({1\over s_{iA_1A_2}}+{1\over s_{iB_1}}\Big)\,{1\over s_{iA_1A_2B_1}}\nn
&\xrightarrow[]{\eref{kine-condi-shuffle-YM}}&\,{\cal V}_2^\rho\,\Big({1\over s_{iA_1A_2}}\,\times\,{1\over s_{iB_1}}\Big)\nn
&=&\Big[\,\Big((\epsilon^{{\cal S}_A}_i\cdot \W{\cal J}_{A_2})\,\W{\cal J}_{A_1}^\rho-{1\over2}\,(\epsilon^{{\cal S}_A}_i\cdot \W{\cal J}_{A_1})\,\W{\cal J}_{A_2}^\rho-{1\over2}\,(\W{\cal J}_{A_1}\cdot \W{\cal J}_{A_2})\,\epsilon_i^{{\cal S}_A;\rho}\Big)\,{1\over s_{iA_1A_2}}\,\Big]\nn
&&\times\,\Big[\,-\sqrt{2}\,(k_i\cdot\W{\cal J}_{B_1})\,{1\over s_{iB_1}}\,\Big]\nn
&&+\Big[-{1\over2}\,(\W{\cal J}_{A_1}\cdot\W{\cal J}_{A_2})\,{1\over s_{iA_1A_2}}\Big]\nn
&&\times\,\Big[\,\sqrt{2}\,\Big(-(\epsilon^{{\cal S}_B}_i\cdot\W{\cal J}_{B_1})\,k_{B_1}^\rho+(\epsilon^{{\cal S}_B}_i\cdot k_{B_1})\,\W{\cal J}_{B_1}^\rho-(k_i\cdot\W{\cal J}_{B_1})\,\epsilon_{i}^{{\cal S}_B;\rho}\Big)\,{1\over s_{iB_1}}\,\Big]\nn
&=&\Big[\,\epsilon_i^{{\cal S}_{A};\mu_1}\W{\cal J}_{A_1}^{\mu_2}\W{\cal J}_{A_2}^{\mu_3}\,V^{4g}_{\mu_1\mu_2\mu_3\mu_4}\,\eta^{\mu_4\rho}\,{1\over s_{iA_1A_2}}\,\Big]\,\times\,\Big[\,\W{\cal J}_{B_1}^\nu\,V_{\nu}^{1g-2\phi}\,{1\over s_{iB_1}}\,\Big]\nn
&&+\Big[\,\W{\cal J}_{A_1}^{\mu_1}\W{\cal J}_{A_2}^{\mu_2}\,V^{2g-2\phi}_{\mu_1\mu_2}\,{1\over s_{iA_1A_2}}\,\Big]\,\times\,\Big[\,\epsilon_i^{{\cal S}_{B};\nu_1}\W{\cal J}_{B_1}^{\nu_2}\,V_{\nu_1\nu_2\nu_3}^{3g}\,\eta^{\nu_3\rho}\,{1\over s_{iB_1}}\,\Big]\,,~~~~\label{result-example2-YM}
\eea
where $V^{1g-2\phi}$ is a YM$\oplus{\rm Tr}(\phi^3)$ vertex coupling one gluon and two scalars.
Therefore, regardless of whether the polarization vector takes $\epsilon_i^{{\cal S}_{A}}$ or $\epsilon_i^{{\cal S}_{B}}$, $F_1+F_2$ always exhibits the factorization structure shown on the r.h.s. of Fig.\ref{Fig19}.

\subsection{Simplest SFASL: third case}
\label{subsec-YM-case3}

\begin{figure}
  \centering
   \includegraphics[width=16cm]{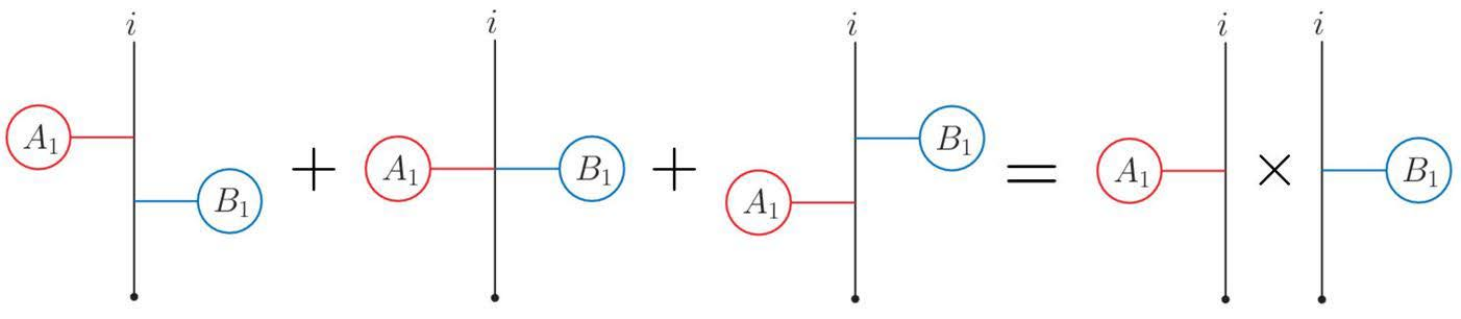} \\
  \caption{Third case of the simplest shuffle permutation.}\label{Fig20}
\end{figure}

The third case of simplest SFASL is shown in Fig.\ref{Fig20}. Two blocks $A_1$ and $B_1$ are connected to $L_{(i,\bullet)}$, and the set of shuffle permutations includes three elements.

The first diagram on the l.h.s. of Fig.\ref{Fig20} can be expressed as
\bea
F_1^\rho&=&\Big[\,\big(\epsilon_i^{\mu_1}\W{\cal J}^{\mu_2}_{A_1}\,V^{3g}_{\mu_1\mu_2\mu_3}\big)\,\eta^{\mu_3\nu_1}\,\W{\cal J}^{\nu_2}_{B_1}\,V^{3g}_{\nu_1\nu_2\nu_3}\,\Big]\,\eta^{\nu_3\rho}\,{1\over s_{iA_1}}\,{1\over s_{iA_1B_1}}\,.~~\label{YM-example3-F1}
\eea
Using the Feynman rule of $V^{3g}$ we get
\bea
J^{\nu_1}_{i|A_1}&=&\epsilon_i^{\mu_1}\W{\cal J}^{\mu_2}_{A_1}\,V^{3g}_{\mu_1\mu_2\mu_3}\,\eta^{\mu_3\nu_1}\nn
&=&{1\over\sqrt{2}}\,\Big[(\epsilon_i\cdot \W{\cal J}_{A_1})\,(k_{A_1}^{\nu_1}-k_i^{\nu_1})+\epsilon_i\cdot(k_{I'}^{\nu_1}-k_{A_1})\,\W{\cal J}^{\nu_1}_{A_1}
+\W{\cal J}_{A_1}\cdot(k_i-k_{I'})\,\epsilon_i^{\nu_1}\Big]\nn
&=&\sqrt{2}\,\Big[(\epsilon_i\cdot \W{\cal J}_{A_1})\,k_{A_1}^{\nu_1}-(\epsilon_i\cdot k_{A_1})\,\W{\cal J}^{\nu_1}_{A_1}
+(k_i\cdot\W{\cal J}_{A_1})\,\epsilon_i^{\nu_1}\Big]\,,~~\label{JiA}
\eea
where $I'$ is the internal line connecting two vertices $V^{3g}$, and we have used the property \eref{BG-gauge} which indicates
\bea
{\cal J}_{\overline{iA_1}}\cdot k_{I'}=0\,,~~~~~~~~{\cal J}_{A_1}\cdot k_{A_1}=0\,,~~~~~~~~\epsilon_i\cdot k_i=0\,.~~\label{for-JiA}
\eea
Substituting it into \eref{YM-example3-F1} yields
\bea
F_1^\rho&=&{1\over\sqrt{2}}\,\Big[\,(J_{i|A_1}\cdot\W{\cal J}_{B_1})\,(k_{iA_1}^\rho-k_{B_1}^\rho)+J_{i|A_1}\cdot(k_{B_1}-k_{I})\,\W{\cal J}_{B_1}^\rho
+\W{\cal J}_{B_1}\cdot(k_I-k_{iA_1})\,J_{i|A_1}^\rho\,\Big]\,{1\over s_{iA_1}}\,{1\over s_{iA_1B_1}}\nn
&=&\sqrt{2}\,\Big[\,-(J_{i|A_1}\cdot\W{\cal J}_{B_1})\,k_{B_1}^\rho+\Big(k_{B_1}+{1\over2}\,k_{iA_1}\Big)\cdot J_{i|A_1}\,\W{\cal J}_{B_1}^\rho
-(k_i\cdot\W{\cal J}_{B_1})\,J_{i|A_1}^\rho\,\Big]\,{1\over s_{iA_1}}\,{1\over s_{iA_1B_1}}\nn
&\xrightarrow[]{\eref{kine-condi-shuffle-YM}}&\,\Big[\,{\cal V}_3^\rho+(\epsilon_i\cdot\W{\cal J}_{A_1})\,(k_{iA_1}\cdot k_{A_1})\,\W{\cal J}^\rho_{B_1}\,\Big]\,{1\over s_{iA_1}}\,{1\over s_{iA_1B_1}}\,,
\eea
where
\bea
{\cal V}_3^\rho&=&2\,\Big[\,-(\epsilon_i\cdot\W{\cal J}_{A_1})\,(k_i\cdot\W{\cal J}_{B_1})\,k_{A_1}^\rho+(\epsilon_i\cdot k_{A_1})\,(k_i\cdot\W{\cal J}_{B_1})\,\W{\cal J}_{A_1}^\rho\nn
&&\,-(\epsilon_i\cdot\W{\cal J}_{B_1})\,(k_i\cdot\W{\cal J}_{A_1})\,k_{B_1}^\rho+(\epsilon_i\cdot k_{B_1})\,(k_i\cdot\W{\cal J}_{A_1})\,\W{\cal J}_{B_1}^\rho\nn
&&\,-(k_i\cdot\W{\cal J}_{A_1})\,(k_i\cdot\W{\cal J}_{B_1})\,\epsilon_i^\rho\,\Big]\,.
\eea
In the above, $I$ is the internal line connecting $V^{3g}$ and $\bullet$, and we have used \eref{BG-gauge}, i.e.,
\bea
{\cal J}_{\overline{iA_1B_1}}\cdot k_I=0\,,~~~~~~~~{\cal J}_{B_1}\cdot k_{B_1}=0\,.~~~\label{condi-2}
\eea
The ${\cal V}_3^\rho$-part is the commuting part, where the contraction occurring at each vertex is completely insensitive to the existence of the other vertex. On the other hand, the $k_{iA_1}\cdot k_{A_1}$-part is the non-commuting part, where the second vertex perceives the momentum $k_{A_1}$ from the subspace ${\cal S}_A$ flowing in at the first vertex.

Note that this time we cannot use the property \eref{BG-gauge} to eliminate the $k_{iA_1}\cdot J_{i|A_1}$ term, as we did in the previous subsection. The reason is that $J_{i|A_1}$ in the last line of \eref{JiA} is not computed solely from the Feynman rules. We have used the relations in \eref{for-JiA} to modify its form. It means, $J_{i|A_1}$ in the last line of \eref{JiA} is not a component of the BG current ${\cal J}_{iA_1}$. Therefore, in order to handle the non-commuting part, we need to apply ${\cal J}_{iA_1}\cdot k_{iA_1}=0$ more carefully.

The $(\epsilon_i\cdot \W{\cal J}_{A_1})\,k^\nu_{A_1}$ in the non-commuting term comes from the first term $\sqrt{2}\,(\epsilon_i\cdot \W{\cal J}_{A_1})\,k^\nu_{A_1}$ in $J^\nu_{i|A_1}$. If we return to the original Feynman rules, this term should be written as
\bea
{1\over\sqrt{2}}\,(\epsilon_i\cdot \W{\cal J}_{A_1})\,(k_{A_1}^\nu-k_i^\nu)\,.~~\label{ka-ki}
\eea
The above term is written via the standard Feynman rule, and is therefore a component of the BG current ${\cal J}_{iA_1}$. Since the BG current satisfies ${\cal J}_{iA_1}\cdot k_{iA_1}=0$, the contribution of $(\epsilon_i\cdot \W{\cal J}_{A_1})\,(k_{A_1}-k_i)\cdot k_{iA_1}$ will be canceled in the summation over Feynman diagrams. This indicates that when the non-commuting term is decomposed as
\bea
(\epsilon_i\cdot\W{\cal J}_{A_1})\,(k_{A_1}\cdot k_{iA_1})\,\W{\cal J}^\rho_{B_1}={1\over2}\,(\epsilon_i\cdot\W{\cal J}_{A_1})\,(k_{A_1}-k_i)\cdot k_{iA_1}\,\W{\cal J}^\rho_{B_1}+{1\over2}\,(\epsilon_i\cdot\W{\cal J}_{A_1})\,(k_{A_1}+k_i)\cdot k_{iA_1}\,\W{\cal J}^\rho_{B_1}\,,
\eea
only the second term is effective. Thus we arrive at
\bea
F_1^\rho\,\sim\,\Big[\,{\cal V}_3^\rho+{k^2_{iA_1}\over2}\,(\epsilon_i\cdot\W{\cal J}_{A_1})\,\W{\cal J}^\rho_{B_1}\,\Big]\,{1\over s_{iA_1}}\,{1\over s_{iA_1B_1}}\,.~~\label{F1-effe}
\eea

Performing the analogous manipulation on the third diagram on the l.h.s. of Fig.\ref{Fig20}, we find
\bea
F_3^\rho\,&\xrightarrow[]{\eref{kine-condi-shuffle-YM}}&\,\Big[\,{\cal V}_3^\rho+{k^2_{iB_1}\over2}\,(\epsilon_i\cdot\W{\cal J}_{B_1})\,\W{\cal J}^\rho_{A_1}\,\Big]\,{1\over s_{iB_1}}\,{1\over s_{iA_1B_1}}\,.~~\label{F3-effe}
\eea
Meanwhile, the second diagram can be computed as
\bea
F_2^\rho\,&\xrightarrow[]{\eref{kine-condi-shuffle-YM}}&\,\Big[\,-{1\over2}\,(\epsilon_i\cdot\W{\cal J}_{A_1})\,\W{\cal J}^\rho_{B_1}-{1\over2}\,(\epsilon_i\cdot\W{\cal J}_{B_1})\,\W{\cal J}^\rho_{A_1}\,\Big]\,{1\over s_{iA_1B_1}}\,,
\eea
where the $\W{\cal J}_{A_1}\cdot\W{\cal J}_{B_1}$ part vanishes due to the kinematic condition \eref{kine-condi-shuffle-YM}.

Putting $F_1^\rho$, $F_2^\rho$ and $F_3^\rho$ together, we see that $F_2^\rho$ is canceled by the non-commuting terms in \eref{F1-effe} and \eref{F3-effe}.
The result of the summation on the l.h.s. of Fig.\ref{Fig20} is then given by
\bea
&&F_1^\rho+F_2^\rho+F_3^\rho\nn
&\xrightarrow[]{\eref{kine-condi-shuffle-YM}}&\,{\cal V}_3^\rho\,\Big({1\over s_{iA_1}}+{1\over s_{iB_1}}\Big)\,{1\over s_{iA_1B_1}}\nn
&\xrightarrow[]{\eref{kine-condi-shuffle-YM}}&\,{\cal V}_3^\rho\,\Big({1\over s_{iA_1}}\,\times\,{1\over s_{iB_1}}\Big)\nn
&=&\Big[\,\sqrt{2}\,\Big((\epsilon^{{\cal S}_A}_i\cdot\W{\cal J}_{A_1})\,\,k_{A_1}^\rho-(\epsilon^{{\cal S}_A}_i\cdot k_{A_1})\,\W{\cal J}_{A_1}^\rho+(k_i\cdot\W{\cal J}_{A_1})\,\epsilon_i^{{\cal S}_A;\rho}\Big)\,{1\over s_{iA_1}}\,\Big]\,\times\,\Big[-\sqrt{2}\,(k_i\cdot\W{\cal J}_{B_1})\,{1\over s_{iB_1}}\Big]\nn
&&+\Big[\sqrt{2}\,(k_i\cdot\W{\cal J}_{A_1})\,{1\over s_{iA_1}}\Big]\,\times\,\Big[\,\sqrt{2}\,\Big(-(\epsilon^{{\cal S}_B}_i\cdot\W{\cal J}_{B_1})\,k_{B_1}^\rho+(\epsilon^{{\cal S}_B}_i\cdot k_{B_1})\,\W{\cal J}_{B_1}^\rho-(k_i\cdot\W{\cal J}_{B_1})\,\epsilon_i^{{\cal S}_A;\rho}\Big)\,{1\over s_{iB_1}}\,\Big]\nn
&=&\Big[\,\epsilon^{{\cal S}_A;\mu_1}_i\W{\cal J}_{A_1}^{\mu_2}\,V^{3g}_{\mu_1\mu_2\mu_3}\,\eta^{\mu_3\rho}\,{1\over s_{iA_1}}\,\Big]\,\times\,\Big[\,\W{\cal J}_{B_1}^\nu\,V^{1g-2\phi}_\nu\,{1\over s_{iB_1}}\,\Big]\nn
&&+\Big[\,\W{\cal J}_{A_1}^\mu\,V^{1g-2\phi}_\mu\,{1\over s_{iA_1}}\,\Big]\,\times\,\Big[\,\epsilon^{{\cal S}_B;\nu_1}_i\W{\cal J}_{B_1}^{\nu_2}\,V^{3g}_{\nu_1\nu_2\nu_3}\,\eta^{\nu_3\rho}\,{1\over s_{iB_1}}\,\Big]\,,~~~~\label{result-example3-YM}
\eea
which is precisely the factorization behavior on the r.h.s. of Fig.\ref{Fig20}, for either $\epsilon_i^{{\cal S}_A}$ or $\epsilon_i^{{\cal S}_B}$.

\subsection{Recursive proof for general SFASL}
\label{subsec-YM-gen}

In the preceding subsections, \ref{subsec-YM-case1}, \ref{subsec-YM-case2} and \ref{subsec-YM-case3}, we have shown that the SFASL expressed in \eref{fac-propa+v-YM} always holds for $(p,q)=(1,1)$, regardless of whether $n_A=n_B=1$, $n_A=n_B=2$, or $n_A=2,\,n_B=1$ ($n_A=1,\,n_B=2$). Meanwhile, it is direct to observe that this SFASL holds for any $(p,0)$ or $(0,q)$. Now we proceed to prove, by recursion, that the SFASL in \eref{fac-propa+v-YM} holds for arbitrary $(p,q)$.

Similar to section \ref{subsec-NLSM-gen}, we assume that the SFASL in \eref{fac-propa+v-YM} holds for $(p,q-1)$, $(p-1,q)$, and $(p-1,q-1)$. According to the number of $A$-lines contained in $\{A\}_p$ and the number of $B$-lines contained in $\{B\}_q$, we still distinguish three cases, namely $n_{A_p}=n_{B_q}=2$, $n_{A_p}=2,\,n_{B_q}=1$, and $n_{A_p}=n_{B_q}=1$. For latter convenience, we denote the first line of \eref{fac-propa+v-YM}
as $J_{i|\underline{\{A\}}\{A\}_p|\underline{\{B\}}\{B\}_q}$, where $\underline{\{A\}}\equiv\{A\}_1\cdots\{A\}_{p-1}$, $\underline{\{B\}}\equiv\{B\}_1\cdots\{B\}_{q-1}$, as defined below \eref{lhs-SFASL-NLSM}.

\noindent\textbf{Case 1: $n_{A_p}=n_{B_q}=2$}

In this case, $\{A\}_p$ contains two $A$-lines and $\{B\}_q$ contains two $B$-lines. Suppose that the two $A$-lines are connected to blocks $A_1$ and $A_2$, and the two $B$-lines are connected to blocks $B_1$ and $B_2$. With the above specifications, to implement the recursion, we express the first line of \eref{fac-propa+v-YM} as
\bea
&&J^{\rho}_{i|\underline{\{A\}}\{A\}_p|\underline{\{B\}}\{B\}_q}\nn
&=&\sum_{\shuffle(p,q)}\,\Big(\epsilon_i\{\W{\cal J}_{\hat{a}}\}\{\W{\cal J}_{\hat{b}}\}\Big)^{\{\mu\}}\,\Big(\prod_{t=1}^{p+q-N_{A|B}}\,{V^{(i,\bullet)}_t\over D_t^{(i,\bullet)}}\Big)_{\{\mu\}\nu}\,\eta^{\nu\rho}\nn
&=&\sum_{\shuffle(p,q-1)}\,\Big(\epsilon_i\{\W{\cal J}_{\hat{a}}\}\{\W{\cal J}_{\hat{b}}\}'\Big)^{\{\mu\}'}\,\Big(\prod_{t=1}^{p+q-1-N_{A|B}}\,{V^{(i,\bullet)}_t\over D_t^{(i,\bullet)}}\Big)_{\{\mu\}'\nu}\,\eta^{\nu\mu_1}\,\W{\cal J}^{\mu_2}_{B_1}\W{\cal J}^{\mu_3}_{B_2}\,{V^{4g}_{\mu_1\mu_2\mu_3\mu_4}\,\eta^{\mu_4\rho}\over s_{i\underline{\{A\}}\{A\}_p\underline{\{B\}}\{B\}_q}}\nn
&&+\sum_{\shuffle(p-1,q)}\,\Big(\epsilon_i\{\W{\cal J}_{\hat{a}}\}'\{\W{\cal J}_{\hat{b}}\}\Big)^{\{\mu\}''}\,\Big(\prod_{t=1}^{p+q-1-N_{A|B}}\,{V^{(i,\bullet)}_t\over D_t^{(i,\bullet)}}\Big)_{\{\mu\}''\nu}\,\eta^{\nu\mu_1}\,\W{\cal J}^{\mu_2}_{A_1}\W{\cal J}^{\mu_3}_{A_2}\,{V^{4g}_{\mu_1\mu_2\mu_3\mu_4}\,\eta^{\mu_4\rho}\over s_{i\underline{\{A\}}\{A\}_p\underline{\{B\}}\{B\}_q}}\nn
&\xrightarrow[]{\eref{kine-condi-shuffle-YM}}&\,\Big[\,\big(J_{i|\underline{\{A\}}|\underline{\{B\}}}^{\nu_1}\W{\cal J}^{\nu_2}_{A_1}\W{\cal J}^{\nu_3}_{A_2}\,V^{4g}_{\nu_1\nu_2\nu_3\nu_4}\big)\,\eta^{\nu_4\mu_1}\,\W{\cal J}^{\mu_2}_{B_1}\W{\cal J}^{\mu_3}_{B_2}\,V^{4g}_{\mu_1\mu_2\mu_3\mu_4}\,\Big]\,\eta^{\mu_4\rho}\,{1\over s_{i\underline{\{A\}}A_1A_2}}\,{1\over s_{i\underline{\{A\}}A_1A_2\underline{\{B\}}B_1B_2}}\nn
&&+\Big[\,\big(J_{i|\underline{\{A\}}|\underline{\{B\}}}^{\nu_1}\W{\cal J}^{\nu_2}_{B_1}\W{\cal J}^{\nu_3}_{B_2}\,V^{4g}_{\nu_1\nu_2\nu_3\nu_4}\big)\,\eta^{\nu_4\mu_1}\,\W{\cal J}^{\mu_2}_{A_1}\W{\cal J}^{\mu_3}_{A_2}\,V^{4g}_{\mu_1\mu_2\mu_3\mu_4}\,\Big]\,\eta^{\mu_4\rho}\,{1\over s_{i\underline{\{B\}}B_1B_2}}\,{1\over s_{i\underline{\{A\}}A_1A_2\underline{\{B\}}B_1B_2}}\,,~~\label{above}
\eea
where the sets $\{\W{\cal J}_{\hat{a}}\}'$ and $\{\W{\cal J}_{\hat{b}}\}'$ are given by
\bea
\{\W{\cal J}_{\hat{a}}\}'=\{\W{\cal J}_{\hat{a}}\}\setminus\{\W{\cal J}_{B_1},\W{\cal J}_{B_2}\}\,,~~~~~~~~
\{\W{\cal J}_{\hat{b}}\}'=\{\W{\cal J}_{\hat{b}}\}\setminus\{\W{\cal J}_{B_1},\W{\cal J}_{B_2}\}\,.
\eea
In the above, we omit some details that are similar to those in section \ref{subsec-NLSM-gen}.
The Lorentz vector $J^\rho_{i|\underline{\{A\}}|\underline{\{B\}}}$ is precisely $J^\rho_{i|\underline{\{A\}}\{A\}_p|\underline{\{B\}}\{B\}_q}$ with $(p,q)\to(p-1,q-1)$, namely,
\bea
J_{i|\underline{\{A\}}|\underline{\{B\}}}^{\rho}=\sum_{\shuffle(p-1,q-1)}\,\Big(\epsilon_i\{\W{\cal J}_{\hat{a}}\}\{\W{\cal J}_{\hat{b}}\}\Big)^{\{\mu\}}\,\Big(\prod_{t=1}^{p+q-2-N_{A|B}}\,{V^{(i,\bullet)}_t\over D_t^{(i,\bullet)}}\Big)_{\{\mu\}\nu}\,\eta^{\nu\rho}
~~\label{define-J}
\eea
Note that in order to obtain such a $J_{i|\underline{\{A\}}|\underline{\{B\}}}$ in \eref{above}, we have inverted the SFASL in \eref{fac-propa+v-YM} for $(p-1,q-1)$. Otherwise, this part will be expressed in a factorized form, analogous to $C_{(p-1,q-1)}$ in \eref{F1-NLSM-gen}. For latter convenience, we also define
\bea
J_{i|\underline{\{A\}}|\underline{\{B\}}}^{{\cal S}_A;\rho}&=&\sum_{\shuffle(p-1,q-1)}\,\Big(\epsilon^{{\cal S}_A}_i\{\W{\cal J}_{\hat{a}}\}\{\W{\cal J}_{\hat{b}}\}\Big)^{\{\mu\}}\,\Big(\prod_{t=1}^{p+q-2-N_{A|B}}\,{V^{(i,\bullet)}_t\over D_t^{(i,\bullet)}}\Big)_{\{\mu\}\nu}\,\eta^{\nu\rho}\,,\nn
J_{i|\underline{\{A\}}|\underline{\{B\}}}^{{\cal S}_B;\rho}&=&\sum_{\shuffle(p-1,q-1)}\,\Big(\epsilon^{{\cal S}_B}_i\{\W{\cal J}_{\hat{a}}\}\{\W{\cal J}_{\hat{b}}\}\Big)^{\{\mu\}}\,\Big(\prod_{t=1}^{p+q-2-N_{A|B}}\,{V^{(i,\bullet)}_t\over D_t^{(i,\bullet)}}\Big)_{\{\mu\}\nu}\,\eta^{\nu\rho}\,,~~\label{defin-JAB}
\eea
therefore,
\bea
J^{\nu_1}_{i|\underline{\{A\}}|\underline{\{B\}}}&=&J^{{\cal S}_A;\nu_1}_{i|\underline{\{A\}}|\underline{\{B\}}}
+J^{{\cal S}_B;\nu_1}_{i|\underline{\{A\}}|\underline{\{B\}}}\,.~~\label{Jpq}
\eea
Since the SFASL in \eref{fac-propa+v-YM} is assumed to be valid for $(p-1,q-1)$, we have
\bea
&&J^{{\cal S}_A;\rho}_{i|\underline{\{A\}}|\underline{\{B\}}}\,\xrightarrow[]{\eref{kine-condi-shuffle-YM}}\nn
&&\Big[\,\Big(\epsilon_i^{{\cal S}_A}\{\W{\cal J}_{\hat{a}}\}'\Big)^{\{\mu\}_1}\,\Big(\prod_{\a=1}^{p-1}\,{V_{\{A\}_\a}\over s_{i\{A\}_1\cdots \{A\}_{\a}}}\Big)_{\{\mu\}_1\nu}\,\eta^{\nu\rho}\,\Big]\,\times\,\Big[\,\Big(\{\W{\cal J}_{\hat{b}}\}'\Big)^{\{\mu\}_2}\,
\Big(\prod_{\b=1}^{q-1}\,{V_{\{B\}_\b}\over s_{i\{B\}_1\cdots \{B\}_{\b}}}\Big)_{\{\mu\}_2}\,\Big]\,,\nn
&&J^{{\cal S}_B;\rho}_{i|\underline{\{A\}}|\underline{\{B\}}}\,\xrightarrow[]{\eref{kine-condi-shuffle-YM}}\nn
&&\Big[\,\Big(\{\W{\cal J}_{\hat{a}}\}'\Big)^{\{\mu\}_1}\,\Big(\prod_{\a=1}^p\,{V_{\{A\}_\a}\over s_{i\{A\}_1\cdots \{A\}_{\a}}}\Big)_{\{\mu\}_1}\,\Big]\,\times\,\Big[\,\Big(\epsilon_i^{{\cal S}_B}\{\W{\cal J}_{\hat{b}}\}'\Big)^{\{\mu\}_2}\,
\Big(\prod_{\b=1}^q\,{V_{\{B\}_\b}\over s_{i\{B\}_1\cdots \{B\}_{\b}}}\Big)_{\{\mu\}_2\nu}\,\eta^{\nu\rho}\,\Big]\,.~~\label{SFASL-JAB}
\eea

Except for the replacement of $\epsilon_i$ by $J_{i|\underline{\{A\}}|\underline{\{B\}}}$, \eref{above} has exactly the same form as \eref{YM-example1-F1}.
Thus, the subsequent calculation is merely a repetition of the procedure from \eref{YM-example1-F1} to \eref{result-example1-YM}, with $\epsilon_i$ replaced by $J_{i|\underline{\{A\}}|\underline{\{B\}}}$. After performing this, we obtain the SFASL in \eref{fac-propa+v-YM},
\bea
&&J^{\rho}_{i|\underline{\{A\}}\{A\}_p|\underline{\{B\}}\{B\}_q}\nn
&\xrightarrow[]{\eref{kine-condi-shuffle-YM}}&\,\Big[J_{i|\underline{\{A\}}|\underline{\{B\}}}^{{\cal S}_A;\mu_1}\W{\cal J}_{A_1}^{\mu_2}\W{\cal J}_{A_2}^{\mu_3}\,V^{4g}_{\mu_1\mu_2\mu_2\mu_4}\,\eta^{\mu_4\rho}\,{1\over s_{i\underline{\{A\}}A_1A_2}}\Big]\,\times\,\Big[\W{\cal J}^{\nu_1}_{B_1}\W{\cal J}^{\nu_2}_{B_2}\,V^{2g-2\phi}_{\nu_1\nu_2}\,{1\over s_{i\underline{\{B\}}B_1B_2}}\Big]\nn
&&+\Big[\W{\cal J}_{A_1}^{\mu_1}\W{\cal J}_{A_2}^{\mu_2}\,V^{2g-2\phi}_{\mu_1\mu_2}\,{1\over s_{i\underline{\{A\}}A_1A_2}}\Big]\,\times\,\Big[J_{i|\underline{\{A\}}|\underline{\{B\}}}^{{\cal S}_B;\nu_1}\W{\cal J}^{\nu_2}_{B_1}\W{\cal J}^{\nu_3}_{B_2}\,V^{4g}_{\nu_1\nu_2\nu_3\nu_4}\,\eta^{\nu_4\rho}\,{1\over s_{i\underline{\{B\}}B_1B_2}}\Big]\nn
&=&\Big[\,\Big(\epsilon_i^{{\cal S}_A}\{\W{\cal J}_{\hat{a}}\}\Big)^{\{\mu\}_1}\,\Big(\prod_{\a=1}^p\,{V_{\{A\}_\a}\over s_{i\{A\}_1\cdots \{A\}_{\a}}}\Big)_{\{\mu\}_1\nu}\,\eta^{\nu\rho}\,\Big]\,\times\,\Big[\,\Big(\{\W{\cal J}_{\hat{b}}\}\Big)^{\{\mu\}_2}\,
\Big(\prod_{\b=1}^q\,{V_{\{B\}_\b}\over s_{i\{B\}_1\cdots \{B\}_{\b}}}\Big)_{\{\mu\}_2}\,\Big]\nn
&&+\Big[\,\Big(\{\W{\cal J}_{\hat{a}}\}\Big)^{\{\mu\}_1}\,\Big(\prod_{\a=1}^p\,{V_{\{A\}_\a}\over s_{i\{A\}_1\cdots \{A\}_{\a}}}\Big)_{\{\mu\}_1}\,\Big]\,\times\,\Big[\,\Big(\epsilon_i^{{\cal S}_B}\{\W{\cal J}_{\hat{b}}\}\Big)^{\{\mu\}_2}\,
\Big(\prod_{\b=1}^q\,{V_{\{B\}_\b}\over s_{i\{B\}_1\cdots \{B\}_{\b}}}\Big)_{\{\mu\}_2\nu}\,\eta^{\nu\rho}\,\Big]\,,~~\label{YM-gen-case1}
\eea
where the last step uses the SFASL of $J_{i|\underline{\{A\}}|\underline{\{B\}}}^{{\cal S}_A}$ and $J_{i|\underline{\{A\}}|\underline{\{B\}}}^{{\cal S}_B}$ in \eref{SFASL-JAB}.

It is worth pointing out that, to obtain the above result, we have used
\bea
&&J_{i|\underline{\{A\}}|\underline{\{B\}}}^{{\cal S}_B;\mu_1}\W{\cal J}_{A_1}^{\mu_2}\W{\cal J}_{A_2}^{\mu_3}\,V^{4g}_{\mu_1\mu_2\mu_2\mu_4}=0\,,~~~~~~~~J_{i|\underline{\{A\}}|\underline{\{B\}}}^{{\cal S}_A;\nu_1}\W{\cal J}^{\nu_2}_{B_1}\W{\cal J}^{\nu_3}_{B_2}\,V^{4g}_{\nu_1\nu_2\nu_3\nu_4}=0\,.~~\label{condi-J}
\eea
The condition in the above may seem rather abstract. However, by returning to the procedure that leads from \eref{F1F2-step1} to \eref{result-example1-YM} and replacing $\epsilon_i$ with $J_{i|\underline{\{A\}}|\underline{\{B\}}}$, one sees that the above condition actually requires that $J^{{\cal S}_A}_{i|\underline{\{A\}}|\underline{\{B\}}}$ annihilate any kinematic variables coming from the subspace ${\cal S}_B$, while $J^{{\cal S}_B}_{i|\underline{\{A\}}|\underline{\{B\}}}$ annihilate any kinematic variables coming from ${\cal S}_A$---that is, they have the same property as $\epsilon_i^{{\cal S}_A}$ and $\epsilon_i^{{\cal S}_B}$, which belong to two orthogonal subspaces. Since the same requirement on
$J^{{\cal S}_A}_{i|\underline{\{A\}}|\underline{\{B\}}}$ and $J^{{\cal S}_B}_{i|\underline{\{A\}}|\underline{\{B\}}}$ also appears in Cases 2 and 3, we will explain at the end of this subsection why the above condition holds.

\noindent\textbf{Case 2: $n_{A_p}=2\,,\,n_{B_q}=1$}

In this case, $\{A\}_p$ contains two $A$-lines and $\{B\}_q$ contains one $B$-line. We assume that the two $A$-lines are connected to blocks $A_1$ and $A_2$, and the $B$-line is connected to the block $B_1$. In this case, the first line in \eref{fac-propa+v-YM} reads
\bea
&&J^{\rho}_{i|\underline{\{A\}}\{A\}_p|\underline{\{B\}}\{B\}_q}\nn
&=&\sum_{\shuffle(p,q)}\,\Big(\epsilon_i\{\W{\cal J}_{\hat{a}}\}\{\W{\cal J}_{\hat{b}}\}\Big)^{\{\mu\}}\,\Big(\prod_{t=1}^{p+q-N_{A|B}}\,{V^{(i,\bullet)}_t\over D_t^{(i,\bullet)}}\Big)_{\{\mu\}\nu}\,\eta^{\nu\rho}\nn
&=&\sum_{\shuffle(p,q-1)}\,\Big(\epsilon_i\{\W{\cal J}_{\hat{a}}\}\{\W{\cal J}_{\hat{b}}\}''\Big)^{\{\mu\}''}\,\Big(\prod_{t=1}^{p+q-1-N_{A|B}}\,{V^{(i,\bullet)}_t\over D_t^{(i,\bullet)}}\Big)_{\{\mu\}''\nu}\,\eta^{\nu\mu_1}\,\W{\cal J}^{\mu_2}_{B_1}\,{V^{3g}_{\mu_1\mu_2\mu_3}\,\eta^{\mu_3\rho}\over s_{i\underline{\{A\}}\{A\}_p\underline{\{B\}}\{B\}_q}}\nn
&&+\sum_{\shuffle(p-1,q)}\,\Big(\epsilon_i\{\W{\cal J}_{\hat{a}}\}'\{\W{\cal J}_{\hat{b}}\}\Big)^{\{\mu\}'}\,\Big(\prod_{t=1}^{p+q-1-N_{A|B}}\,{V^{(i,\bullet)}_t\over D_t^{(i,\bullet)}}\Big)_{\{\mu\}'\nu}\,\eta^{\nu\mu_1}\,\W{\cal J}^{\mu_2}_{A_1}\W{\cal J}^{\mu_3}_{A_2}\,{V^{4g}_{\mu_1\mu_2\mu_3\mu_4}\,\eta^{\mu_4\rho}\over s_{i\underline{\{A\}}\{A\}_p\underline{\{B\}}\{B\}_q}}\nn
&\xrightarrow[]{\eref{kine-condi-shuffle-YM}}&\,\Big[\,\big(J_{i|\underline{\{A\}}|\underline{\{B\}}}^{\nu_1}\W{\cal J}^{\nu_2}_{A_1}\W{\cal J}^{\nu_3}_{A_2}\,V^{4g}_{\nu_1\nu_2\nu_3\nu_4}\big)\,\eta^{\nu_4\mu_1}\,\W{\cal J}^{\mu_2}_{B_1}\,V^{3g}_{\mu_1\mu_2\mu_3}\,\Big]\,\eta^{\mu_3\rho}\,{1\over s_{i\underline{\{A\}}A_1A_2}}\,{1\over s_{i\underline{\{A\}}A_1A_2\underline{\{B\}}B_1}}\nn
&&+\Big[\,\big(J_{i|\underline{\{A\}}|\underline{\{B\}}}^{\nu_1}\W{\cal J}^{\nu_2}_{B_1}\,V^{3g}_{\nu_1\nu_2\nu_3}\big)\,\eta^{\nu_3\mu_1}\,\W{\cal J}^{\mu_2}_{A_1}\W{\cal J}^{\mu_3}_{A_2}\,V^{4g}_{\mu_1\mu_2\mu_3\mu_4}\,\Big]\,\eta^{\mu_4\rho}\,{1\over s_{i\underline{\{B\}}B_1}}\,{1\over s_{i\underline{\{A\}}A_1A_2\underline{\{B\}}B_1}}\,,~~~\label{above2}
\eea
where
\bea
\{\W{\cal J}_{\hat{b}}\}''=\{\W{\cal J}_{\hat{b}}\}\setminus\W{\cal J}_{B_1}\,.
\eea

Except that $\epsilon_i$ is replaced by $J_{i|\underline{\{A\}}|\underline{\{B\}}}$, the form of \eref{above2} is identical to that of \eref{YM-example2-F1}. By repeating the process from \eref{YM-example2-F1} to \eref{result-example2-YM}, we get the SFASL in \eref{fac-propa+v-YM},
\bea
&&J^{\rho}_{i|\underline{\{A\}}\{A\}_p|\underline{\{B\}}\{B\}_q}\nn
&\xrightarrow[]{\eref{kine-condi-shuffle-YM}}&\,\Big[\,J_{i|\underline{\{A\}}|\underline{\{B\}}}^{{\cal S}_A;\mu_1}\W{\cal J}_{A_1}^{\mu_2}\W{\cal J}_{A_2}^{\mu_3}\,V^{4g}_{\mu_1\mu_2\mu_3\mu_4}\,\eta^{\mu_4\rho}\,{1\over s_{i\underline{\{A\}}A_1A_2}}\,\Big]\,\times\,\Big[\,\W{\cal J}_{B_1}^\nu\,V_{\nu}^{1g-2\phi}\,{1\over s_{i\underline{\{B\}}B_1}}\,\Big]\nn
&&+\Big[\,\W{\cal J}_{A_1}^{\mu_1}\W{\cal J}_{A_2}^{\mu_2}\,V^{2g-2\phi}_{\mu_1\mu_2}\,{1\over s_{i\underline{\{A\}}A_1A_2}}\,\Big]\,\times\,\Big[\,J_{i|\underline{\{A\}}|\underline{\{B\}}}^{{\cal S}_B;\nu_1}\W{\cal J}_{B_1}^{\nu_2}\,V_{\nu_1\nu_2\nu_3}^{3g}\,\eta^{\nu_3\rho}\,{1\over s_{i\underline{\{B\}}B_1}}\,\Big]\nn
&=&\Big[\,\Big(\epsilon_i^{{\cal S}_A}\{\W{\cal J}_{\hat{a}}\}\Big)^{\{\mu\}_1}\,\Big(\prod_{\a=1}^p\,{V_{\{A\}_\a}\over s_{i\{A\}_1\cdots \{A\}_{\a}}}\Big)_{\{\mu\}_1\nu}\,\eta^{\nu\rho}\,\Big]\,\times\,\Big[\,\Big(\{\W{\cal J}_{\hat{b}}\}\Big)^{\{\mu\}_2}\,
\Big(\prod_{\b=1}^q\,{V_{\{B\}_\b}\over s_{i\{B\}_1\cdots \{B\}_{\b}}}\Big)_{\{\mu\}_2}\,\Big]\nn
&&+\Big[\,\Big(\{\W{\cal J}_{\hat{a}}\}\Big)^{\{\mu\}_1}\,\Big(\prod_{\a=1}^p\,{V_{\{A\}_\a}\over s_{i\{A\}_1\cdots \{A\}_{\a}}}\Big)_{\{\mu\}_1}\,\Big]\,\times\,\Big[\,\Big(\epsilon_i^{{\cal S}_B}\{\W{\cal J}_{\hat{b}}\}\Big)^{\{\mu\}_2}\,
\Big(\prod_{\b=1}^q\,{V_{\{B\}_\b}\over s_{i\{B\}_1\cdots \{B\}_{\b}}}\Big)_{\{\mu\}_2\nu}\,\eta^{\nu\rho}\,\Big]\,,~~\label{YM-gen-case2}
\eea
where the final step uses the SFASL of $J_{i|\underline{\{A\}}|\underline{\{B\}}}^{{\cal S}_A}$ and $J_{i|\underline{\{A\}}|\underline{\{B\}}}^{{\cal S}_B}$ in \eref{SFASL-JAB}.

A subtlety arises here: in \eref{onshell}, we used the on-shell condition $\epsilon_i\cdot k_i=0$ for $\epsilon_i$. Thus, we need to guarantee that
$J_{i|\underline{\{A\}}|\underline{\{B\}}}$ also obeys a similar condition $J_{i|\underline{\{A\}}|\underline{\{B\}}}\cdot k_{i\underline{\{A\}}\underline{\{B\}}}=0$ upon replacing $\epsilon_i$ with $J_{i|\underline{\{A\}}|\underline{\{B\}}}$; otherwise, the derivation in section \ref{subsec-YM-case2} cannot be repeated. The solution is that the condition $J_{i|\underline{\{A\}}|\underline{\{B\}}}\cdot k_{i\underline{\{A\}}\underline{\{B\}}}=0$ holds effectively, owing to the fact that
$J_{i|\underline{\{A\}}|\underline{\{B\}}}$ is a component of the BG current ${\cal J}_{i\underline{\{A\}}\underline{\{B\}}}$, and the BG current obeys \eref{BG-gauge}. Consequently, when summing over Feynman diagrams, the contribution from $J_{i|\underline{\{A\}}|\underline{\{B\}}}\cdot k_{i\underline{\{A\}}\underline{\{B\}}}$ cancels out; that is, $J_{i|\underline{\{A\}}|\underline{\{B\}}}\cdot k_{i\underline{\{A\}}\underline{\{B\}}}$ is effectively zero.
The reason why the property \eref{BG-gauge} can be used here is that, since $J_{i|\underline{\{A\}}|\underline{\{B\}}}$ is defined by the first line of \eref{fac-propa+v-YM} (or defined by \eref{define-J}), it is evident that
$J_{i|\underline{\{A\}}|\underline{\{B\}}}$ is computed according to the original Feynman rules and is therefore a component of the BG current. In the first step of \eref{YM-gen-case2}, we have not made any modification to the form of
$J_{i|\underline{\{A\}}|\underline{\{B\}}}$; hence, the property of the BG current can be safely applied. The problem discussed below \eref{condi-J} is not present here. It is only in the second step of \eref{YM-gen-case2}, where we substitute the SFASL in \eref{SFASL-JAB}, that the original form of
$J_{i|\underline{\{A\}}|\underline{\{B\}}}$ is altered.

\noindent\textbf{Case 3: $n_{A_p}=n_{B_q}=1$}

In this case, $\{A\}_p$ contains one $A$-line and $\{B\}_q$ contains one $B$-line. Suppose that these two lines are connected to blocks $A_1$ and $B_1$, respectively. Then, the first line in \eref{fac-propa+v-YM} becomes
\bea
&&J^{\rho}_{i|\underline{\{A\}}\{A\}_p|\underline{\{B\}}\{B\}_q}\nn
&=&\sum_{\shuffle(p,q)}\,\Big(\epsilon_i\{\W{\cal J}_{\hat{a}}\}\{\W{\cal J}_{\hat{b}}\}\Big)^{\{\mu\}}\,\Big(\prod_{t=1}^{p+q-N_{A|B}}\,{V^{(i,\bullet)}_t\over D_t^{(i,\bullet)}}\Big)_{\{\mu\}\nu}\,\eta^{\nu\rho}\nn
&=&\sum_{\shuffle(p,q-1)}\,\Big(\epsilon_i\{\W{\cal J}_{\hat{a}}\}\{\W{\cal J}_{\hat{b}}\}''\Big)^{\{\mu\}'}\,\Big(\prod_{t=1}^{p+q-1-N_{A|B}}\,{V^{(i,\bullet)}_t\over D_t^{(i,\bullet)}}\Big)_{\{\mu\}'\nu}\,\eta^{\nu\mu_1}\,\W{\cal J}^{\mu_2}_{B_1}\,{V^{3g}_{\mu_1\mu_2\mu_3}\,\eta^{\mu_3\rho}\over s_{i\underline{\{A\}}\{A\}_p\underline{\{B\}}\{B\}_q}}\nn
&&+\sum_{\shuffle(p-1,q)}\,\Big(\epsilon_i\{\W{\cal J}_{\hat{a}}\}''\{\W{\cal J}_{\hat{b}}\}\Big)^{\{\mu\}''}\,\Big(\prod_{t=1}^{p+q-1-N_{A|B}}\,{V^{(i,\bullet)}_t\over D_t^{(i,\bullet)}}\Big)_{\{\mu\}''\nu}\,\eta^{\nu\mu_1}\,\W{\cal J}^{\mu_2}_{A_1}\,{V^{3g}_{\mu_1\mu_2\mu_3}\,\eta^{\mu_3\rho}\over s_{i\underline{\{A\}}\{A\}_p\underline{\{B\}}\{B\}_q}}\nn
&&+\sum_{\shuffle(p-1,q-1)}\,\Big(\epsilon_i\{\W{\cal J}_{\hat{a}}\}''\{\W{\cal J}_{\hat{b}}\}''\Big)^{\{\mu\}'''}\,\Big(\prod_{t=1}^{p+q-1-N_{A|B}}\,{V^{(i,\bullet)}_t\over D_t^{(i,\bullet)}}\Big)_{\{\mu\}'''\nu}\,\eta^{\nu\mu_1}\,\W{\cal J}^{\mu_2}_{A_1}\W{\cal J}^{\mu_3}_{B_1}\,{V^{4g}_{\mu_1\mu_2\mu_3\mu_4}\,\eta^{\mu_4\rho}\over s_{i\underline{\{A\}}\{A\}_p\underline{\{B\}}\{B\}_q}}\nn
&\xrightarrow[]{\eref{kine-condi-shuffle-YM}}&\,\Big[\,\big(J_{i|\underline{\{A\}}|\underline{\{B\}}}^{\nu_1}\W{\cal J}^{\nu_2}_{A_1}\,V^{3g}_{\nu_1\nu_2\nu_3}\big)\,\eta^{\nu_3\mu_1}\,\W{\cal J}^{\mu_2}_{B_1}\,V^{3g}_{\mu_1\mu_2\mu_3}\,\Big]\,\eta^{\mu_3\rho}\,{1\over s_{i\underline{\{A\}}A_1}}\,{1\over s_{i\underline{\{A\}}A_1\underline{\{B\}}B_1}}\nn
&&+\Big[\,\big(J_{i|\underline{\{A\}}|\underline{\{B\}}}^{\nu_1}\W{\cal J}^{\nu_2}_{B_1}\,V^{3g}_{\nu_1\nu_2\nu_3}\big)\,\eta^{\nu_3\mu_1}\,\W{\cal J}^{\mu_2}_{A_1}\,V^{3g}_{\mu_1\mu_2\mu_3}\,\Big]\,\eta^{\mu_3\rho}\,{1\over s_{i\underline{\{B\}}B_1}}\,{1\over s_{i\underline{\{A\}}A_1\underline{\{B\}}B_1}}\nn
&&+\Big[\,J_{i|\underline{\{A\}}|\underline{\{B\}}}^{\mu_1}\W{\cal J}^{\mu_2}_{A_1}\,\W{\cal J}^{\mu_3}_{B_1}\,V^{4g}_{\mu_1\mu_2\mu_3\mu_4}\,\Big]\,\eta^{\mu_3\rho}\,{1\over s_{i\underline{\{A\}}A_1\underline{\{B\}}B_1}}\,.~~\label{above3}
\eea

Apart from that $\epsilon_i$ is replaced by $J_{i|\underline{\{A\}}|\underline{\{B\}}}$, \eref{above3} is identical in form to \eref{YM-example3-F1}.
Repeating the manipulation from \eref{YM-example3-F1} to \eref{result-example3-YM}, and using the effective property $J_{i|\underline{\{A\}}|\underline{\{B\}}}\cdot k_{i\underline{\{A\}}\underline{\{B\}}}=0$ discussed earlier, we again arrive at the SFASL in \eref{fac-propa+v-YM},
\bea
&&J^{\rho}_{i|\underline{\{A\}}\{A\}_p|\underline{\{B\}}\{B\}_q}\nn
&\xrightarrow[]{\eref{kine-condi-shuffle-YM}}&\,\Big[\,J_{i|\underline{\{A\}}|\underline{\{B\}}}^{{\cal S}_A;\mu_1}\W{\cal J}_{A_1}^{\mu_2}\,V^{3g}_{\mu_1\mu_2\mu_3}\,\eta^{\mu_3\rho}\,{1\over s_{i\underline{\{A\}}A_1A_2}}\,\Big]\,\times\,\Big[\,\W{\cal J}_{B_1}^\nu\,V_{\nu}^{1g-2\phi}\,{1\over s_{i\underline{\{B\}}B_1}}\,\Big]\nn
&&+\Big[\,\W{\cal J}_{A_1}^{\mu_1}\,V^{1g-2\phi}_{\mu_1\mu_2}\,{1\over s_{i\underline{\{A\}}A_1A_2}}\,\Big]\,\times\,\Big[\,J_{i|\underline{\{A\}}|\underline{\{B\}}}^{{\cal S}_B;\nu_1}\W{\cal J}_{B_1}^{\nu_2}\,V_{\nu_1\nu_2\nu_3}^{3g}\,\eta^{\nu_3\rho}\,{1\over s_{i\underline{\{B\}}B_1}}\,\Big]\nn
&=&\Big[\,\Big(\epsilon_i^{{\cal S}_A}\{\W{\cal J}_{\hat{a}}\}\Big)^{\{\mu\}_1}\,\Big(\prod_{\a=1}^p\,{V_{\{A\}_\a}\over s_{i\{A\}_1\cdots \{A\}_{\a}}}\Big)_{\{\mu\}_1\nu}\,\eta^{\nu\rho}\,\Big]\,\times\,\Big[\,\Big(\{\W{\cal J}_{\hat{b}}\}\Big)^{\{\mu\}_2}\,
\Big(\prod_{\b=1}^q\,{V_{\{B\}_\b}\over s_{i\{B\}_1\cdots \{B\}_{\b}}}\Big)_{\{\mu\}_2}\,\Big]\nn
&&+\Big[\,\Big(\{\W{\cal J}_{\hat{a}}\}\Big)^{\{\mu\}_1}\,\Big(\prod_{\a=1}^p\,{V_{\{A\}_\a}\over s_{i\{A\}_1\cdots \{A\}_{\a}}}\Big)_{\{\mu\}_1}\,\Big]\,\times\,\Big[\,\Big(\epsilon_i^{{\cal S}_B}\{\W{\cal J}_{\hat{b}}\}\Big)^{\{\mu\}_2}\,
\Big(\prod_{\b=1}^q\,{V_{\{B\}_\b}\over s_{i\{B\}_1\cdots \{B\}_{\b}}}\Big)_{\{\mu\}_2\nu}\,\eta^{\nu\rho}\,\Big]\,.~~\label{YM-gen-case3}
\eea

We have completed the recursive proof of the SFASL in all cases. Now we return to the premise on which the proof relies, namely \eref{condi-J}. If \eref{condi-J} does not hold, then the difference between $J_{i|\underline{\{A\}}|\underline{\{B\}}}$ and $\epsilon_i$ would prevent us from replicating the manipulations in sections \ref{subsec-YM-case1} through \ref{subsec-YM-case3}. As discussed earlier, the essence of \eref{condi-J} is that $J^{{\cal S}_A}_{i|\underline{\{A\}}|\underline{\{B\}}}$ ($J^{{\cal S}_B}_{i|\underline{\{A\}}|\underline{\{B\}}}$) annihilates any kinematic variables from ${\cal S}_B$ (${\cal S}_A$). First, from \eref{result-example1-YM}, \eref{result-example2-YM} and \eref{result-example3-YM}, one sees that this obviously holds when
$(p,q)=(1,1)$. Meanwhile, it is straightforward to verify that this condition also holds for $(p,q)=(1,0)$ and $(p,q)=(0,1)$. The recursive pattern shown in \eref{YM-gen-case1}, \eref{YM-gen-case2} and \eref{YM-gen-case3} indicates that, $J_{i|\underline{\{A\}}|\underline{\{B\}}}$ for $(p+1,q+1)$ can be obtained by replacing $\epsilon_i$ with $J_{i|\underline{\{A\}}|\underline{\{B\}}}$ for $(p,q)$ in \eref{result-example1-YM} or \eref{result-example2-YM} or \eref{result-example3-YM}. Therefore, the validity of condition \eref{condi-J} is ensured recursively.

\subsection{From SFASL to hidden zero and $2$-split}
\label{subsec-YM-0andsplit}

The procedure for reproducing the hidden zeros and $2$-split of YM amplitudes from the perspective of SFASL is similar to the ${\rm Tr}(\phi^3)$ and NLSM cases. Therefore, we will omit many details.

\noindent\textbf{Hidden zeros}

The kinematic condition for hidden zeros of YM amplitudes is given as
\bea
\{\epsilon_a,\,k_a\}\cdot\{\epsilon_b\cdot k_b\}=0\,,~~~~{\rm for}~\forall\,a\in\pmb A\,,~b\in\pmb B\,,~~~~\label{kine-condi-0-YM}
\eea
where two sets, $\pmb A$ and $\pmb B$, are defined as before. Clearly, condition \eref{kine-condi-0-YM} implies that the lines attached to $L_{(i,j)}$ from the $\pmb A$-side and $\pmb B$-side are $A$-lines and $B$-lines, respectively, satisfying the SFASL condition \eref{kine-condi-shuffle-YM}.
Based on the SFALS in \eref{fac-propa+v-YM}, we can extend the formula \eref{0-NLSM} for NLSM amplitudes to
\bea
&&\Big[\,\sum_{\shuffle(p,q)}\,\Big(\epsilon_i\{\W{\cal J}_{\hat{a}}\}\{\W{\cal J}_{\hat{b}}\}\Big)^{\{\mu\}}\,\Big(\prod_{t=1}^{p+q-N_{A|B}}\,{V^{(i,\bullet)}_t\over D_t^{(i,\bullet)}}\Big)_{\{\mu\}\nu}\,\eta^{\nu\rho}\,\Big]\,s_{i\{A\}_1\cdots\{A\}_p\{B\}_1\cdots\{B\}_q}\nn
&\xrightarrow[]{\eref{kine-condi-0-YM}}&\,\Big\{\Big[\,\Big(\epsilon_i^{{\cal S}_A}\{\W{\cal J}_{\hat{a}}\}\Big)^{\{\mu\}_1}\,\Big(\prod_{\a=1}^p\,{V_{\{A\}_\a}\over s_{i\{A\}_1\cdots \{A\}_{\a}}}\Big)_{\{\mu\}_1\nu}\,\eta^{\nu\rho}\,\Big]\,\times\,\Big[\,\Big(\{\W{\cal J}_{\hat{b}}\}\Big)^{\{\mu\}_2}\,
\Big(\prod_{\b=1}^q\,{V_{\{B\}_\b}\over s_{i\{B\}_1\cdots \{B\}_{\b}}}\Big)_{\{\mu\}_2}\,\Big]\Big\}\,k_j^2\nn
&&+\Big\{\Big[\,\Big(\{\W{\cal J}_{\hat{a}}\}\Big)^{\{\mu\}_1}\,\Big(\prod_{\a=1}^p\,{V_{\{A\}_\a}\over s_{i\{A\}_1\cdots \{A\}_{\a}}}\Big)_{\{\mu\}_1}\,\Big]\,\times\,\Big[\,\Big(\epsilon_i^{{\cal S}_B}\{\W{\cal J}_{\hat{b}}\}\Big)^{\{\mu\}_2}\,
\Big(\prod_{\b=1}^q\,{V_{\{B\}_\b}\over s_{i\{B\}_1\cdots \{B\}_{\b}}}\Big)_{\{\mu\}_2\nu}\,\eta^{\nu\rho}\,\Big]\Big\}\,k_j^2\,.\nn~~\label{0-YM}
\eea
Thus, the hidden zeros of YM amplitudes also follow from the on-shell condition $k_j^2=0$.

\noindent\textbf{$2$-split}

The kinematic condition for $2$-split of YM amplitude can be generated from the kinematic condition for hidden zeros by removing $k\in\pmb B$ (or $k\in\pmb A$), and restrict the three polarization vectors $\epsilon_i$, $\epsilon_j$ and $\epsilon_k$ to the subspace ${\cal S}_A$ or ${\cal S}_B$. In this paper we choose ${\cal S}_A$; the treatment for another choice is extremely similar. Thus, the kinematic condition for $2$-split is,
\bea
\{\epsilon_i,\,\epsilon_j,\,\epsilon_k,\,\epsilon_a,\,k_a\}\cdot\{\epsilon_b\cdot k_b\}=0\,,~~~~{\rm for}~\forall\,a\in\pmb A\,,~b\in\pmb B\setminus k\,.~~~~\label{kine-condi-2split-YM}
\eea

Analogous to \eref{NLSM-forsplit}, a YM amplitude can be expressed as
\bea
{\cal A}_{2n}^{\rm YM}(1,\cdots,n)&=&\sum_{{\rm div}\pmb A}\,\sum_{{\rm div}\pmb B}\,\sum_{{\cal P}_{\pmb A}}\,\sum_{{\cal P}_{\pmb B}}\,\Big(\sum_{\shuffle(p,q)}\,\prod_{t=1}^{p+q-N_{A|B}}\,{V_t^{(i,v)}\over D_t^{(i,v)}}\Big)_{\{\mu\}_1}\,\Big(\sum_{\shuffle(m,l)}\,\prod_{t=1}^{m+l-N'_{A|B}}\,{V_t^{(j,v)}\over D_t^{(j,v)}}\Big)_{\{\mu\}_2}\nn
&&\big[V_v\big]_{\{\mu\}_3}\,
\big[\,f^{\rm YM}(R)\,\big]^{\{\mu\}_1\{\mu\}_2\{\mu\}_3}\,,~~\label{YM-forsplit}
\eea
where $v$ is the vertex common to $L_{(i,v)}$, $L_{(j,v)}$ and $L_{k,v}$.
For the given divisions in \eref{div-AB-NLSM}, and given partitions ${\cal P}_{\pmb A}$ and ${\cal P}_{\pmb B}$, the SFASL in \eref{fac-propa+v-YM} leads to
\bea
\Big(\prod_{t=1}^{p+q-N_{A|B}}\,{V^{(i,\bullet)}_t\over D_t^{(i,\bullet)}}\Big)_{\{\mu\}_1}\,&\xrightarrow[]{\eref{kine-condi-2split-YM}}&\,
\Big(\prod_{\a=1}^p\,{V_{\{A\}_\a}\over s_{i\{A\}_1\cdots \{A\}_{\a}}}\Big)_{\mu_i\{\mu\}_{11}}\,\times\,
\Big(\prod_{\b=1}^q\,{V_{\{B\}_\b}\over s_{i\{B\}_1\cdots \{B\}_{\b}}}\Big)_{\{\mu\}_{12}}\,,\nn
\Big(\prod_{t=1}^{m+l-N'_{A|B}}\,{V^{(i,\bullet)}_t\over D_t^{(i,\bullet)}}\Big)_{\{\mu\}_2}\,&\xrightarrow[]{\eref{kine-condi-2split-YM}}&\,
\Big(\prod_{\a=1}^m\,{V_{\{A\}_\a}\over s_{i\{A\}_1\cdots \{A\}_{\a}}}\Big)_{\mu_j\{\mu\}_{21}}\,\times\,
\Big(\prod_{\b=1}^l\,{V_{\{B\}'_\b}\over s_{i\{B\}'_1\cdots \{B\}'_{\b}}}\Big)_{\{\mu\}_{22}}\,,~~\label{YM-2split-propa}
\eea
where the indices $\mu_i$ and $\mu_j$ are contracted with $\epsilon_i^{\mu_i}$ and $\epsilon_j^{\mu_j}$, respectively.
Notice that the kinematic condition \eref{kine-condi-2split-YM} which restricts $\epsilon_{i,j,k}$ to lie in ${\cal S}_A$ removes one of two parts in \eref{fac-propa+v-YM}, therefore yields the factorization structure in \eref{YM-2split-propa}.
For the given divisions \eref{div-AB-NLSM}, the tensor $f^{\rm YM}(R)$ takes the form
\bea
\big[\,f^{\rm YM}(R)\,\big]^{\{\mu\}_1\{\mu\}_2\{\mu\}_3}&=&\epsilon_i^{\mu_i}\,\epsilon_{j}^{\mu_j}\,\Big(\prod_{\a=1}^r\,\W{\cal J}^{\rm YM}_{A_\a}\Big)^{\{\mu\}_a}\,\Big(\prod_{\b=1}^h\,\W{\cal J}^{\rm YM}_{B_\b}\Big)^{\{\mu\}_b}\,,~~\label{f(R)-YM}
\eea
which automatically factorizes as
\bea
\big[\,f^{\rm YM}(R)\,\big]^{\{\mu\}_1\{\mu\}_2\{\mu\}_3}=\big[f^{\rm YM}_A(R)\big]^{\mu_i\mu_j\{\mu\}_a}\,\times\,\big[f^{\rm YM}_B(R)\big]^{\{\mu\}_b}\,,~~\label{fac-f-YM}
\eea
where
\bea
\big[f^{\rm YM}_A(R)\big]^{\mu_i\mu_j\{\mu\}_a}=\epsilon_i^{\mu_i}\,\epsilon_{j}^{\mu_j}\,\Big(\prod_{\a=1}^r\,\W{\cal J}^{\rm YM}_{A_\a}\Big)^{\{\mu\}_a}\,,~~~~~~~~\big[f^{\rm YM}_B(R)\big]^{\{\mu\}_b}=\Big(\prod_{\b=1}^h\,\W{\cal J}^{\rm YM}_{B_\b}\Big)^{\{\mu\}_b}\,.
\eea
The factorization behaviors in \eref{YM-2split-propa} and \eref{fac-f-YM} are parallel to the ${\rm Tr}(\phi^3)$ and NLSM cases, as illustrated in Fig.\ref{Fig6}, but with cubic vertices replaced by general vertices carrying $A$-sets and $B$-sets.

To complete the reconstruction for $2$-split, we now study the behavior of the term $V_v$ from the special vertex $v$.
There are three vectors $J_{i|\{A\}_1\cdots\{A\}_p|\{B\}_1\cdots\{B\}_q}$, $J_{j|\{A\}'_1\cdots\{A\}'_m|\{B\}'_1\cdots\{B\}'_l}$, and $\W{\cal J}_{B(k)}$ all contracted at the vertex $v$. We begin by analyzing some properties of these three vectors. Under the constraint of the kinematic condition \eref{kine-condi-2split-YM}, $\epsilon_i$ and $\epsilon_j$ are equivalent to $\epsilon_i^{{\cal S}_A}$ and $\epsilon_j^{{\cal S}_A}$, therefore
\bea
&&J_{i|\{A\}^\rho_1\cdots\{A\}_p|\{B\}_1\cdots\{B\}_q}\,\sim\,J^{{\cal S}_A;\rho}_{i|\{A\}_1\cdots\{A\}_p|\{B\}_1\cdots\{B\}_q}\,,\nn
&&J^\rho_{j|\{A\}'_1\cdots\{A\}'_m|\{B\}'_1\cdots\{B\}'_l}\,\sim\,J^{{\cal S}_A;\rho}_{j|\{A\}'_1\cdots\{A\}'_m|\{B\}'_1\cdots\{B\}'_l}\,.~~\label{identify-J}
\eea
From the discussion in the previous subsection, we know that the vectors $J^{{\cal S}_A}_{i|\{A\}_1\cdots\{A\}_p|\{B\}_1\cdots\{B\}_q}$ and $J^{{\cal S}_A}_{j|\{A\}'_1\cdots\{A\}'_m|\{B\}'_1\cdots\{B\}'_l}$ annihilate all kinematic variables from the subspace ${\cal S}_B$. Then \eref{identify-J} implies that the vectors $J_{i|\{A\}_1\cdots\{A\}_p|\{B\}_1\cdots\{B\}_q}$ and $J_{j|\{A\}'_1\cdots\{A\}'_m|\{B\}'_1\cdots\{B\}'_l}$ also have this property. On the other hand, the kinematic condition \eref{kine-condi-2split-YM} together with the on-shell condition $\epsilon_k\cdot k_k=0$ implies that the polarization vector $\epsilon_k$ cannot be contracted at any vertex on $L_{(k,v)}$ other than $v$. Consequently, $\W{\cal J}_{B(k)}$ takes the form
\bea
\W{\cal J}^\mu_{B(k)}\,&\xrightarrow[]{\eref{kine-condi-2split-YM}}&\,j_{B(k)}\,\epsilon_k^\mu\,,~~\label{decom-JBk}
\eea
where $j_{B(k)}$ is a Lorentz scalar. This means that the particle propagating in $L_{(k,v)}$ behaves like a scalar. This is precisely in line with the dimensional reduction perspective: from the viewpoint of the subspace ${\cal S}_B$, the polarization vector $\epsilon_k$ exists in the extra dimensions, so the particle $k$ behaves like a scalar.

\begin{figure}
  \centering
   \includegraphics[width=14cm]{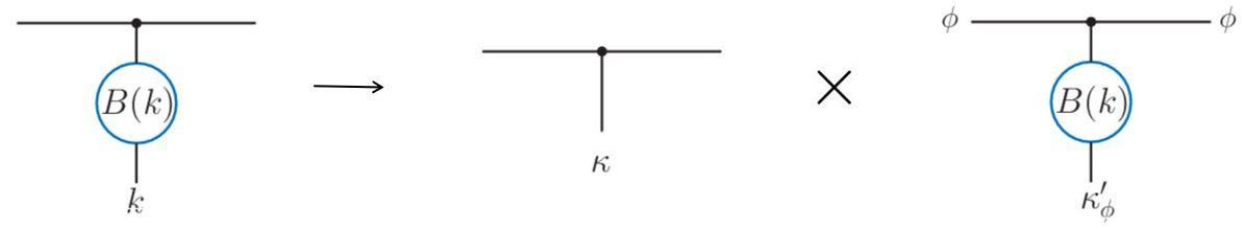} \\
  \caption{First case of $V_v$. The scalars are labeled by $\phi$.}\label{Fig21}
\end{figure}

\begin{figure}
  \centering
   \includegraphics[width=14cm]{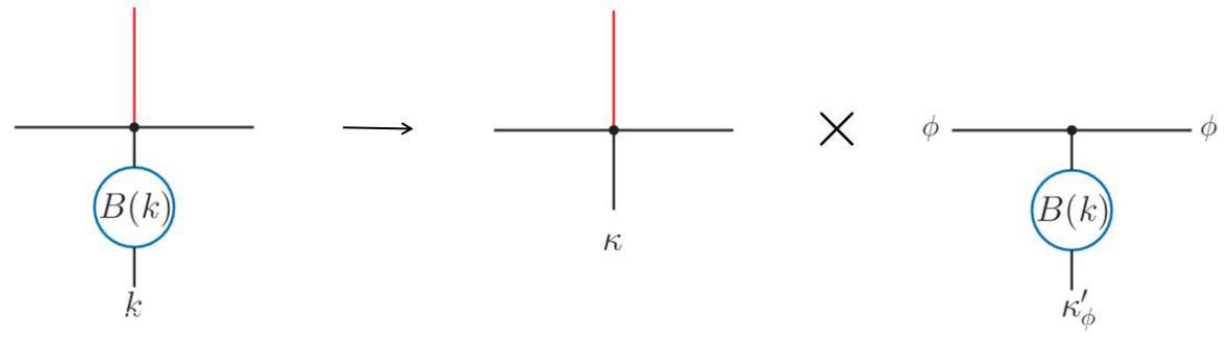} \\
  \caption{Second case of $V_v$. The scalars are labeled by $\phi$.}\label{Fig22}
\end{figure}

\begin{figure}
  \centering
   \includegraphics[width=11.5cm]{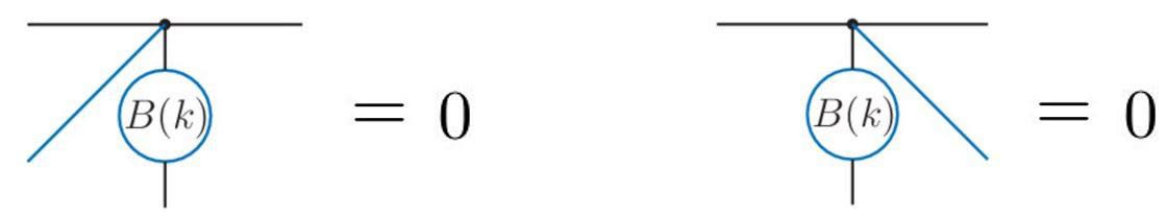} \\
  \caption{Third case of $V_v$. The scalars are labeled by $\phi$.}\label{Fig23}
\end{figure}

Based on the above discussion of the three vectors, we now analyze the three cases of the vertex $v$ shown in Fig.\ref{Fig21}, Fig.\ref{Fig22}, and Fig.\ref{Fig23} in turn. In the first case, where $v$ is a cubic vertex receives only $L_{(i,v)}$, $L_{(j,v)}$ and $L_{k,v}$, we obtain
\bea
&&J^\tau_{i|\{A\}_1\cdots\{A\}_p|\{B\}_1\cdots\{B\}_q}\,J^\nu_{j|\{A\}'_1\cdots\{A\}'_m|\{B\}'_1\cdots\{B\}'_l}\,\W{\cal J}^{\mu}_{B(k)}\,\big[V^{3g}_{v}\big]_{\tau\nu\mu}\nn
&\xrightarrow[]{\eref{kine-condi-2split-YM}}&\,\Big(\,J^\tau_{i|\{A\}_1\cdots\{A\}_p|\{B\}_1\cdots\{B\}_q}\,
J^\nu_{j|\{A\}'_1\cdots\{A\}'_m|\{B\}'_1\cdots\{B\}'_l}\,\epsilon_k^{\mu}\,
\big[V^{3g}_{v}\big]_{\tau\nu\mu}\,\Big)\,\times\,\Big(\,j_{B(k)}\,\Big)\nn
&=&\Big(\,J^\tau_{i|\{A\}_1\cdots\{A\}_p|\{B\}_1\cdots\{B\}_q}\,J^\nu_{j|\{A\}'_1\cdots\{A\}'_m|\{B\}'_1\cdots\{B\}'_l}\,\epsilon_k^{\mu}\,
\big[V^{3g}_{v}\big]_{\tau\nu\mu}\,\Big)\,\times\,\Big(\,j_{B(k)}\,V^{3\phi}_v\Big)\,,
\eea
as illustrated in Fig.\ref{Fig21}. At first glance, $V_v^{3g}$ contracts with $\W{\cal J}_{B(k)}$, so information from the subspace ${\cal S}_A$ becomes entangled with information from the subspace ${\cal S}_B$ at the vertex $v$. However, the decomposition \eref{decom-JBk} renders the vertex $v$ no longer able to perceive information from the subspace ${\cal S}_B$.

In the second case, $v$ is a quartic vertex, and the fourth line is an $A$-line connecting to a block $A_\a$. For this case, we have
\bea
&&J^\tau_{i|\{A\}_1\cdots\{A\}_p|\{B\}_1\cdots\{B\}_q}\W{\cal J}^\rho_{A_\a}\,J^\nu_{j|\{A\}'_1\cdots\{A\}'_m|\{B\}'_1\cdots\{B\}'_l}\,\W{\cal J}^{\mu}_{B(k)}\,\big[V^{4g}_{v}\big]_{\tau\rho\nu\mu}\nn
&\xrightarrow[]{\eref{kine-condi-2split-YM}}&\,\Big(\,J^\tau_{i|\{A\}_1\cdots\{A\}_p|\{B\}_1\cdots\{B\}_q}\,\W{\cal J}^\rho_{A_\a}\,J^\nu_{j|\{A\}'_1\cdots\{A\}'_m|\{B\}'_1\cdots\{B\}'_l}\,\epsilon_k^{\mu}\,
\big[V^{4g}_{v}\big]_{\tau\rho\nu\mu}\,\Big)\,\times\,\Big(\,j_{B(k)}\,V^{3\phi}_v\Big)\,,
\eea
as illustrated in Fig.\ref{Fig22}. Again, information from ${\cal S}_B$ is decoupled from the vertex $v$.

The third case is shown in Fig.\ref{Fig23}. In this case $v$ is a quartic vertex, and the fourth line is a $B$-line connecting to a block $B_\b$. In this case, we have
\bea
J^\tau_{i|\{A\}_1\cdots\{A\}_p|\{B\}_1\cdots\{B\}_q}\,J^\nu_{j|\{A\}'_1\cdots\{A\}'_m|\{B\}'_1\cdots\{B\}'_l}\,\W{\cal J}^{\mu}_{B(k)}\,\W{\cal J}^\rho_{B_\b}\,\big[V^{4g}_{v}\big]_{\tau\nu\mu\rho}\,&\xrightarrow[]{\eref{kine-condi-2split-YM}}&\,0\,,
\eea
since
\bea
J_{i|\{A\}_1\cdots\{A\}_p|\{B\}_1\cdots\{B\}_q}\cdot\W{\cal J}_{B_\b}=0\,,~~~~J_{j|\{A\}'_1\cdots\{A\}'_m|\{B\}'_1\cdots\{B\}'_l}\cdot\W{\cal J}_{B_\b}=0\,,~~~~\epsilon_k\cdot\W{\cal J}_{B_\b}=0\,.
\eea

In summary, the behavior of vertex v always satisfies
\bea
V_v\,=\,V^{{\cal S}_A}\,\times\,V^{3\phi}\,,~~\label{fac-V-YM}
\eea
where $V^{{\cal S}_A}_v$ in the result only contracts with vectors from the subspace ${\cal S}_A$.
Plugging the factorization structures in \eref{YM-2split-propa}, \eref{fac-f-YM} and \eref{fac-V-YM} into \eref{YM-forsplit}, we ultimately get the $2$-split,
\bea
{\cal A}_{2n}^{\rm YM}(1,\cdots,n)\,&\xrightarrow[]{\eref{kine-condi-2split-YM}}&\,\epsilon_k^{\mu_k}\big[{\cal J}^{\rm YM}_{n_1}(i,\pmb A,j,\kappa)\big]_{\mu_k}\,
\times\,{\cal J}^{{\rm YM}\oplus{\rm Tr}(\phi^3)}_{n+3-n_1}(j_\phi,\pmb B(\kappa'_{\phi}),i_\phi)\,,
\eea
where
\bea
\big[{\cal J}^{\rm YM}_{n_1}(i,\pmb A,j,\kappa)\big]_{\mu}\,&=&\,\sum_{{\rm div}\pmb A}\,\sum_{{\cal P}_{\pmb A}}\,\epsilon_i^{\mu_i}\,\epsilon_{j}^{\mu_j}\,\Big(\prod_{\a=1}^r\,\W{\cal J}^{\rm YM}_{A_\a}\Big)^{\{\mu\}_a}\nn
&&\Big(\prod_{\a=1}^p\,{V_{\{A\}_\a}\over s_{i\{A\}_1\cdots \{A\}_{\a}}}\Big)_{\mu_i\{\mu\}_{11}}\,
\Big(\prod_{\a=1}^m\,{V_{\{A\}_\a}\over s_{i\{A\}_1\cdots \{A\}_{\a}}}\Big)_{\mu_j\{\mu\}_{21}}\,\big[V_v\big]_{\{\mu\}_3}\,,\nn
{\cal J}^{{\rm YM}\oplus{\rm Tr}(\phi^3)}_{n+3-n_1}(j_\phi,\pmb B(\kappa'_{\phi}),i_\phi)&=&\sum_{{\rm div}\pmb B}\,\sum_{{\cal P}_{\pmb B}}\,\Big(\prod_{\b=1}^h\,\W{\cal J}^{\rm YM}_{B_\b}\Big)^{\{\mu\}_b}\Big|_{\W{\cal J}^{\mu_k}_{B(k)}\to j_{B(k)}}\nn
&&\Big(\prod_{\b=1}^q\,{V_{\{B\}_\b}\over s_{i\{B\}_1\cdots \{B\}_{\b}}}\Big)_{\{\mu\}_{12}}\,\Big(\prod_{\b=1}^l\,{V_{\{B\}'_\b}\over s_{i\{B\}'_1\cdots \{B\}'_{\b}}}\Big)_{\{\mu\}_{22}}\,,
\eea
where the lower index $\mu$ of the first current originates from $[V_v]_{\{\mu\}_3}$.

\section{Conclusion and discussion}
\label{sec-conclu}

In this paper, we generalized a mechanism of Feynman diagrams---denoted as SFASL---to the cases of NLSM and YM. This mechanism was previously found in our earlier work \cite{Zhou:2024ddy} to be satisfied by the Feynman diagrams of ${\rm Tr}(\phi^3)$ model, and the generalization was achieved by extending the pattern of shuffle permutations. Through this generalized SFASL, we further interpreted the hidden zeros and $2$-split structures of tree-level amplitudes in NLSM and YM. In the extended shuffle permutations, mixed vertices $V_{\{A\}|\{B\}}$ emerged. We have showed that these mixed vertices are canceled by the non-commuting parts of the unmixed vertices. Ultimately, the SFASL reduces the hidden zeros to the on-shell condition $k_j^2=0$, and causes the $2$-splits by separating each Feynman diagram along $L_{(i,v)}$ and $L_{(j,v)}$. Although the verification procedure was rather involved and lengthy, the underlying picture of the SFASL and the cancellation of mixed vertices is remarkably concise. Synthesizing our present results with previous work \cite{Zhou:2024ddy}, we now know that the hidden zeros and $2$-splits of tree amplitudes for the three models---${\rm Tr}(\phi^3)$, NLSM, and YM---can all be universally interpreted from the SFASL.

We naturally expect that the SFASL can also be used to interpret the hidden zeros and $2$-splits in other models, including SG, DBI, GR, as well as YM and GR with specific higher-derivative corrections. At the very least, it is easy to verify that the mass dimensions of the vertices in these models all satisfy the required constraints \eref{mass-dim-v}. Furthermore, it is straightforward to see that for amplitudes without ordering, the shuffle permutations along $L_{(i,\bullet)}$ discussed in this paper also applies, provided that the sets $\pmb A$ and $\pmb B$ (unordered) are appropriately chosen. This is because, one can always fix the order of $A$-sets and the order of $B$-sets, while absorbing the permutations among $A$-sets and the permutations among $B$-sets into the summation over different divisions. For standard GR amplitudes, and YM amplitudes with specific higher-derivative corrections, we have made some progress, but several technical difficulties remain unresolved. Therefore, we will report these results in a forthcoming paper.

Suppose we prove that all currently known models exhibiting hidden zeros and $2$-splits satisfy the SFASL, and that their hidden zeros and $2$-splits can be explained by it. Then, although logically the SFASL is only a sufficient condition for these structures rather than a necessary one, we can at least conclude that the SFASL governs the presence of hidden zeros and $2$-split in a large class of physical models.
As discussed in section \ref{sec-intro}, this allows us to reverse the logical direction. Instead of starting from a model and testing the SFASL, we can ask what constraints the SFASL imposes on physical models---for example, on their Lagrangians or Feynman rules. For instance, we may ask: what requirements must the interaction vertices satisfy so that mixed vertices are canceled by the non-commuting parts of unmixed vertices? If these questions can be answered, then when encountering a new physical model, we may be able to quickly determine whether its tree-level amplitudes possess hidden zeros and $2$-split structures.

Finally, as noted in section \ref{sec-intro}, our recent discovery of hidden zeros and $2$-split in loop-level Feynman integrands of ${\rm Tr}(\phi^3)$ model was achieved by a method that directly generalizes the tree-level SFASL for ${\rm Tr}(\phi^3)$ \cite{Zhou:2026ukg}. Given that the tree-level SFASL can be extended to other models such as NLSM and YM, it is reasonable to expect that a similar generalization also holds at loop-level. This suggests that, through the SFASL, we may be able to extend the hidden zeros and $2$-split of ${\rm Tr}(\phi^3)$ Feynman integrands to other physical models.

\section*{Acknowledgments}

This work is supported by NSFC under Grant No. 11805163.

\bibliographystyle{JHEP}

\bibliography{reference}

\end{document}